\documentclass[12pt, draftclsnofoot, onecolumn]{IEEEtran}

\usepackage{cite}
\usepackage{amsmath}
\usepackage{array}
\usepackage{graphicx}
\usepackage{amsfonts}
\usepackage{caption}
\usepackage{algorithm}
\usepackage{algorithmic}
\usepackage{subfigure}
\usepackage{bm}
\usepackage{color}
\usepackage{booktabs}
\usepackage{float}
\usepackage{lipsum}

\makeatletter
\newenvironment{breakablealgorithm}
{\begin{center}
		\refstepcounter{algorithm}
		\hrule height.8pt depth0pt \kern2pt
   		\renewcommand{\caption}[2][\relax]{
		{\raggedright\textbf{\ALG@name~\thealgorithm} ##2\par}
    	\ifx\relax##1\relax
		\addcontentsline{loa}{algorithm}{\protect\numberline{\thealgorithm}##2}
		\else
		\addcontentsline{loa}{algorithm}{\protect\numberline{\thealgorithm}##1}
		\fi
		\kern2pt\hrule\kern2pt}}{
		\kern2pt\hrule\relax
\end{center}}
\makeatother

\newtheorem{remark}{Remark}
\newtheorem{prop}{Proposition}

\begin{document}

\title{Next-Generation URLLC with Massive Devices: \!\!\! \\ A Unified Semi-Blind Detection Framework for Sourced and Unsourced Random Access}

\author{Malong~Ke, Zhen~Gao,~\IEEEmembership{Member,~IEEE,} Mingyu Zhou, Dezhi Zheng,\\ Derrick Wing Kwan Ng,~\IEEEmembership{Fellow, IEEE,} and H. Vincent Poor,~\IEEEmembership{Life Fellow, IEEE}
\thanks{Part of the paper has been presented at the 2022 IEEE International Wireless Communications and Mobile Computing Conference (IWCMC), where the proposed solution is limited to grant-free sourced RA for mMTC \cite{Ref_IWCMC_Ke'22}.}
\thanks{M. Ke, Z. Gao, and D. Zheng are with the School of Information and Electronics, Beijing Institute of Technology, 100081 Beijing, China (e-mails: kemalong@bit.edu.cn; gaozhen16@bit.edu.cn; zhengdezhi@bit.edu.cn).
M. Zhou is with Baicells Technologies Co. Ltd., Beijing 100089, China (e-mail: zhoumingyu@baicells.com).	
D. W. K. Ng is with the School of Electrical Engineering and Telecommunications, University of New South Wales, 2052 Sydney, Australia (e-mail: w.k.ng@unsw.edu.au).
H. Vincent Poor is with the Department of Electrical and Computer Engineering, Princeton University, NJ 08542 Princeton, USA (e-mail: poor@princeton.edu).}}

% the paper headers
\markboth{Journal of \LaTeX\ Class Files, Vol. xx, No. xx, January 2023}
{Shell \MakeLowercase{\textit{et al.}}: Bare Demo of IEEEtran.cls for IEEE Journals}



\maketitle

\vspace{-14mm}
\begin{abstract}
This paper proposes a unified semi-blind detection framework for sourced and unsourced random access (RA), which enables next-generation ultra-reliable low-latency communications (URLLC) with massive devices.
Specifically, the active devices transmit their uplink access signals in a grant-free manner to realize ultra-low access latency.
Meanwhile, the base station aims to achieve ultra-reliable data detection under severe inter-device interference without exploiting explicit channel state information (CSI).
We first propose an efficient transmitter design, where a small amount of reference information (RI) is embedded in the access signal to resolve the inherent ambiguities incurred by the unknown CSI.
At the receiver, we further develop a successive interference cancellation-based semi-blind detection scheme, where a bilinear generalized approximate message passing algorithm is utilized for joint channel and signal estimation (JCSE), while the embedded RI is exploited for ambiguity elimination.
Particularly, a rank selection approach and a RI-aided initialization strategy are incorporated to reduce the algorithmic computational complexity and to enhance the JCSE reliability, respectively.
Besides, four enabling techniques are integrated to satisfy the stringent latency and reliability requirements of massive URLLC.
Numerical results demonstrate that the proposed semi-blind detection framework offers a better scalability-latency-reliability tradeoff than the state-of-the-art detection schemes dedicated to sourced or unsourced RA.
\end{abstract}

\begin{IEEEkeywords}
\vspace{-2mm}
Massive URLLC, grant-free, sourced/unsourced random access, semi-blind detection, approximate message passing.
 \end{IEEEkeywords}

\section{Introduction}
\label{Section I}

\subsection{Background and Motivation}
\label{Section I-A}

\IEEEPARstart{T}{he} emerging Internet-of-Things (IoT) applications in various vertical sectors have driven the massive machine-type communication (mMTC) and ultra-reliable low-latency communication (URLLC) services in the fifth-generation (5G) cellular systems, which pursue scalability and reliability with low user plane latency, respectively \cite{Ref_IWCMC_Ke'22, Ref_IOTJ_Nguyen'22, Ref_Tcom_Popovski'19}.
{\color{red} Motivated by the grander Internet-of-Everything (IoE) that is envisioned to connect millions of people and billions of machines, the next-generation, i.e., Beyond 5G or sixth-generation (6G), cellular systems must further scale the classical URLLC across the device dimension, leading to a new massive URLLC service that merges legacy mMTC and URLLC \cite{Ref_Network_Saad'20}.
The application scenarios range from extended reality (XR) services to flying vehicles, brain-computer interfaces, and connected autonomous systems.
Although the conventional network slicing is effective in supporting a simple mixture of mMTC and URLLC, it is still very challenging to simultaneously satisfy the stringent scalability, latency, and reliability requirements (e.g., 10$^6$ devices/km$^2$, 1 ms user plane latency, and 99.99999$\%$ reliability) of massive URLLC \cite{Ref_Access_Popovski'18, Ref_Access_Pokhrel'20}.}

More specifically, the unprecedentedly high density of wireless devices has already posed great challenges in random access (RA), which is essential for ensuring ubiquitous IoE connectivity \cite{Ref_JSAC_Chen'21, Ref_WCM_Wu'20, Ref_CS&T_Laya'18}.
In legacy cellular systems, the widely adopted grant-based RA protocol requires multiple signaling interactions to facilitate the scheduling of interference-free transmissions \cite{Ref_WCM_Wu'20}.
Despite its simplicity and reliability, this protocol would become inefficient or even impractical in the context of massive URLLC due to its extremely high access latency resulting from severe access collisions among the massive devices \cite{Ref_CS&T_Laya'18}.
To tackle this issue, the promising grant-free RA protocol has been recently proposed as a key enabler to achieve ultra-low access latency, where the active devices directly transmit their access signals to the base station (BS) without any scheduling in advance \cite{Ref_CL_Zhang'16}.
However, since the signals of all the active devices are transmitted via the same physical resources, the inter-device interference becomes a severely limiting factor for realizing ultra-reliable data detection.
Therefore, the key challenge of massive URLLC lies in the improvement of data detection reliability for grant-free massive RA (MRA) \cite{Ref_IoTJ_Shao'19}.

{\color{red} In general, grant-free MRA can be classified into two paradigms, i.e., sourced and unsourced RA, which focus on two practical RA scenarios having different access requirements \cite{Ref_TSP_Shao'20}.
For sourced RA, the BS is interested in both the transmitted messages and the identities (IDs) of the devices that generated them.
Hence, some reference information (RI), such as pilot sequence, should be transmitted along with the payload data for device identification.
While for unsourced RA, the BS is solely interested in estimating a list of sent messages, without any interest in the identities of the transmitters.
Therefore, the payload efficiency can be improved by omitting the device ID information in the transmission.
Considering their different access requirements, the research community has developed two independent lines of research to study the reliable data detection for grant-free sourced and unsourced RA, respectively.
However, designing a unified data detection framework for incorporating both RA paradigms is still an open issue, which is indispensable to satisfy the heterogeneous access requirements of future IoE applications \cite{Ref_TSP_Shao'20}.}
Meanwhile, the previous works generally focus on the traditional mMTC and fail to support the emerging massive URLLC that simultaneously pursues the stringent scalability, latency, and reliability requirements \cite{Ref_Access_Popovski'18, Ref_Access_Pokhrel'20}.

\vspace{-4mm}
\subsection{Related Work}
\label{Section I-B}

Grant-free sourced RA has been intensively investigated in the literature, e.g., \cite{Ref_IoTJ_Shao'19, Ref_TSP_Shao'20, Ref_WCM_Ke'21, Ref_CL_Shim'12, Ref_TSP_Liu'18, Ref_TSP_Shao'20-2, Ref_TSP_Ke'20}, where the non-orthogonal pilot-based coherent detection framework is generally considered.
Specifically, each active device transmits a non-orthogonal pilot sequence along with its payload data to the BS in a grant-free manner.
Meanwhile, the BS first performs active device detection (ADD) and channel estimation (CE) based on the received pilot signal, then the acquired results are adopted for the subsequent coherent data detection \cite{Ref_IoTJ_Shao'19}.
A key feature of massive URLLC is the sporadic uplink traffic, i.e., for any given time interval, only a small number of devices are activated by external events and desire to access the network~\cite{Ref_TSP_Shao'20, Ref_WCM_Ke'21}.
By leveraging the sparse device activity, the authors in \cite{Ref_CL_Shim'12} formulated the joint ADD and CE design as a compressive sensing (CS) problem and an orthogonal matching pursuit-based algorithm was developed for the related sparse signal recovery.
However, this work assumes only a single-antenna receiver at the BS and the solution is not applicable to multi-antenna systems.  
Also, the work in \cite{Ref_TSP_Liu'18} revealed that the detection error probability of ADD can be driven to zero as the number of BS antennas is sufficiently large.
On the other hand, to reduce the computational complexity in the case of large numbers of devices and BS antennas, a dimension reduction-based joint ADD and CE approach was further proposed in~\cite{Ref_TSP_Shao'20-2}.
Particularly, the massive multiple-input multiple-output (MIMO) channels between the devices and the BS usually exhibit clustered sparsity in the virtual angular domain \cite{Ref_TSP_Gao'15}.
In this context, the authors in \cite{Ref_TSP_Ke'20} developed an approximate message passing (AMP)-based ADD and CE scheme to leverage the angular-domain clustered sparsity for further enhanced MRA performance.
Overall, the previous works on grant-free sourced RA generally focus on the scalability of the traditional mMTC service, where the transmission latency (or pilot length) must increase linearly with the number of active devices to guarantee the reliable data detection \cite{Ref_TSP_Ke'20}.
Therefore, it is challenging for them to simultaneously satisfy the stringent latency and reliability requirements of massive URLLC.
%The works in \cite{Ref_CL_Shim'12, Ref_ISWCS_Schepker'13, Ref_TSP_Liu'18, Ref_TSP_Shao'20-2, Ref_TSP_Ke'20} focus on single-cell MRA, which is limiting to serve massive IoT devices with broad distribution. 
%In \cite{Ref_JSAC_Ke'21}, the authors proposed to employ cell-free massive MIMO to achieve massive IoT connectivity, where multiple access points cooperate to provide ubiquitous and uniformly good services for all devices.

Recent studies on grant-free unsourced RA mainly rely on the common codebook-based non-coherent detection framework introduced in \cite{Ref_ISIT_Polyanskiy'17}.
Specifically, according to the payload data bits to be transmitted, each active device sends a codeword selected from a common codebook.
Unlike the sourced RA counterpart, the BS in this case is solely interested in estimating a list of sent messages without any interests in the identities of the transmitters, i.e., the estimated messages have an unknown permutation.
The main obstacle of realizing the scheme stems from the extremely large size of the codebook, i.e., the number of codewords, which grows exponentially with respect to the payload data length and causes prohibitive computational complexity \cite{Ref_ISIT_Polyanskiy'17}.
To overcome this limitation, the first low-complexity coding scheme for unsourced RA was proposed in \cite{Ref_ISIT_Ordentlich'17}, where the transmission period was divided into multiple small sub-blocks and each active device randomly chose a sub-block to transmit its codeword.
Relying on a similar transmission structure, the subsequent work in \cite{Ref_Tcom_Vem'19} further proposed a close-to-optimal coding strategy, where user-independent successive interference cancellation (SIC) was applied for improved decoding performance.
Subsequently, the authors in \cite{Ref_TIT_Amalladinne'20} proposed another efficient approach, which leveraged recent advances in the CS field to further reduce the decoding complexity.
For this scheme, the message of each active device is split into several sub-messages and the coding scheme is divided into two parts, i.e., inner and outer encoder/decoder.
Here, a CS-based inner encoder/decoder is adopted to map a sub-message into a codeword at the devices and estimate the transmitted sub-messages at the BS, as in~\cite{Ref_ISIT_Polyanskiy'17}.
Meanwhile, a tree-based outer decoder is employed to acquire the original messages by stitching the estimated sub-messages together.
The works in \cite{Ref_ISIT_Polyanskiy'17, Ref_ISIT_Ordentlich'17, Ref_Tcom_Vem'19, Ref_TIT_Amalladinne'20} consider a Gaussian multiple access channel model, where the BS is equipped with a single-antenna and the channel gains between the devices and the BS are assumed to be unity.
Although this assumption facilitates the performance analysis of the proposed coding scheme, it hinders the practical application of the results.
Moreover, the authors in~\cite{Ref_arXiv_Fenler'19}, \cite{Ref_TIT_Fengler'21} revealed that the required transmit power-per-bit can be driven to an arbitrarily small value as the number of BS antennas grows sufficiently large.
Considering the emerging massive MIMO systems, an uncoupled CS-based unsourced RA solution was proposed, which exploited the rich spatial dimensionality offered by the large-scale antenna array to enhance the decoding performance \cite{Ref_JSAC_Shyianov'21}.
%Besides, the work was extended to the cell-free massive MIMO network and a cooperative activity detection scheme was developed to address the inner decoding problem \cite{Ref_TSP_Shao'20}.
The strong common characteristic of the aforementioned works lies in the employment of the coding scheme based on a common codebook.
%  in \cite{Ref_ISIT_Polyanskiy'17, Ref_ISIT_Ordentlich'17, Ref_Tcom_Vem'19, Ref_TIT_Amalladinne'20, Ref_arXiv_Fenler'19, Ref_TIT_Fengler'21, Ref_JSAC_Shyianov'21}
%Although many previous works have confused unsourced RA and the common codebook-based coding scheme, we aim to clarify in this paper that they are not exactly equivalent.
%Indeed, the unsourced RA mainly refers to the RA paradigm where the BS is solely interested in recovering the collection of sent messages without regard for the identities of individual sources.
%While the common codebook-based coding scheme is an attractive transmission scheme to realize unsourced RA.
It is also challenging for them to simultaneously satisfy the stringent latency and reliability requirements of massive URLLC due to the low payload efficiency or the high computational complexity resulting from the employment of the common codebook-based coding scheme \cite{Ref_JSAC_Shyianov'21, Ref_Access_Pokhrel'20}. 

{\color{red} In previous works, the traditional sourced and unsourced RA paradigms generally adopt their dedicated data detection frameworks, i.e., coherent and non-coherent detection, respectively, which rely on different transceiver designs, cf. \cite{Ref_TSP_Ke'20, Ref_TIT_Amalladinne'20}.
The authors in \cite{Ref_TSP_Shao'20} have tried to support both sourced and unsourced RA services in the same IoE system.
However, the two RA paradigms still adopt their dedicated data detection frameworks, which rely on different transmission schemes, signal models, and data detection schemes.
Here, only the related activity detection algorithm is unified.
In this context, we have to integrate two different transceivers into the same system, allowing the network to switch between sourced and unsourced RA modes according to practical access requirements.
This solution is unattractive in terms of device size, hardware complexity, and overall cost \cite{Ref_TSP_Shao'20}.
Therefore, a more beneficial unified detection framework is needed, where both RA paradigms can share almost the same RA procedure, transceiver hardware design, and receive algorithm.}

\vspace{-4mm}
\subsection{Main Contributions}
\label{Secrion I-C}
 
In this paper, we design a unified semi-blind detection framework for grant-free sourced and unsourced RA, which pursues the ultra-reliable and low-latency requirements of massive URLLC.
Specifically, the active devices directly transmit their uplink access signals exploiting the same physical resources, where a small amount of RI is embedded in the access signals.
Based on the overlapped received signal, the BS jointly estimates the channels and detects the signals of the active devices, then the embedded RI is exploited to eliminate the inherent ambiguities.
For sourced RA, the RI contains device ID bits, cyclic redundancy check (CRC) bits, and a scalar pilot symbol, which are adopted for eliminating the phase and permutation ambiguities.
While for unsourced RA, only CRC bits and a scalar pilot symbol are transmitted for phase ambiguity elimination, and thus higher payload efficiency can be achieved.
In summary, our main contributions are listed as follows:

\begin{itemize}
\item{{\color{red} We propose a unified semi-blind detection framework for enabling grant-free sourced and unsourced RA, under which both RA paradigms share almost the same RA procedure, transceiver hardware design, and receive algorithm.
Moreover, in contrast to the existing non-orthogonal pilot-based coherent detection for sourced RA \cite{Ref_IoTJ_Shao'19, Ref_TSP_Shao'20, Ref_WCM_Ke'21, Ref_CL_Shim'12, Ref_TSP_Liu'18, Ref_TSP_Shao'20-2, Ref_TSP_Ke'20}, the proposed detection framework results in a significant transmission latency reduction when the same detection reliability is considered.
Furthermore, compared to the common codebook-based non-coherent detection for unsourced RA \cite{Ref_ISIT_Polyanskiy'17, Ref_ISIT_Ordentlich'17, Ref_Tcom_Vem'19, Ref_TIT_Amalladinne'20, Ref_arXiv_Fenler'19, Ref_TIT_Fengler'21, Ref_JSAC_Shyianov'21}, the proposed detection framework dramatically reduces the processing latency by circumventing the common codebook-based coding scheme.}
Due to the reduced transmission and processing latencies, the proposed detection framework achieves a much lower user plane latency than its counterparts \cite{Ref_WCM_Ke'21}.}

%\footnote{Different from fully-blind detection without transmitting any reference information, the proposed framework embeds little reference information in the access signals. Meanwhile, the occupied resources for the reference information is far less than that for the pilot sequence in coherent detection. Hence, the proposed framework is referred to as semi-blind detection framework.}

\item{We propose an SIC-based semi-blind detection scheme at the BS, which mitigates the inter-device interference iteratively. 
In each SIC iteration, the channels and the signals of the active devices are jointly inferred from the overlapped received signal, while the embedded RI is exploited for ambiguity elimination.
Moreover, the signal components of reliably detected active devices are removed from the received signal to alleviate the inter-device interference in the following iterations.}

\item{We propose a bilinear generalized AMP (BiG-AMP)-based joint channel and signal estimation (JCSE) algorithm, where the JCSE is formulated as a matrix factorization problem based on the Bayesian theory and the advanced BiG-AMP algorithm is employed to obtain a low-complexity approximate solution.
{\color{red} Particularly, we develop a rank selection approach to estimate the unknown number of active devices, which facilitates the computational complexity reduction of the BiG-AMP algorithm. 	
Moreover, a RI-aided initialization strategy is further incorporated for improved JCSE reliability.
The proposed algorithm significantly outperforms the classic BiG-AMP algorithm adopting the random initialization strategy~\cite{Ref_TSP_Parker'14}.}}

\item{We introduce four enabling techniques that can be flexibly integrated into the proposed semi-blind detection framework to further reduce the user plane latency and enhance the detection reliability. The obtained URLLC-enhanced version of the proposed detection framework is capable of simultaneously satisfying the stringent scalability, latency, and reliability requirements of massive URLLC.}
\end{itemize}

\vspace{-4mm}
\subsection{Notations}
\label{Section I-D}

We adopt normal-face letters to denote scalars and lowercase (uppercase) boldface letters to denote column vectors (matrices).
The $(n,k)$th element, the $n$th row vector, and the $k$th column vector of the matrix ${\bf G} \in {\mathbb C}^{N \times K}$ are denoted as $g_{n,k}$, $[{\bf G}]_{n,:}$, and $[{\bf G}]_{:,k}$, respectively, where ${\mathbb C}$ is the set of complex numbers.
${\mathbb B}$ is the set of binary numbers and ${\bf 0}_{N \times K}$ is the zero matrix of size ${N \times K}$.
The superscripts $(\cdot)^{\rm T}$, $(\cdot)^*$, $(\cdot)^{\rm H}$ and $(\cdot)^{\dagger}$ represent the transpose, complex conjugate, conjugate transpose, and pseudo-inverse operators, respectively.
$[K]$ denotes the set of integers $\{1, 2, \cdots, K\}$, $|{\cal A}|_c$ is the cardinal number of the set $\cal A$, $\emptyset$ is an empty set, and ${\rm supp}\{\cdot\}$ denotes the support set of a sparse vector or matrix.
$\left\|{\bf G}\right\|_{\rm F}$ denotes the Frobenius-norm of the matrix ${\bf G}$ and $\left\|{\bf G}\right\|_0$ denotes the zero-norm of ${\bf G}$, i.e., the number of non-zero elements in ${\bf G}$.
{\color{red} $\left[{\bf G}\right]_{:,{\cal A}}$ represents the matrix that stacks the columns of ${\bf G}$ indexed by the set ${\cal A}$, while $\left[{\bf G}\right]_{{\cal A},:}$ is the matrix that stacks the rows of ${\bf G}$ indexed by the set ${\cal A}$.}
${\cal R}(\cdot)$ is the real part of a complex number.
$\lceil b \rceil$ rounds $b$ to the nearest integer greater than or equal to $b$.
${\cal U}(x;a,b)$ denotes that the variable $x$ follows the uniform distribution between $a$ and $b$.
Finally, ${\cal CN}\left(x; \mu, v\right)$ denotes the complex Gaussian distribution of a random variable $x$ with mean $\mu$ and variance $v$.
${\mathbb    E}[\cdot]$ and ${\mathbb V}[\cdot]$ denote statistical expectation and variance operators, respectively.

\section{System Model}
\label{Sec_sysModel}

\begin{figure}[!t]
     \centering
     \includegraphics[width=0.5\columnwidth, keepaspectratio]{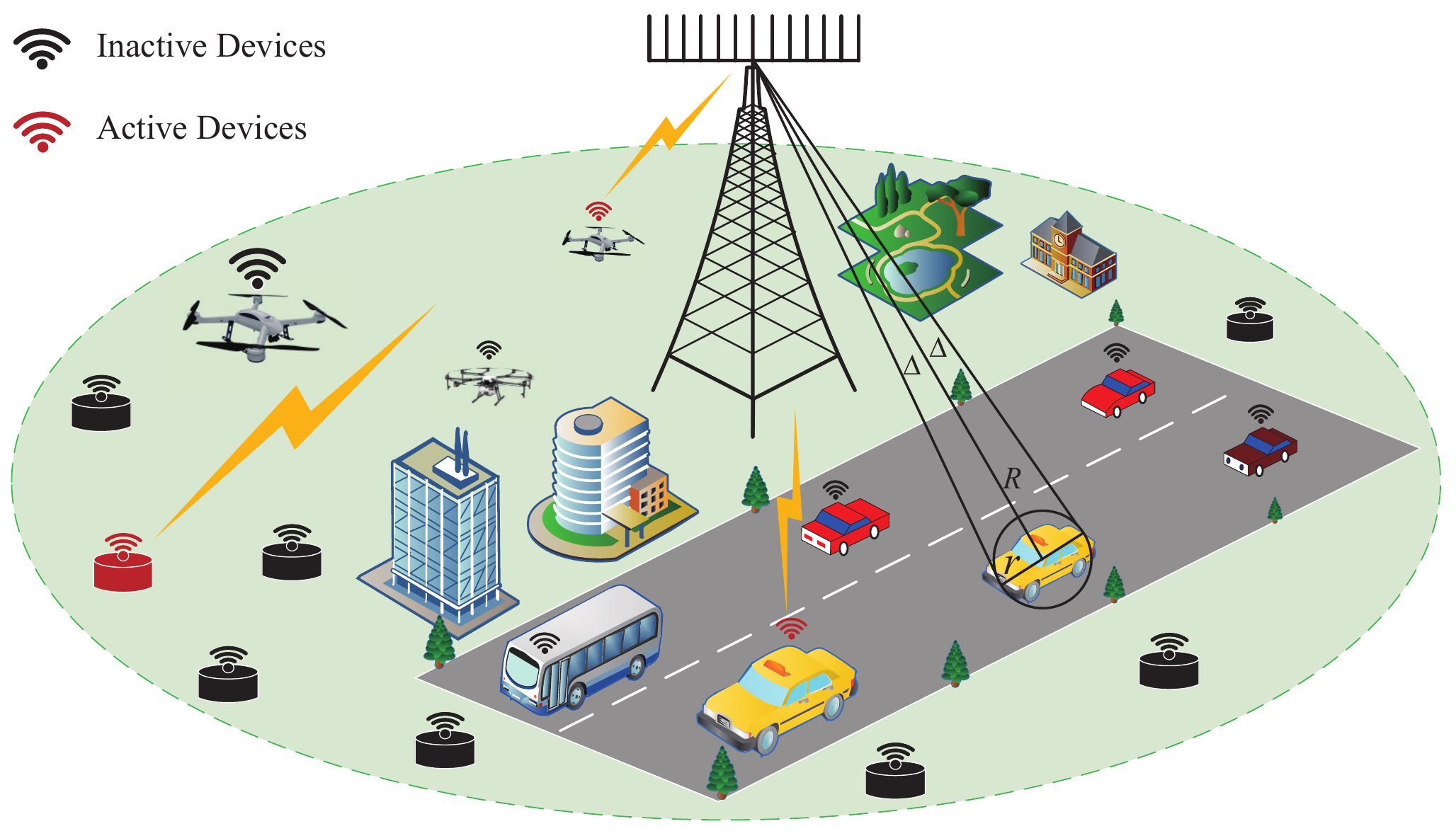}
     \caption{An illustration of future massive URLLC scenarios with sparse device activity. A one-ring channel model is considered between the devices and the massive MIMO BS.}
     \label{Fig1}
     \vspace{-3mm}
\end{figure}

Consider the uplink of a typical massive URLLC scenario in massive MIMO systems, as depicted in Fig.~\ref{Fig1}.
Here, we employ a BS equipped with an $N$-antenna uniform linear array (ULA) to provide access service for $K$ synchronized single-antenna devices.
Due to the sporadic uplink traffic of IoE, it is assumed that only $K_a$ $(K_a \ll K)$ out of the total $K$ devices are activated by external events and desire to access the network \cite{Ref_TSP_Ke'20}.
To avoid the complicated access scheduling for ultra-low access latency, the promising grant-free RA protocol is adopted for uplink transmission, where the active devices directly transmit their access signals to the BS via the same time-frequency resources.
At the BS, the signal ${\bf r}_l \in {\mathbb C}^{N \times 1}$ received in the $l$th symbol duration is expressed as
\begin{equation}\label{Eq_rxSlot}
	\begin{aligned}
		{\bf r}_l = \sum\limits_{k = 1}^K {\bf g}_k{\alpha_k}x_{k,l} + {\bf w}_l = {\bf G}{\bf x}_l + {\bf w}_l,
	\end{aligned}
\end{equation}
where ${\bf g}_k \in {\mathbb C}^{N \times 1}$ denotes the uplink channel between the $k$th device and the BS, the binary variable $\alpha_k$ indicates the device activity, i.e., $\alpha_k = 1$ for active and 0 otherwise, $x_{k,l} \in {\mathbb C}$ is the transmitted signal (i.e., modulated symbol) of the $k$th device in the $l$th symbol duration, ${\bf w}_l \sim {\cal CN}\left({\bf 0}, \sigma^2{\bf I}\right)$ is the additive white Gaussian noise (AWGN), and $\sigma^2$ is the noise variance.
Incorporating both channel response and device activity, ${\bf G} = \left[\alpha_1{\bf g}_1, \alpha_2{\bf g}_2, \cdots, \alpha_K{\bf g}_K\right] \in {\mathbb C}^{N \times K}$ is referred to as the MRA channel matrix and ${\bf x}_l = \left[x_{1,l}, x_{2,l}, \cdots, x_{K,l}\right]^{\rm T} \in {\mathbb C}^{K \times 1}$.
{\color{red} Further focusing on small data packets, the length of the symbol frame $L$ is usually far smaller than the channel coherence time.
Meanwhile, the device activity remains constant during the frame.
In this context, the number of active devices is fixed within each frame but may change across different frames.
For a specific frame, the received signal over $L$ successive symbol durations is given as}
\begin{equation}\label{Eq_rxFrame}
	\begin{aligned}
		{\bf R} = {\bf G}{\bf X} + {\bf W},
	\end{aligned}
\end{equation}   
where ${\bf R} = \left[{\bf r}_1, {\bf r}_2, \cdots, {\bf r}_L\right] \in {\mathbb C}^{N \times L}$, ${\bf X} = \left[{\bf x}_1, {\bf x}_2, \cdots, {\bf x}_L\right] \in {\mathbb C}^{K \times L}$, and ${\bf W} = \left[{\bf w}_1, {\bf w}_2, \cdots, {\bf w}_L\right]$.

Considering the widely studied spatial channel model \cite{Ref_TSP_Ke'20}, the channel between the $k$th device and the BS is modeled as
\begin{equation}\label{Eq_chModel}
	\begin{aligned}
		{\bf g}_k = \rho_k \sum\limits_{p = 1}^{P} \beta_{k,p}{\bf a}_R\left(\phi_{k,p}\right),
	\end{aligned}
\end{equation}    
where $\rho_k$ is the large-scale fading parameter, $P$ is the number of multi-path components (MPCs), $\beta_{k,p}$ denotes the complex gain of the MPC, and ${\bf a}_R\left(\phi_{k,p}\right) = \left[1, e^{-j2\pi\phi_{k,p}}, \cdots, e^{-j2\pi(N-1)\phi_{k,p}}\right]/\sqrt{N}$ is the array response vector at the BS.
Here, $\phi_{k,p} = \frac{d}\lambda{\rm sin}\left(\psi_{k,p}\right)$, where $\psi_{k,p}$ is the physical angle-of-arrival (AoA) associated with the $k$th device and the $p$th MPC, $d$ is the antenna spacing, and $\lambda$ is the wavelength.

\begin{figure}[!t]
	\centering
	\includegraphics[width=0.5\columnwidth, keepaspectratio]{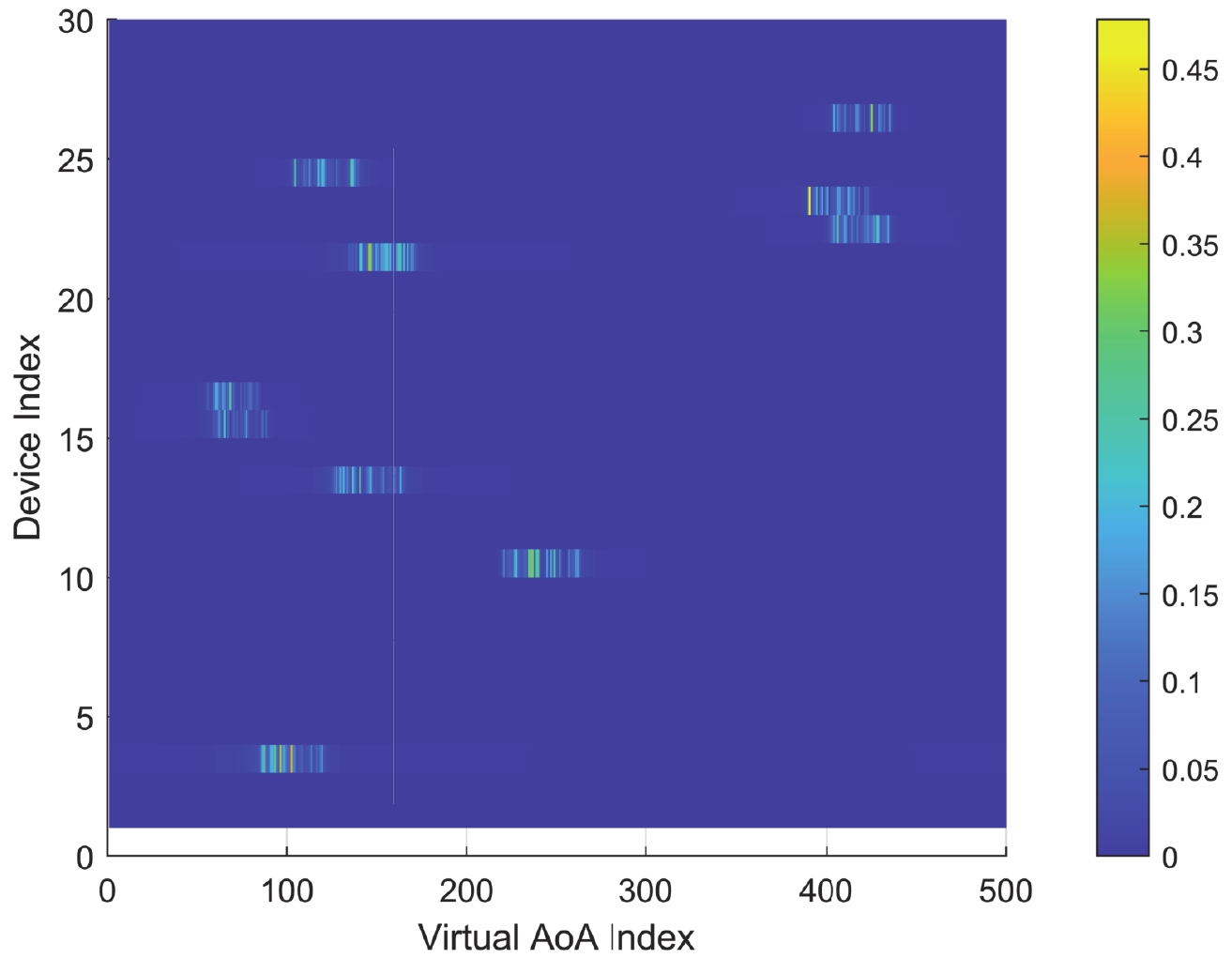}
	\caption{\small{The clustered sparsity of the angular-domain MRA channel matrix ${\bf H}$, where $K = 30$ devices, $K_a = 10$ active devices, and $N = 500$ BS antennas are considered.}}
	\label{Fig2}
	\vspace{-3mm}
\end{figure}

%l\footnote{The rationality of the one-ring channel model lies in the different elevations and surrounding scattering characteristics between devices and the BS \cite{Ref_TSP_Ke'20}.}
{\color{red} For a typical network deployment, the spatial propagation characteristics of the channels between the devices and the BS can be modeled as an one-ring channel model, see Fig.~\ref{Fig1}~\cite{Ref_TSP_Ke'20}.
Here, the MPCs only can be observed within a small angular window at the BS, i.e., $\psi_{k,p} \in \left[\psi_0-\Delta, \psi_0+\Delta\right]$, where $\psi_0$ is the central AoA and $2\Delta \ll 180^{\circ}$  is the angular spread.
Define ${\bf H} = {\bf A}_R{\bf G}$ as the angular-domain representation of the MRA channel matrix ${\bf G}$, where ${\bf A}_R$ denotes the transformation matrix and becomes a discrete Fourier transform matrix for ULA with $d = \lambda/2$.
The limited AoA spread leads to the clustered angular-domain sparsity of massive MIMO channels}, i.e.,
\begin{equation}\label{Eq_chSupp}
	\begin{aligned}
		1< \left|{\rm supp}\left\{\left[{\bf H}\right]_{:,k}\right\}\right|_c \ll N, \forall k \in \left[K\right].
	\end{aligned}
\end{equation}
Moreover, considering the sparse device activity, we further have 
\begin{equation}\label{Eq_chSupp2}
	\begin{aligned}
		\left|{\rm supp}\left\{\left[{\bf H}\right]_{n,:}\right\}\right|_c \ll K_a, \forall n \in \left[N\right].
	\end{aligned}
\end{equation} 
By combining the sparsity features presented in (\ref{Eq_chSupp}) and (\ref{Eq_chSupp2}), the clustered sparsity of the angular-domain MRA channel matrix ${\bf H}$ is illustrated in Fig. \ref{Fig2}, which will be exploited to facilitate the development of a semi-blind detection scheme at the BS.

{\color{red} \begin{remark}
It should be noted that the received signal model in (\ref{Eq_rxFrame}) is identical for both the coherent detection framework dedicated to sourced RA and the non-coherent detection framework dedicated to unsourced RA.
The major differences between two detection frameworks lie in the transmitted signal ${\bf X}$ and the receive algorithm, which will be detailed in \textit{Section III}.
\end{remark}}

\section{Traditional Detection Frameworks for Sourced and Unsourced RA}
\label{Sec_LegacyDet}

As described in \emph{Section \ref{Sec_sysModel}}, the key idea of grant-free RA protocol is to avoid complicated signaling interactions between the devices and the BS, thus achieving the ultra-low access latency, but at the expense of severe inter-device interference. 
Without access scheduling in advance, the uplink signals of all the active devices are overlapped on the same time-frequency resources, which makes reliable data detection at the BS a challenging problem. 
In this section, we first introduce two state-of-the-art detection frameworks for grant-free sourced and unsourced RA, respectively, which focus on different access requirements. 
Moreover, the related merits and faults are discussed.

\subsection{Non-Orthogonal Pilot-Based Coherent Detection for Sourced RA}
\label{Sub_coheDet}

The non-orthogonal pilot-based coherent detection framework for sourced RA adopts a two-phase transmission scheme \cite{Ref_IoTJ_Shao'19, Ref_TSP_Shao'20, Ref_WCM_Ke'21, Ref_CL_Shim'12, Ref_TSP_Liu'18, Ref_TSP_Shao'20-2, Ref_TSP_Ke'20}, where each frame is divided into the pilot and payload data phases, i.e., ${\bf X} = \left[{\bf X}_p, {\bf X}_d\right]$ with ${\bf X}_p \in {\mathbb C}^{K \times L_p}$ and ${\bf X}_d \in {\mathbb C}^{K \times L_d}$, respectively.
Here, the first $L_p$ symbol durations are used to transmit the non-orthogonal pilot sequences of active devices and the remaining $L_d = L-L_p$ symbol durations are reserved for payload data transmission. 
Similarly, the received signal can be expressed as ${\bf R} = \left[{\bf R}_p, {\bf R}_d\right]$, where ${\bf R}_p \in {\mathbb C}^{N \times L_p}$ and ${\bf R}_d \in {\mathbb C}^{N \times L_d}$ correspond to the received pilot and data signals, respectively.
At the receiver, the BS first performs joint ADD and CE based on the received pilot signal ${\bf R}_p = {\bf G}{\bf X}_p + {\bf W}_p$, which is equivalent to estimating ${\bf G}$ based on the known ${\bf X}_p$ and ${\bf R}_p$.
By leveraging the sparse device activity, the problem can be formulated as a CS problem and the advanced AMP algorithm in \cite{Ref_TSP_Ke'20} can be employed to acquire the solution.
With the estimated active device set ${\widehat {\cal A}}$ and channel matrix ${\widehat {\bf G}}$, the coherent data detection is then achieved as 
\begin{equation}\label{Eq_dataDet}
	\begin{aligned}
		{\widehat {\bf X}_d} = \left[{\widehat {\bf G}}\right]_{:,{\widehat {\cal A}}}^{\dagger}{\bf R}_d,
	\end{aligned}
\end{equation}
where ${\bf R}_d = {\bf G}{\bf X}_d + {\bf W}_d$.
At this point, the inter-device interference can be effectively resolved as long as the reliable estimates of the active  device set and the MRA channel matrix, i.e., ${\widehat {\cal A}}$ and ${\widehat {\bf G}}$, respectively, are obtained.
However, according to the CS theory, the pilot length $L_p \ge K_a{\rm log}_2\left(K\right)$ is required to obtain the satisfactory ADD and CE performance, which significantly degrades the payload efficiency, especially in the scenarios of massive URLLC conveying small data packets \cite{Ref_TSP_Ke'20}.
By further utilizing the angular-domain sparsity of massive MIMO channels, i.e.,
{\color{red} \begin{equation}\label{Eq_rxFrameAng}
	\begin{aligned}
		{\bf Y} = \left[{\bf Y}_p, {\bf Y}_d\right] = {\bf A}_R{\bf R} = {\bf H}\left[{\bf X}_p, {\bf X}_d\right] + {\bf N}, 	
	\end{aligned}
\end{equation} 
with ${\bf N} = {\bf A}_R{\bf W}$ denoting the noise matrix,} the authors in \cite{Ref_TSP_Ke'20} revealed that the minimum pilot overhead can be reduced to $S_a{\rm log}_2\left(K\right)$ with $S_a = {\rm max}\left\{{\rm supp}\left\{{\bf H}_{n,:}\right\}, \forall n \in \left[N\right]\right\}$ and $S_a \ll K_a$.
Yet, the payload efficiency is still limited when $K$ is extremely large.
Note that given the fixed payload data length, a lower payload efficiency indicates a higher transmission latency.
%On the other hand, the coherent detection in (\ref{Eq_dataDet}) is highly dependent on the accurate CSI, which becomes inefficient in time-varying channels.
%Because the devices have to frequently transmit the non-orthogonal pilot sequences for the update of channel state information (CSI). 

\subsection{Common Codebook-Based Non-Coherent Detection for Unsourced RA}
\label{Sub_nonCoherentDet}  

The common codebook-based non-coherent detection framework is dedicated to unsourced RA, where each active device delivers $B$-bit information using a common codebook ${\cal C} = \left\{{\bf c}_1, {\bf c}_2, \cdots, {\bf c}_{2^B}\right\} \subset {\mathbb C}^{L \times 1}$. 
Specifically, the $B$-bit information ${\bf b}_k \in {\mathbb B}^{B \times 1}$ produced by the active device $k$ is mapped to an integer $b_k \in \left\{1, 2, \cdots, 2^B\right\}$.
Then, the active device simply sends the $b_k$th codeword of the common codebook, i.e., ${\bf c}_{b_k}$, to the BS.
We can model the codeword selection by a set of $2^BK$ Bernoulli random variables $\delta_{b,k}$, $\forall b \in \left[2^B\right]$ and $\forall k \in \left[K\right]$.
Here, $\delta_{b,k} = 1$ if the $k$th device is active and transmits the code ${\bf c}_b$, and $\delta_{b,k} = 0$ otherwise.
On this basis, the transmitted signal of the $k$th device can be expressed as $\left[{\bf X}\right]_{k:,} = \sum\nolimits_{b=1}^{2^B} \delta_{b,k}{\bf c}_b^{\rm T}$, and the signal model in (\ref{Eq_rxFrame}) can be re-formulated as 
 \begin{equation}\label{Eq_rxUnsourced}
 	\begin{aligned}
 		{\bf R} &= {\bf G}\left[{\bm \delta}_1, {\bm \delta}_2, \cdots, {\bm \delta}_K\right]^{\rm T}{\bf C} + {\bf W} \\
 		        &= {\bf G}{\bm \Delta}{\bf C} + {\bf W} = {\widetilde {\bf G}}{\bf C} + {\bf W},
 	\end{aligned}
 \end{equation}  
where ${\bf C} = \left[{\bf c}_1, {\bf c}_2, \cdots, {\bf c}_{2^B}\right]^{\rm T} \in {\mathbb C}^{2^B \times L}$ is the common codebook, ${\bm \delta}_k = \left[\delta_{1,k}, \delta_{2,k}, \cdots, \delta_{2^B,k}\right]^{\rm T} \in {\mathbb B}^{2^B \times 1}$, and ${\widetilde {\bf G}}$ is the matrix combining the spatial-domain MRA channel matrix ${\bf G}$ and the codeword selection matrix ${\bm \Delta} \in {\mathbb B}^{K \times 2^B}$.
The matrix ${\bm \Delta}$ contains only $K_a$ non-zero rows, each of which has a single non-zero entry.
With this formulation, each active device contributes a single non-zero coefficient in $[{\widetilde {\bf G}}]_{n,:}$, thereby resulting in a $K_a$-sparse $2^B$-dimensional vector.
Considering the BS with $N$ receive antennas, the problem can be formulated as a multiple measurement vectors (MMV) support detection problem, where the different rows of ${\widetilde {\bf G}}$ have a common sparsity pattern.
The problem can be effectively addressed by the CS recovery algorithm such as AMP \cite{Ref_TSP_Ke'20}, but the computational complexity scales exponentially with $B$, which is prohibitive even for short packets with dozens of bits.
The prohibitive computational complexity leads to an extremely high processing latency at the BS.
Although several low-complexity solutions have been proposed \cite{Ref_Tcom_Vem'19, Ref_TIT_Amalladinne'20}, the payload efficiency is dramatically degraded due to the introduced redundant coding.

The user plane latency accounts for the one-way latency from the beginning of the packet processing at the transmitter to the successful detection at the receiver.
In grant-free MRA, the transmission and receive processing latencies are the two most dominant components contributing to the user plane latency \cite{Ref_IoTJ_Ding'22}.
Therefore, it is generally challenging for the traditional detection frameworks to satisfy the ultra-low latency requirement of massive URLLC due to the low payload efficiency or the high data detection complexity.
Moreover, their applications are limited to either sourced or unsourced RA, which is not conductive to accommodating future massive URLLC with heterogeneous access requirements.

\section{Proposed Unified Semi-Blind Detection Framework: Transmitter Design}
\label{Sec_txDesign}

To overcome the limitations of conventional coherent and non-coherent detection frameworks, this paper develops a unified semi-blind detection framework for supporting both sourced and unsourced RA.
Particularly, our goal is to jointly infer the sparse MRA channel matrix ${\bf H}$ and the signal matrix ${\bf X}$ from the received signal ${\bf Y}$ in (\ref{Eq_rxFrameAng}), based on which the payload data of active devices can be further detected.
By avoiding the pilot phase, an extremely high payload efficiency can be achieved, which leads to an ultra-low transmission latency.
{\color{red} However, the JCSE problem suffers from the inherent phase and permutation ambiguities.
Specifically, define ${\bf \Sigma}$ and ${\bf \Pi}$ as a diagonal matrix with phase shifts in the diagonal and a permutation matrix, respectively.
The ambiguities are caused by the fact that if $\left({\widehat {\bf H}}, {\widehat {\bf X}}\right)$ is a solution to the JCSE problem based on (\ref{Eq_rxFrameAng}), then $\left({\widehat {\bf H}}{\bf \Sigma}^{-1}{\bf \Pi}^{-1}, {\bf \Pi}{\bf \Sigma}{\widehat {\bf X}}\right)$ is also a valid solution.
In fact, the cost function $\left\|{\bf Y} - {\widehat {\bf H}}{\widehat {\bf X}}\right\|_{\rm F}^2$ is invariant to any phase shifts and permutations of the rows of ${\bf X}$.
The phase shift will lead to the demodulation error of estimated signals, while the row permutation will lead to the identification error of active devices.
To tackle this issue, we propose to insert a small amount of RI in the access signal ${\bf X}$ to eliminate the ambiguities.}

\begin{figure}[!t]
	\centering
	\includegraphics[width=0.7\columnwidth, keepaspectratio]
	{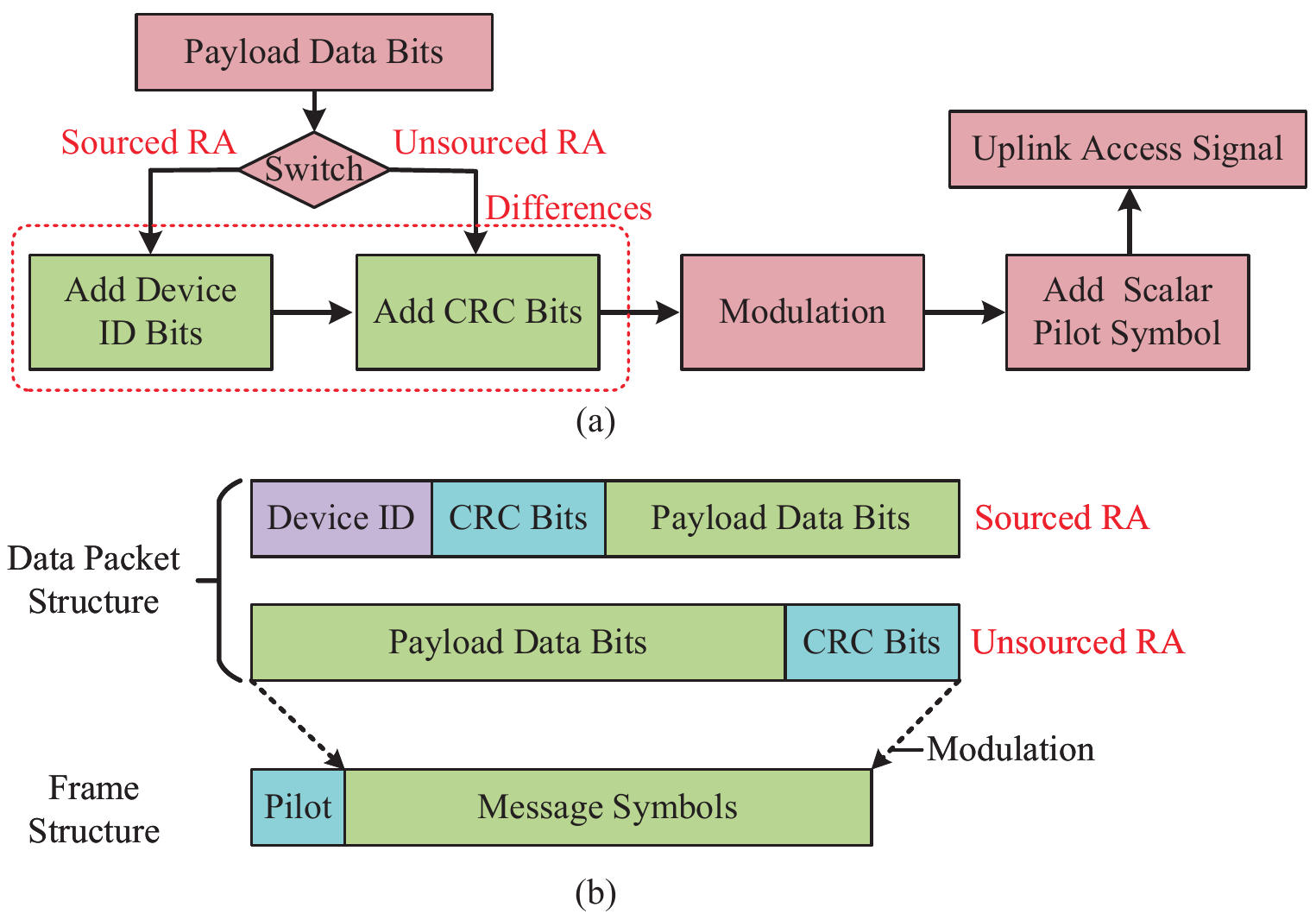}
	\caption{\small{The proposed unified transmitter design for sourced and unsourced RA: (a) Block diagram; (b) Data packet structure and frame structure.}}
	\label{Fig_txDesign}
	\vspace{-3mm}
\end{figure}

The proposed detection framework involves the transmitter design at the devices and the SIC-based semi-blind detection scheme at the BS.
This section first introduces a unified transmitter design for sourced and unsourced RA, where the required modules are almost identical for both RA paradigms, as illustrated in Fig. \ref{Fig_txDesign}. 
Therefore, our explanation mainly focuses on the sourced RA and the major differences between the two RA paradigms will be further clarified.

\subsection{Transmitter Design for Sourced RA}
For arbitrary active device with index $k$, its uplink access signal is generated based on the following key steps.
\begin{itemize}
\item{\textbf{Step 1:} To eliminate the permutation ambiguity, a binary device ID sequence of $B_i = \lceil {\rm log}_2\left(K\right)\rceil$ bits is inserted at the head of the payload data packet to identify the $K$ devices.
For the $k$th device, its ID sequence is provided as ${\bf b}_k^i = {\rm dec2bin}\left(k\right)$, where the operator ${\rm dec2bin}\left(\cdot\right)$ converts a decimal integer to its binary representation.}
	
\item{\textbf{Step 2:} To verify the correctness of the detected ID bits, a $B_c$-bit CRC code ${\bf b}_k^c$ is added to the end of the device ID sequence, as ${\bf b}_k = \left[{\bf b}_k^i; {\bf b}_k^c; {\bf b}_k^d\right]\in {\mathbb C}^{B \times 1}$, where ${\bf b}_k^d \in {\mathbb C}^{B_d \times 1}$ is the payload data packet and $B = B_i + B_c + B_d$.
The CRC code is generated as
\begin{equation}\label{Eq_crcCodeSourced}
	\begin{aligned}  
		{\bf b}_k^c = f\left([{\bf b}_k^i; {\bf 0}_{B_c \times 1}] \div {\bf p}_c\right),   
	\end{aligned}
\end{equation}
where $\div$ denotes the binary (modulo-2) division, ${\bf p}_c$ is the generator polynomial of CRC, and $f(\cdot)$ is the function to compute the remainder of the binary division.}
	
\item{\textbf{Step 3:} The overall data packet ${\bf b}_k$ is modulated by an $M$-order phase shift keying (PSK) modulator, where the modulated symbol sequence is defined as ${\bf x}_k^d \in {\mathbb C}^{\left(L-1\right) \times 1}$ with $L = \lceil B/{\rm log}_2\left(M\right) \rceil + 1$.}
	
{\color{red} \item{\textbf{Step 4:} To eliminate the phase ambiguity, a known scalar pilot symbol $x_p$ is inserted at the head of the modulated symbol sequence, i.e., ${\bf x}_k = [x_p; {\bf x}_k^d] \in {\mathbb C}^{L \times 1}$, where ${\bf x}_k$ is the uplink access signal of the $k$th device to be transmitted.
Here, $x_p$ is drawn from the constellation set of the adopted modulation scheme and is identical for all active devices.
Note that since ${\bf \Sigma}$ is a diagonal matrix, the phase shifts of phase ambiguity are identical for all the transmitted symbols of a specific active device, but different for the symbol frames of different active devices.
In this case, only one pilot symbol in each ${\bf x}_k$ is sufficient to estimate the phase shift matrix ${\bf \Sigma}$.}}
\end{itemize}

\subsection{Extension to Unsourced RA}

The aforementioned transmitter design for sourced RA can be further extended to the unsourced RA, where the major difference lies in the structure of the data packet, see Fig.~\ref{Fig_txDesign}.
For unsourced RA, the BS is solely interested in the list of the sent messages, without regard for the identities of individual sources, i.e., the permutation ambiguity could be ignored.
Therefore, the device ID sequence is removed from the data packet for improved payload efficiency.
Meanwhile, the CRC code is attached to the end of the payload data packet, as ${\bf b}_k = \left[{\bf b}_k^d; {\bf b}_k^c\right]\in {\mathbb C}^{B \times 1}$ with $B = B_c + B_d$, and the generation of the CRC code is modified to
\begin{equation}\label{Eq_crcCodeUnsourced}
	\begin{aligned}
		{\bf b}_k^c = f\left([{\bf b}_k^d; {\bf 0}_{B_c \times 1}] \div {\bf p}_c\right).
	\end{aligned}
\end{equation}
{\color{red} Different from unsourced RA, the CRC code in sourced RA is mainly used for evaluating the reliability of the detected device ID bits, which effectively avoids the whole packet loss due to the detection error of few payload data bits, thus dramatically reducing the probability of miss detection.}
Based on the proposed transmitter design, both sourced and unsourced RA could share the same hardware modules and only a software-defined switch is required to determine which data packet structure is adopted.
Compared to the traditional detection frameworks detailed in {\em Section III}, the proposed unified transmitter design is more beneficial to satisfying the ultra-low latency requirement of massive URLLC due to the significantly improved payload efficiency.

\section{Proposed Unified Semi-Blind Detection Framework: Receiver Design}
\label{Sec_rxDesign}

Adopting the transmitter design proposed in \emph{Section \ref{Sec_txDesign}}, the inserted RI is insufficient to achieve reliable ADD and CE, which significantly degrades the performance of traditional coherent detection.
In this section, we develop an SIC-based semi-blind detection scheme at the BS, where the payload data of active devices is directly detected from the overlapped received signal without exploiting explicit channel state information. 
Specifically, we first propose a BiG-AMP-based JCSE algorithm, where the channel and signal matrices are jointly estimated by factorizing the noisy received signal, without regard for the phase and permutation ambiguities.
In particular, a singular value decomposition (SVD)-based rank selection approach and a RI-aided initialization strategy are incorporated to reduce the computational complexity and to enhance the JCSE reliability, respectively, for the conventional BiG-AMP algorithm. 
Finally, the SIC-based semi-blind data detection scheme is developed, where the inserted RI is exploited to resolve the ambiguities and the SIC technique is utilized to mitigate the inter-device interference iteratively.

\subsection{SVD-Based Rank Selection}
\label{SubSec_rankSel}

As clarified in \emph{Section \ref{Sec_sysModel}}, only $K_a$ ($K_a \ll K$) active devices contribute to the received signal ${\bf Y}$, thus the signal model in (\ref{Eq_rxFrameAng}) can be re-expressed as 
\begin{equation} \label{Eq_rxAngAct}
	\begin{aligned}
		{\bf Y} = \left[{\bf H}\right]_{:,{\cal A}}\left[{\bf X}\right]_{{\cal A},:} + {\bf N} = {\bf H}_{\rm act}{\bf X}_{\rm act} + {\bf N}.
	\end{aligned}
\end{equation}
Here, ${\cal A}$ is the active device set, ${\bf H}_{\rm act} = \left[{\bf H}\right]_{:,{\cal A}} \in {\mathbb C}^{N \times K_a}$ and ${\bf X}_{\rm act} = \left[{\bf X}\right]_{{\cal A},:} \in {\mathbb C}^{K_a \times L}$ represent the MRA channel matrix and the transmitted signal matrix associated with the active devices, respectively.
For JCSE, our goal is to jointly infer the channel matrix ${\bf H}_{\rm act}$ and the signal matrix ${\bf X}_{\rm act}$ based on ${\bf Y}$.
By exploiting the angular-domain sparsity of massive MIMO channels, as well as the statistical information of ${\bf H}_{\rm act}$ and ${\bf X}_{\rm act}$, the efficient BiG-AMP algorithm derived in \cite{Ref_TSP_Parker'14} can be employed to achieve the goal, where the concerned problem is formulated as a matrix factorization problem.
In practice, since the number of active devices $K_a$ is generally unknown in advance, a straightforward solution is to apply the BiG-AMP algorithm to the model (\ref{Eq_rxFrameAng}), where ${\bf H}$ and ${\bf X}$ can be jointly estimated.
Then, the estimates of ${\bf H}_{\rm act}$ and ${\bf X}_{\rm act}$ are obtained by removing the channels and the signals of the devices whose channel gains are smaller than a predefined threshold.
However, this solution poses stringent requirements on the number of BS antennas and the length of uplink access signal, i.e., $N > K$ and $L >  K$, which is impractical in massive URLLC with small data packets \cite{Ref_TSP_Parker'14-2}.
Meanwhile, the resulting computational complexity at each BiG-AMP iteration scales with the number of the total devices, i.e., ${\cal O}\left(NK + KL + NL\right)$~\cite{Ref_Access_Yan'19}.

{\color{red} \begin{prop}
When $N > K_a$ and $L > K_a$, the rank of the noiseless received signal ${\bf Z} = {\bf H}_{\rm act}{\bf X}_{\rm act}$ is $K_a$.
\end{prop}

\textit{Proof:} Due to ${\bf Z} = {\bf H}_{\rm act}{\bf X}_{\rm act}$, the rank of ${\bf Z}$ satisfies the following inequalities, as
\begin{equation} \label{InEq_rank1}
	\begin{aligned}
		{\rm rank}\left({\bf Z}\right) \le {\rm min}\left\{{\rm rank}\left({\bf H}_{\rm act}\right), {\rm rank}\left({\bf X}_{\rm act}\right)\right\}
	\end{aligned}
\end{equation}
and
\begin{equation} \label{InEq_rank2}
	\begin{aligned}
		{\rm rank}\left({\bf H}_{\rm act}\right) + {\rm rank}\left({\bf X}_{\rm act}\right) - K_a \le {\rm rank}\left({\bf Z}\right),
	\end{aligned}
\end{equation} 
where ${\rm rank}(\cdot)$ denotes the rank of a matrix.
On the one hand, the assumptions $N > K_a$ and $L > K_a$ lead to ${\rm min}\left\{{\rm rank}\left({\bf H}_{\rm act}\right), {\rm rank}\left({\bf X}_{\rm act}\right)\right\} \le K_a$.
Thus, the inequality (\ref{InEq_rank1}) can be re-expressed as ${\rm rank}\left({\bf Z}\right) \le K_a$.
On the other hand, since the access signals of different active devices are generated independently, we have ${\rm rank}\left({\bf X}_{\rm act}\right) = K_a$.
Meanwhile, since the active devices are independently distributed in the BS coverage, their channels are linearly independent, which results in ${\rm rank}\left({\bf H}_{\rm act}\right) = K_a$.
Therefore, the inequality (\ref{InEq_rank2}) can be re-expressed as $K_a \le {\rm rank}\left({\bf Z}\right)$.
At the point, the rank of ${\bf Z}$ is proofed to be ${\rm rank}\left({\bf Z}\right) = K_a$ by combining the inequalities in (\ref{InEq_rank1}) and (\ref{InEq_rank2}).

\begin{figure}[t]
	\centering
	\includegraphics[width=0.5\columnwidth, keepaspectratio]
	{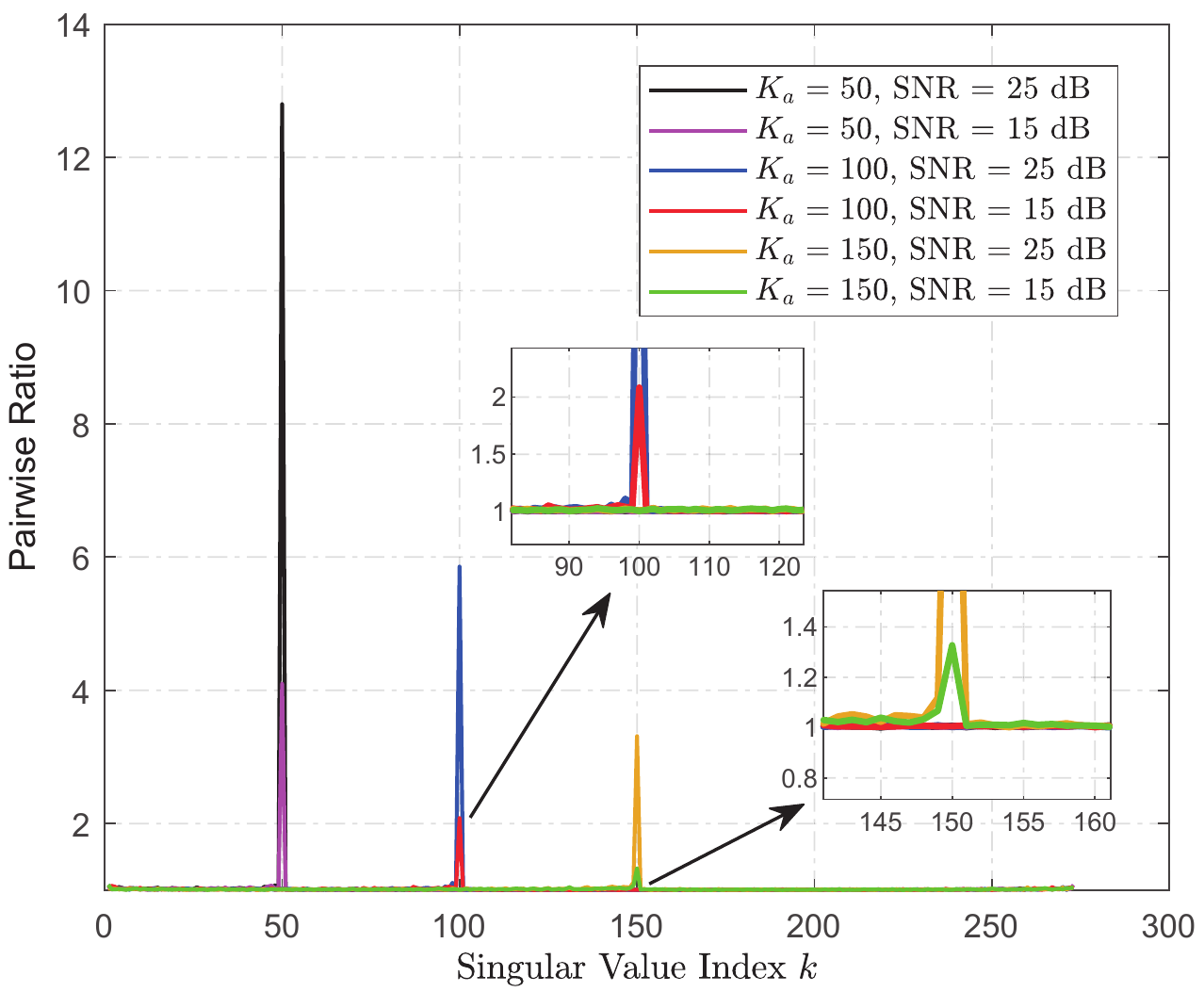}
	\caption{\color{red} {\small{ The noisy received signal ${\bf Y}$ has a prominent peak in the pairwise ratios of its adjacent descending singular values, where $N = 512$ and $L = 274$ are considered.}}}
	\label{Fig_rankSel}
	\vspace{-3mm}
\end{figure}

\quad With ${\rm rank}\left({\bf Z}\right) = K_a$, the authors in [16] have revealed that the space of ${\bf Y} = {\bf Z} + {\bf N}$ can be divided into a noisy signal subspace and a pure noise subspace in high signal-to-noise ratio (SNR) cases.
Specifically, by exploiting SVD, the noisy received signal is re-expressed as ${\bf Y} = {\bf U}{\bf \Sigma}{\bf V}^{\rm H}$, where ${\bf U} \in \mathbb{C}^{N \times N}$ and ${\bf V} \in \mathbb{C}^{L \times L}$ are unitary matrices, ${\bf \Sigma} \in \mathbb{C}^{N \times L}$ is a diagonal matrix with $K_{\rm max}$ non-zero real numbers, i.e., the singular values of ${\bf Y}$, on the diagonal, and $K_{\rm max} = {\rm min}\left(N, L\right) > K_a$.
Then, the signal subspace is constructed as ${\bf Y}_s = \left[{\bf U}\right]_{:,{\cal K}_s}\left[{\bf \Sigma}\right]_{{\cal K}_s, {\cal K}_s}\left[{\bf V}\right]_{{\cal K}_s,:}^{\rm H} = {\bf Z} + {\bf N}_s$, with ${\cal K}_s = \left\{1, 2, \cdots, K_a\right\}$ and ${\bf N}_s$ denoting the noise incorporated in the signal space.
Meanwhile, the noise subspace is constructed as ${\bf Y}_n = \left[{\bf U}\right]_{:,{\cal K}_n}\left[{\bf \Sigma}\right]_{{\cal K}_n, {\cal K}_n}\left[{\bf V}\right]_{{\cal K}_n,:}^{\rm H} = {\bf N}_n$ with ${\cal K}_n = \left\{K_a+1, \cdots, K_{\rm max}\right\}$ and ${\bf N}_n$ denoting the noise incorporated in the noise space.
Particularly, the singular values of ${\bf Y}_s$ are considerably larger than those of ${\bf Y}_n$ as a relatively high SNR is considered.
Therefore, the received signal ${\bf Y}$ has a prominent peak in the pairwise ratios of its adjacent descending singular values, i.e., $\left\{\left[{\bf \Sigma}\right]_{k,k}/\left[{\bf \Sigma}\right]_{k+1, k+1} | \forall k \in \left[K_{\rm max} - 1 \right] \right\}$, as illustrated in Fig. 4.
Moreover, the singular value index corresponding to the maximum ratio is exactly $K_a$.
Based on this remarkable characteristic, the number of active devices $K_a$ can be estimated via the following rank selection procedure,
\begin{equation} \label{Eq_rankSel}
	{\widehat K_a} = \mathop {\arg \max}\limits_{k \in [K_{\rm max}-1]} \left[{\bf \Sigma}\right]_{k,k}/\left[{\bf \Sigma}\right]_{k+1, k+1}.
\end{equation}}
{\color{red} In this context, we can apply the BiG-AMP algorithm to model (\ref{Eq_rxAngAct}) to jointly estimate ${\bf H}_{\rm act}$ and ${\bf X}_{\rm act}$, where the dimension constraint relaxes to $N > K_a$ and $L > K_a$ with $K_a \ll K$, i.e., the considered problem is independent of the number of potential devices.}
Meanwhile, the computational complexity of each iteration of the BiG-AMP algorithm reduces to  ${\cal O}\left(NK_a + K_aL + NL\right)$.
This leads to the dramatically reduced processing latency, which is another key to guarantee the ultra-low user plane latency of massive URLLC. 

{\color{red} \begin{remark}
In this paper, the considered JCSE problem is formulated based on the angular-domain signal model (7), rather than the spatial-domain model (2).
Compared with the spatial-domain channel matrix ${\bf G}$, the angular-domain channel matrix ${\bf H}$ exhibits an enhanced sparsity, which dramatically reduces the number of unknown channel coefficients to be estimated.
In this case, for a given number of measurements, the JCSE performance can be significantly improved by further leveraging the angular-domain sparsity of massive MIMO channels.
The authors in \cite{Ref_Tcom_Zhang'18} have revealed that the performance can be very close to the ideal case with perfect CSI as long as the channel matrix is sufficiently sparse.
\end{remark}}

{\color{red} \begin{remark}
For the cases with an extremely low SNR (e.g., ${\rm SNR} < 0$ dB) or an extremely large number of active devices (e.g., $K_a > 500$), the singular values of ${\bf Y}$ will decay smoothly, which makes the signal and noise subspaces indistinguishable.
In this context, the proposed SVD-based rank selection approach fails to work.
However, due to the sporadic uplink traffic of massive URLLC and the adaptive transmit power control, such extreme cases are rare to occur in practice.
\end{remark}}

\subsection{BiG-AMP-Based JCSE Algorithm}
\label{SubSec_jcseAlg}

{\color{red} Next, we utilize the BiG-AMP algorithm to address the aforementioned matrix factorization problem, where the expectation maximization (EM) algorithm is incorporated to learn the unknown hyper-parameters and a RI-based initialization strategy is proposed to improve the estimation accuracy.}
Under the Bayesian inference framework, the detailed description of the BiG-AMP algorithm begins with the probabilistic model of the problem.
Specifically, the minimum mean-square-error (MMSE) estimates of ${\bf H}_{\rm act}$ and ${\bf X}_{\rm act}$, denoted by ${\widehat {\bf H}_{\rm act}}$ and ${\widehat {\bf X}_{\rm act}}$, respectively, are expressed as
\begin{equation} \label{Eq_postMeanMtx}
	\begin{aligned}
		\left({\widehat {\bf H}_{\rm act}}, {\widehat {\bf X}_{\rm act}}\right) = {\mathbb E}\left[{\bf H}_{\rm act}, {\bf X}_{\rm act}|{\bf Y}\right] 
			= \iint p\left({\bf H}_{\rm act}, {\bf X}_{\rm act}|{\bf Y}\right) d{\bf H}_{\rm act}d{\bf X}_{\rm act},
	\end{aligned}
\end{equation}
where the joint posterior distribution is given as
\begin{equation} \label{Eq_jointPost}
	\begin{aligned}
		p\left({\bf H}_{\rm act}, {\bf X}_{\rm act}|{\bf Y}\right) &= \frac{p\left({\bf Y}|{\bf H}_{\rm act}, {\bf X}_{\rm act}\right)p\left({\bf H}_{\rm act}\right)p\left({\bf X}_{\rm act}\right)} {p\left({\bf Y}\right)} \\
			&\propto p\left({\bf Y}|{\bf H}_{\rm act}, {\bf X}_{\rm act}\right)p\left({\bf H}_{\rm act}\right)p\left({\bf X}_{\rm act}\right),  
	\end{aligned}
\end{equation}
with the notation $\propto$ denoting an equality up to a constant scaling factor.
It is assumed that the elements of the noise matrix ${\bf N}$ are independently drawn from ${\cal CN}\left(0, \sigma^2\right)$.
Hence, given ${\bf H}_{\rm act}$ and ${\bf X}_{\rm act}$, the likelihood function can be factorized into
\begin{equation} \label{Eq_likelihood}
	\begin{aligned}
		p\left({\bf Y}|{\bf H}_{\rm act}, {\bf X}_{\rm act}\right) &= \prod\limits_{n=1}^N\prod\limits_{l=1}^L p\left(y_{n,l}|z_{n,l} = \sum\limits_{k=1}^{K_a}h_{n,k}x_{k,l}\right) \\
			&= \prod\limits_{n=1}^N\prod\limits_{l=1}^L \frac{1}{\pi\sigma^2} {\rm exp}\left(-\frac{1}{\sigma^2}\left|y_{n,l} - z_{n,l}\right|^2\right),
	\end{aligned}
\end{equation}
where the subscript ``act" is omitted in $h_{n,k}$ and $x_{k,l}$ for notational simplicity.
Meanwhile, we adopt the well-studied spike-and-slab \emph{a priori} distribution to capture the sparse feature of the angular-domain channel matrix ${\bf H}_{\rm act}$, i.e.,
\begin{equation} \label{Eq_chPri}
	\begin{aligned}
		p\left({\bf H}_{\rm act}\right) = \prod\limits_{n=1}^N\prod\limits_{k=1}^{K_a} p\left(h_{n,k}\right) 
		= \prod\limits_{n=1}^N\prod\limits_{k=1}^{K_a} \left[\left(1 - \gamma_{n,k}\right)\delta\left(h_{n,k}\right) + \gamma_{n,k}{\tilde f}\left(h_{n,k}\right)\right],
	\end{aligned}
\end{equation}
where $0 \le \gamma_{n,k} \le 1$ denotes the sparsity ratio, i.e., the probability of $h_{n,k}$ being non-zero, $\delta\left(\cdot\right)$ is the Dirac delta function, ${\tilde f}\left(h_{n,k}\right) = {\cal CN}\left(h_{n,k}; \mu_{n,k}, \tau_{n,k}\right)$ is the \emph{a priori} distribution of non-zero channel coefficients.
This distribution has been widely applied in the literature for AMP-based MIMO channel estimation \cite{Ref_TSP_Ke'20}, which shows its effectiveness in modeling the \emph{a priori} distribution of real-world MIMO channels
{\color{red} Here, the channel coefficients associated with different BS antennas are assumed to be mutually independent. This assumption simplifies the considered problem and facilitates the application of the efficient AMP inference framework with acceptable performance loss, as discussed in \cite{Ref_TSP_Ke'20}. Note that although taking into account the correlation of different antennas may further enhance the performance, the corresponding algorithm would be much more involved.}
In addition, since the transmitted signals are randomly drawn from a finite constellation set ${\cal S}$, the \emph{a priori} distribution of ${\bf X}_{\rm act}$ is provided as
\begin{equation} \label{Eq_sigPri}
	\begin{aligned}
		p\left({\bf X}_{\rm act}\right) = \prod\limits_{k=1}^{K_a}\prod\limits_{l=1}^{L} p\left(x_{k,l}\right)
		= \prod\limits_{k=1}^{K_a}\prod\limits_{l=1}^{L} \frac{1}{M} \sum\limits_{m= 1}^M \delta\left(x_{k,l} - s_m\right),
	\end{aligned}
\end{equation}
where $s_m \in {\cal S}, \forall m \in [M]$ are the constellation symbols.
Benefitting from the factorizability of the likelihood function and \emph{a priori} distributions, as in (\ref{Eq_likelihood})-(\ref{Eq_sigPri}), the joint posterior distribution in (\ref{Eq_jointPost}) can be represented by a factor graph.
In this context, the standard sum-product algorithm can operate to compute the means of the marginal posterior distributions $p\left(h_{n,k}|{\bf Y}\right)$ and $p\left(x_{k,l}|{\bf Y}\right)$ for all pairs $\left(n,k\right)$ and $\left(k,l\right)$, i.e., the solution of the problem in (\ref{Eq_postMeanMtx}) \cite{Ref_TIT_Frank'01}.
However, for massive URLLC in massive MIMO systems, the exact implementation of the sum-product algorithm is impractical, as the large numbers of BS antennas and active devices make the related computational complexity prohibitive.
To overcome this obstacle, the key idea of the BiG-AMP algorithm is to provide a low-complexity approximation of the sum-product algorithm by applying the central-limit theorem and Taylor-series approximations in the large system limits \cite{Ref_TSP_Parker'14}.
Intuitively, with the approximations, the matrix estimation problem in (\ref{Eq_postMeanMtx}) can be decoupled into multiple independent scalar estimation problems, which avoids high-dimensional integrals and facilitates the practical implementation of the algorithm.

\begin{breakablealgorithm}
	\caption{BiG-AMP-Based JCSE Algorithm}
	\label{Alg_BiG-AMP}
	\begin{algorithmic}[1]
		\REQUIRE Angular-domain received signal ${\bf Y}$, the maximum number of iterations $U$, and termination threshold $\epsilon_{\rm amp}$.
		\ENSURE The estimates of the channel matrix ${\bf H}_{\rm act}$ and transmitted signal matrix ${\bf X}_{\rm act}$ associated with the active devices.
		\STATE Determine the problem dimensions, i.e., $\left[N, L\right] = {\rm size}\left({\bf Y}\right)$, and $K_a$ is estimated by (\ref{Eq_rankSel}).
		\STATE ${\widehat {\bf H}_{\rm act}} = {\bf 0}_{N \times {\widehat K_a}}$, ${\widehat {\bf X}_{\rm act}} = {\bf 0}_{{\widehat K_a} \times L}$.
		\STATE $\forall n, k$: Initialize the hyper-parameters $\sigma^2\left(1\right)$, $\mu_{n,k}\left(1\right)$, $\tau_{n,k}\left(1\right)$, and $\gamma_{n,k}\left(1\right)$.
		\label{Step_initEM}
		\STATE $\forall n, k, l$: Initialize the marginal posterior means and variances of ${\bf H}_{\rm act}$ and ${\bf X}_{\rm act}$, i.e., ${\widehat h}_{n,k}\left(1\right)$, $v_{n,k}^h\left(1\right)$, ${\widehat x}_{k,l}\left(1\right)$, and $v_{k,l}^x\left(1\right)$.
		\label{Step_initBiGAMP}
		\FOR {$u = 1, \cdots, U$}
		\STATE // BiG-AMP Updates:
		\STATE $\forall n, l$: ${\overline v}_{n,l}^p\left(u\right) = \sum\nolimits_{k=1}^{K_a} \left|{\widehat h}_{n,k}\left(u\right)\right|^2v_{k,l}^x\left(u\right) + v_{n,k}^h\left(u\right)\left|{\widehat x}_{k,l}\left(u\right)\right|^2$
		\label{Step_varP1}
		\STATE $\forall n, l$: ${\overline p}_{n,l}\left(u\right) = \sum\nolimits_{k=1}^{K_a} {\widehat h}_{n,k}\left(u\right){\widehat x}_{k,l}\left(u\right)$
		\label{Step_meanP1}
		\STATE $\forall n, l$: $v_{n,l}^p\left(u\right) = {\overline v}_{n,l}^p\left(u\right) + \sum\nolimits_{k=1}^{K_a} v_{n,k}^h\left(u\right)v_{k,l}^x\left(u\right)$
		\label{Step_varP2}
		\vspace{0.5mm}
		\STATE $\forall n, l$: ${\widehat p}_{n,l}\left(u\right) = {\overline p}_{n,l}\left(u\right) - {\widehat g}_{n,l}\left(u-1\right){\overline p}_{n,l}\left(u\right)$
		\label{Step_meanP2}
		\vspace{0.5mm}
		\STATE $\forall n, l$: $v_{n,l}^z\left(u\right) = {\mathbb V}\left[z_{n,l} | {\widehat p}_{n,l}\left(u\right), v_{n,l}^p\left(u\right)\right]$
		\label{Step_varZ}
		\vspace{0.5mm}
		\STATE $\forall n, l$: ${\widehat z}_{n,l}\left(u\right) = {\mathbb E}\left[z_{n,l} | {\widehat p}_{n,l}\left(u\right), v_{n,l}^p\left(u\right)\right]$
		\label{Step_meanZ}
		\vspace{0.5mm}
		\STATE $\forall n, l$: $v_{n,l}^g\left(u\right) = \left(1-v_{n,l}^z\left(u\right)/v_{n,l}^p\left(u\right)\right) / v_{n,l}^p\left(u\right)$
		\label{Step_varG}
		\vspace{0.5mm}
		\STATE $\forall n, l$: ${\widehat g}_{n,l}\left(u\right) = \left({\widehat z}_{n,l}\left(u\right) - {\widehat p}_{n,l}\left(u\right)\right) / v_{n,l}^p\left(u\right)$
		\label{Step_meanG}
		\STATE $\forall n, k$: $v_{n,k}^q\left(u\right) = \left(\sum\nolimits_{l=1}^{L} \left|{\widehat x}_{k,l}\left(u\right)\right|^2v_{n,l}^g\left(u\right)\right)^{-1}$
		\label{Step_varQ}
		\STATE $\forall n, k$: ${\widehat q}_{n,k}\left(u\right) = v_{n,k}^q\left(u\right)\sum\nolimits_{l=1}^{L}{\widehat x}_{k,l}^*\left(u\right){\widehat g}_{n,l}\left(u\right) + {\widehat h}_{n,k}\left(u\right)\left(1-v_{n,k}^q\left(u\right)\sum\nolimits_{l=1}^{L} v_{k,l}^x\left(u\right)v_{n,l}^g\left(u\right)\right)$\!\!\!\!\!\!\!\!
		\label{Step_meanQ}
		\STATE $\forall n, k$: $v_{n,k}^h\left(u+1\right) = {\mathbb V}\left[h_{n,k} | {\widehat q}_{n,k}\left(u\right), v_{n,k}^q\left(u\right)\right]$
		\label{Step_varH}
		\vspace{0.5mm}
		\STATE $\forall n, k$: ${\widehat h}_{n,k}\left(u+1\right) = {\mathbb E}\left[h_{n,k} | {\widehat q}_{n,k}\left(u\right), v_{n,k}^q\left(u\right)\right]$
		\label{Step_meanH}
		\vspace{0.5mm}
		\STATE $\forall k, l$: $v_{k,l}^r\left(u\right) = \left(\sum\nolimits_{n=1}^{N} \left|{\widehat h}_{n,k}\left(u\right)\right|^2v_{n,l}^g\left(u\right)\right)^{-1}$
		\label{Step_varR}
		\STATE $\forall k, l$: ${\widehat r}_{k,l}\left(u\right) = v_{k,l}^r\left(u\right)\sum\nolimits_{n=1}^{N}{\widehat h}_{n,k}^*\left(u\right){\widehat g}_{k,l}\left(u\right) + {\widehat x}_{k,l}\left(u\right)\left(1-v_{k,l}^r\left(u\right)\sum\nolimits_{n=1}^{N} v_{n,k}^h\left(u\right)v_{k,l}^g\left(u\right)\right)$\!\!\!\!\!\!\!\!
		\label{Step_meanR}
		\STATE $\forall k, l$: $v_{k,l}^x\left(u+1\right) = {\mathbb V}\left[x_{k,l} | {\widehat r}_{k,l}\left(u\right); v_{k,l}^r\left(u\right)\right]$
		\label{Step_varX}
		\vspace{0.5mm}
		\STATE $\forall k, l$: ${\widehat x}_{k,l}\left(u+1\right) = {\mathbb E}\left[x_{k,l} | {\widehat r}_{k,l}\left(u\right); v_{k,l}^r\left(u\right)\right]$
		\label{Step_meanX}
		\vspace{1mm}
		\STATE // EM Updates:
		\STATE $\sigma^2\left(u+1\right) = \frac{1}{NL} \sum\nolimits_{n=1}^{N}\sum\nolimits_{l=1}^{L} \left[\frac{\left|y_{n,l}-{\widehat p}_{n,l}\left(u\right)\right|^2}{1+v_{n,l}^p\left(u\right)/\sigma^2\left(u\right)}\right. + \left.\frac{\sigma^2\left(u\right)v_{n,l}^p\left(u\right)}{\sigma^2\left(u\right)+v_{n,l}^p\left(u\right)}\right]$
		\label{Step_varN}
		\vspace{0.5mm}
		\STATE $\tau_{n,k} \left(u+1\right) = \frac{\pi_{n,k}\left(u\right) \left[\left|\mu_{n,k}\left(u\right) - {\widehat d}_{n,k}\left(u\right)\right|^2 + v_{n,k}^d\left(u\right)\right]}{\pi_{n,k}\left(u\right)}$
		\label{Step_varChGain}
		\vspace{0.5mm}
		\STATE $\mu_{n,k} \left(u+1\right) = \frac{\pi_{n,k}\left(u\right){\widehat d}_{n,k}\left(u\right)}{\pi_{n,k}\left(u\right)}$
		\label{Step_meanChGain}
		\vspace{0.5mm}
		\IF{$\frac{\sum\nolimits_{n=1}^{N}\sum\nolimits_{l=1}^{L} \left|{\widehat p}_{n,l}\left(u\right) - {\widehat p}_{n,l}\left(u-1\right)\right|^2}{\sum\nolimits_{n=1}^{N}\sum\nolimits_{l=1}^{L} \left|{\widehat p}_{n,l}\left(u\right)\right|^2} \le \epsilon_{\rm amp}$}
		\label{tolStart1}
		\vspace{0.5mm}
		\STATE break;
		\ENDIF
		\label{Step_tolEnd1}
		\ENDFOR
		\RETURN  ${\widehat {\bf H}_{\rm act}}$, ${\widehat {\bf X}_{\rm act}}$
	\end{algorithmic}
\end{breakablealgorithm}

The overall steps of the BiG-AMP-based JCSE algorithm are summarized in \emph{Algorithm \ref{Alg_BiG-AMP}}.
For completeness, we provide more detailed descriptions as follows.
\emph{Lines \ref{Step_varP1}} and \emph{\ref{Step_meanP1}} acquire a plug-in\footnote{The plug-in principle is a technique used in the probability theory and statistics to approximately estimate a feature of a distribution (e.g., the expected value and the variance) that cannot be computed exactly. It is widely used in the theories of Monte Carlo simulation and bootstrapping \cite{Ref_Book_Vaart'98}.} estimate of the noiseless received signal ${\bf Z} = {\bf H}_{\rm act}{\bf X}_{\rm act}$, where the corresponding means $\left\{{\overline p}_{n,l}\right\}$ and variances $\left\{{\overline v}_{n,l}^p\right\}$ are computed in element-wise.
\emph{Lines \ref{Step_varP2}} and \emph{\ref{Step_meanP2}} introduce the so called Onsager reaction term\footnote{The Onsager reaction term has been extensively discussed in the context of AMP. For more details, please refer to reference \cite{Ref_ITW_Rangan'10}.} (i.e., the last term on the right-hand side of the equation) to correct the the rough plug-in estimates, which further improves the estimation accuracy.
With the obtained quantities $\left\{{\widehat p}_{n,l}\right\}$ and $\left\{v_{n,l}^p\right\}$, \emph{lines \ref{Step_varZ}} and \emph{\ref{Step_meanZ}} compute the marginal posterior means $\left\{{\widehat z}_{n,l}\right\}$ and variances $\left\{v_{n,l}^z\right\}$ of ${\bf Z}$.
Specifically, the MMSE estimation of ${\bf Z}$ is decoupled into $NL$ independent scalar inference problems, i.e., ${\widehat p}_{n,l}  = z_{n,l} + w_{n,l}^z, \forall n, l$, with
$z_{n,l} \sim {\cal CN}\left(z_{n,l}; y_{n,l}, \sigma^2\right)$ and $w_{n,l}^z \sim {\cal CN}\left(w_{n,l}^z; 0, v_{n,l}^p\right)$.
Therefore, \emph{lines \ref{Step_varZ}} and \emph{\ref{Step_meanZ}} are explicitly computed as
\begin{align}
	v_{n,l}^z\left(u\right) &= \frac{\sigma^2\left(u\right)v_{n,l}^p\left(u\right)}{\sigma^2\left(u\right) + v_{n,l}^p\left(u\right)}, \label{Eq_postVarRx} \\
	{\widehat z}_{n,l}\left(u\right)  &= \frac{y_{n,l}v_{n,l}^p\left(u\right) + \sigma^2\left(u\right){\widehat p}_{n,l}\left(u\right)}{\sigma^2\left(u\right) + v_{n,l}^p\left(u\right)}, \label{Eq_postMeanRx}
\end{align}
respectively.
Subsequently, \emph{lines \ref{Step_varG}} and \emph{\ref{Step_meanG}} use the related posterior moments to compute the scaled residual $\left\{{\widehat g}_{n,l}\right\}$ and its inverse variances $\left\{v_{n,l}^g\right\}$.
Finally, \emph{lines \ref{Step_varQ}} and \emph{\ref{Step_meanQ}} obtain an equivalent AWGN corrupted observation of the true $h_{n,k}$, i.e., ${\widehat q}_{n,k}  = h_{n,k} + w_{n,k}^h, \forall n, k$, with $w_{n,k}^q \sim {\cal CN}\left(w_{n,k}^q; 0, v_{n,k}^q\right)$.
Adopting the \emph{a priori} distribution $p\left(h_{n,k}\right)$ given in (\ref{Eq_chPri}), the posterior distribution of $h_{n,k}$ is computed as
\begin{equation} \label{chMarPost}
	\begin{aligned}
		p\left(h_{n,k}|{\widehat q}_{n,k}\left(u\right), v_{n,k}^q\left(u\right)\right) = \left(1-\pi_{n,k}\left(u\right)\right)\delta\left(h_{n,k}\right) + \pi_{n,k}\left(u\right){\cal CN}\left(h_{n,k}; {\widehat d}_{n,k}\left(u\right), v_{n,k}^d\left(u\right)\right),
	\end{aligned}
\end{equation}
where
\begin{align}
	v_{n,k}^d\left(u\right) &= \frac{\tau_{n,k}\left(u\right)v_{n,k}^q\left(u\right)}{\tau_{n,k}\left(u\right)+v_{n,k}^q\left(u\right)}, \label{Eq_varD} \\
	{\widehat d}_{n,k}\left(u\right) &= \frac{\mu_{n,k}\left(u\right)v_{n,k}^q\left(u\right) + \tau_{n,k}\left(u\right){\widehat q}_{n,k}\left(u\right)}{\tau_{n,k}\left(u\right)+v_{n,k}^q\left(u\right)}, \label{Eq_meanD} \\
	\hspace{-12mm}{\cal L} &= {\rm ln}\frac{v_{n,k}^q\left(u\right)}{\tau_{n,k}\left(u\right)+v_{n,k}^q\left(u\right)} + \frac{\left|{\widehat q}_{n,k}\left(u\right)\right|^2}{v_{n,k}^q\left(u\right)}
	+ \frac{\left|{\widehat q}_{n,k}\left(u\right)-\mu\left(u\right)\right|^2}{\tau_{n,k}\left(u\right)+v_{n,k}^q\left(u\right)}, \label{Eq_interVarL}\\
	\pi_{n,k}\left(u\right) &= \frac{\gamma_{n,k}\left(u\right)}{\gamma_{n,k}\left(u\right) + \left(1-\gamma_{n,k}\left(u\right)\right){\rm exp}\left(-{\cal L}\right)}. \label{Eq_postSparseRatio}
\end{align}
Then, the posterior mean ${\widehat h_{n,k}}$ and variance $v_{n,k}^h$ of $h_{n,k}$ in \emph{lines \ref{Step_varH}} and \emph{\ref{Step_meanH}}, respectively, are explicitly given as
\begin{align} 
		{\widehat h}_{n,k}\left(u+1\right) &= \pi_{n,k}\left(u\right){\widehat d}_{n,k}\left(u\right), \label{Eq_postMeanCh} \\
		v_{n,k}^h\left(u+1\right) &= \pi_{n,k}\left(u\right)\left[\left|{\widehat d}_{n,k}\left(u\right)\right|^2+v_{n,k}^d\left(u\right)\right] - \left|{\widehat h}_{n,k}\left(u+1\right)\right|^2 \label{Eq_postVarCh},
\end{align}
%\begin{align}
%	{\widehat h}_{n,k}\left(u+1\right) &= \pi_{n,k}\left(u\right){\widehat d}_{n,k}\left(u\right), \label{Eq_postMeanCh}\\
%	v_{n,k}^h\left(u+1\right) &= \pi_{n,k}\left(u\right)\left[\left|{\widehat d}_{n,k}\left(u\right)\right|^2+v_{n,k}^d\left(u\right)\right] \\
%	&\hspace{5mm}- \left|{\widehat h}_{n,k}\left(u+1\right)\right|^2. \label{Eq_postVarCh}
%\end{align}
respectively.
Similarly, the AWGN corrupted observation of the true $x_{k,l}$ and its variance are computed in \emph{lines \ref{Step_varR}} and \emph{\ref{Step_meanR}}.
Based on ${\widehat r}_{k,l}  = x_{k,l} + w_{k,l}^x, \forall k, l$, with $w_{k,l}^x \sim {\cal CN}\left(w_{k,l}^x; 0, v_{k,l}^r\right)$ and $p\left(x_{k,l}\right)$ in (\ref{Eq_sigPri}), the posterior mean ${\widehat x}_{k,l}$ and variance $v_{k,l}^x$ of $x_{k,l}$ in \emph{lines \ref{Step_varX}} and \emph{\ref{Step_meanX}}, respectively, are explicitly computed as
\begin{align}
	{\widehat x}_{k,l}\left(u+1\right) &= \frac{\sum\limits_{m=1}^M s_m{\rm exp}\left[-\left|s_m\right|^2 - \frac{2{\cal R}\left({\widehat r}_{k,l}^*\left(u\right)s_m\right)}{v_{k,l}^r\left(u\right)}\right]}{\sum\limits_{m=1}^M {\rm exp}\left[-\left|s_m\right|^2 - \frac{2{\cal R}\left({\widehat r}_{k,l}^*\left(u\right)s_m\right)}{v_{k,l}^r\left(u\right)}\right]}, \label{Eq_postMeanSig}\\
	v_{k,l}^x\left(u+1\right) &= \frac{\sum\limits_{m=1}^M \left|s_m\right|^2{\rm exp}\left[-\left|s_m\right|^2 - \frac{2{\cal R}\left({\widehat r}_{k,l}^*\left(u\right)s_m\right)}{v_{k,l}^r\left(u\right)}\right]}{\sum\limits_{m=1}^M {\rm exp}\left[-\left|s_m\right|^2 - \frac{2{\cal R}\left({\widehat r}_{k,l}^*\left(u\right)s_m\right)}{v_{k,l}^r\left(u\right)}\right]}, \label{Eq_postVarSig}
	\vspace{-1.5mm}
\end{align}
respectively.
Note that \emph{lines \ref{Step_varP1}-\ref{Step_meanX}} of \emph{Algorithm \ref{Alg_BiG-AMP}} constitute the basic version of the BiG-AMP algorithm developed in \cite{Ref_TSP_Parker'14}.
Here, based on the likelihood function and the \emph{a priori} distributions provided in (\ref{Eq_likelihood})-(\ref{Eq_sigPri}), we re-derived the explicit expressions of the MMSE estimates of ${\bf H}_{\rm act}$ and ${\bf X}_{\rm act}$, i.e., (\ref{Eq_postVarRx})-(\ref{Eq_postVarSig}).
In this paper, we further introduce the following two mechanisms to improve the realizability and the estimation reliability of the algorithm.

\subsubsection{EM-Based Hyper-Parameter Learning}
The implementation of the BiG-AMP algorithm requires the full knowledge of the likelihood function $p\left({\bf Y}|{\bf H}_{\rm act}, {\bf X}_{\rm act}\right)$ and the \emph{a priori} distributions $p\left({\bf H}_{\rm act}\right)$ and $p\left({\bf X}_{\rm act}\right)$.
In practice, only the families of these distributions are known in advance and the governing hyper-parameters ${\bm \xi} = \left\{\sigma^2, \mu_{n,k}, \tau_{n,k}, \gamma_{n,k}, \forall n, k\right\}$ are generally unknown to the BS.
Therefore, the EM algorithm proposed in~\cite{Ref_JRSC_Dempster'77} is incorporated to iteratively learn the unknown hyper-parameters.
Intuitively, each iteration of the EM algorithm consists of two steps: E-step computes the joint distribution of all involved variables given the current estimate of the hyper-parameters ${\widehat {\bm \xi}}\left(u\right)$; M-step re-estimates the hyper-parameters with the goal of maximizing the likelihood, as
\vspace{-0.5mm}
\begin{equation} \label{Eq_EM1}
	\begin{aligned}
		{\widehat {\bm \xi}}\left(u+1\right) &= \mathop {\arg \max}\limits_{\bm \xi} {\mathbb E}\left[{\rm ln} p\left({\bf H}_{\rm act}, {\bf X}_{\rm act}, {\bf Z}, {\bf Y}; {\bm \xi}\right)| {\bf Y}; {\widehat {\bm \xi}}\left(u\right)\right] \\
		&= \mathop {\arg \max}\limits_{\bm \xi} \left\{\sum\limits_{n=1}^{N}\sum\limits_{k=1}^{K_a}{\mathbb E}\left[{\rm ln} p\left(h_{n,k}; {\bm \xi}\right)| {\bf Y}; {\widehat {\bm \xi}}\left(u\right)\right] \right. \\
		&\hspace{5mm} + \sum\limits_{k=1}^{K_a}\sum\limits_{l=1}^{L}{\mathbb E}\left[{\rm ln} p\left(x_{k,l}; {\bm \xi}\right)| {\bf Y}; {\widehat {\bm \xi}}\left(u\right)\right] \\
		&\hspace{5mm} + \left. \sum\limits_{n=1}^{N}\sum\limits_{l=1}^{L}{\mathbb E}\left[{\rm ln} p\left(y_{n,l}|z_{n,l}; {\bm \xi}\right)| {\bf Y}; {\widehat {\bm \xi}}\left(u\right)\right] \right\}.
	\end{aligned}
\end{equation}
Here, the factorizability of $p\left({\bf H}_{\rm act}\right)$, $p\left({\bf X}_{\rm act}\right)$, and $p\left({\bf Y}|{\bf Z}\right)$ simplifies the computation of the joint distribution in (\ref{Eq_EM1}).
Moreover, instead of jointly optimizing all parameters in ${\bm \xi}$, we adopt the incremental update strategy from \cite{Ref_Book_Neal'98}, where ${\bm \xi}$ is updated one element at a time and the other parameters are held constant.
By setting the derivative of (\ref{Eq_EM1}) with respect to one element of ${\bm \xi}$ to zero, the estimates of the hyper-parameters are provided in \emph{lines \ref{Step_varN}-\ref{Step_meanChGain}} of \emph{Algorithm \ref{Alg_BiG-AMP}}.

\subsubsection{RI-Aided Initialization Strategy} 
For \emph{lines \ref{Step_initEM}} and \emph{\ref{Step_initBiGAMP}} of \emph{Algorithm \ref{Alg_BiG-AMP}}, the traditional random initialization strategy may lead the algorithm to converge to a local extremum of the mean-square-error function \cite{Ref_TSP_Ke'20}.
To avoid this situation, the authors in \cite{Ref_Access_Yan'19} proposed to initialize the algorithm multiple times and select the optimal pair of solutions as the final estimates, which improves the estimation accuracy but significantly increases the computational complexity. 
%In this paper, we introduce a more efficient RI-aided initialization strategy, where the embedded RI for eliminating phase and permutation ambiguities also serves as a short pilot sequence to acquire an initial estimate of the MRA channel matrix.
%Specifically, by treating the modulated device ID bits and CRC bits, as well as the scalar pilot symbol as the pilot matrix, the angular-domain joint ADD and CE scheme proposed in \cite{Ref_TSP_Ke'20} is employed to roughly estimate the active device set and the MRA channel matrix.
%On this basis, the transmitted signal matrix is further estimated by the least squares method.
{\color{red} In this paper, we propose a more efficient RI-aided initialization strategy, where the transmitted reference signal for eliminating phase and permutation ambiguities also serves as a short pilot sequence to acquire an initial estimate of the MRA channel matrix.
Specifically, the reference signal is composed of the modulated symbols of device ID bits and CRC bits, as well as the scalar pilot symbol.
By stacking all devices' reference signals in rows as the pilot matrix, the angular-domain joint ADD and CE scheme proposed in \cite{Ref_TSP_Ke'20}  is employed to acquire the coarse estimates of the active device set and MRA channel matrix.
On this basis, the transmitted signal matrix of active devices can be further estimated by the least squares (LS) method.
The aforementioned processing has been detailed in \textit{Section III-A}, i.e., non-orthogonal pilot-based coherent detection.
Note that the hyper-parameters can be simultaneously estimated with the incorporated EM algorithm, as in \cite{Ref_TSP_Ke'20} .
In this context, the estimated channel matrix, signal matrix, and hyper-parameters are exploited as the initial estimates of the proposed BiG-GAMP-based JCSE algorithm.}

\subsection{SIC-Based Semi-Blind Detection Scheme}
\label{SubSec_sicSemiBlind}

With the BiG-AMP-based JCSE algorithm developed in \emph{Section \ref{SubSec_jcseAlg}}, we further propose an SIC-based semi-blind detection scheme, where the embedded RI is utilized for ambiguity elimination.
Meanwhile, the SIC technique is incorporated to mitigate the inter-device interference iteratively, as summarized in \emph{Algorithm \ref{Alg_sicSemiBlind}}.
%Similarly, we first focus on sourced RA here.
In the $j$th SIC iteration, \emph{line \ref{Step_computeRes}} computes the residual received signal ${\widetilde {\bf Y}}^j$ and the residual number of active devices ${\widehat K}_a^j$, where ${\widehat {\cal A}}^{j-1}$, ${\widehat {\bf H}}^{j-1}$, and ${\widehat {\bf X}}^{j-1}$ are the estimated active device set, channel matrix, and signal matrix in the last iteration, respectively.
If all the active devices have been detected or the power of the residual received signal is small enough, the processing is terminated to avoid unnecessary iterations, see \emph{lines \ref{Step_tolSIC1}-\ref{Step_tolSIC2}}.
In \emph{line \ref{Step_JCSE}}, without regard for the phase and permutation ambiguities, we employ the BiG-AMP-based JCSE algorithm to jointly infer the residual channel and signal matrices, i.e., $\left[{\bf H}\right]_{:,{\cal A}^j}$ and $\left[{\bf X}\right]_{{\cal A}^j,:}$, respectively, based on the following model
\begin{equation} \label{Eq_sigRes}
	\begin{aligned}
		{\widetilde {\bf Y}}^j = {\bf Y} - {\widehat {\bf H}}^{j-1}{\widehat {\bf X}}^{j-1} 
			= \left[{\bf H}\right]_{:,{\cal A}^j}\left[{\bf X}\right]_{{\cal A}^j,:} + {\widehat {\bf N}} + {\bf N},
	\end{aligned}
\end{equation}
with the estimation error of the last iteration given as
\begin{equation} \label{Eq_estErr}
	\begin{aligned}
		{\widehat {\bf N}} = \left(\left[{\bf H}\right]_{:,{\widehat {\cal A}}^{j-1}}-\left[{\widehat {\bf H}}^{j-1}\right]_{:,{\widehat {\cal A}}^{j-1}}\right) \left(\left[{\bf X}\right]_{{\widehat {\cal A}}^{j-1},:}-\left[{\widehat {\bf X}}^{j-1}\right]_{{\widehat {\cal A}}^{j-1},:}\right).
	\end{aligned}
\end{equation}
Here, ${\cal A}^j = {\cal A} - {\widehat {\cal A}}^{j-1}$ denotes the residual active devices to be detected.
%Particularly, we propose a clustered sparsity-based \emph{a prior} refinement strategy and a re-initialization mechanism to improve the performance of the conventional BiG-AMP algorithm.
%The details of the updated algorithm will be provided in the next subsection.

\begin{breakablealgorithm}
	\caption{SIC-Based Semi-Blind Detection Scheme}
	\label{Alg_sicSemiBlind}
	\begin{algorithmic}[1]
		\REQUIRE Angular-domain received signal ${\bf Y}$, the maximum number of SIC iterations $J$, and termination threshold $\epsilon_{\rm sic}$.
		\ENSURE The estimated active device set ${\widehat {\cal A}}$, channel matrix ${\widehat {\bf H}}$, signal matrix ${\widehat {\bf X}}$, and binary data matrix ${\widehat {\bf B}}$.
		\STATE Estimate the number of active devices based on (\ref{Eq_rankSel}).
		\STATE // SIC Initializations:
		\STATE ${\widehat {\cal A}}^0 = \emptyset$, ${\widehat {\bf H}}^0 = {\bf 0}_{N \times K}$, ${\widehat {\bf X}}^0 = {\bf 0}_{K \times L}$, ${\widehat {\bf B}}^0  = {\bf 0}_{K \times B}$.\!\!\!
		\vspace{0.5mm}
		\label{Step_initSIC}
		\STATE // SIC Loops:
		\label{Step_startSIC}
		\FOR {$j = 1, \cdots, J$}
		\vspace{0.5mm}
		\STATE ${\widetilde {\bf Y}}^j = {\bf Y} - {\widehat {\bf H}}^{j-1}{\widehat {\bf X}}^{j-1}$, ${\widehat K}_a^j = {\widehat K}_a - \left|{\widehat {\cal A}}^{j-1}\right|_c$
		\vspace{0.5mm}
		\label{Step_computeRes}
		\IF {${\widehat K}_a^j \le 0$ $||$ ${\left\|{\widetilde {\bf Y}}^j\right\|_{\rm F}^2}/{\left\|{\bf Y}\right\|_{\rm F}^2} < \epsilon_{\rm sic}$}
		\vspace{0.5mm}
		\label{Step_tolSIC1}
		\STATE break;
		\ENDIF
		\label{Step_tolSIC2}
		\STATE Employ the BiG-AMP-based JCSE algorithm to acquire the estimates of residual channel and residual signal matrices, i.e., ${\widehat {\bf H}}_{\rm act}^j$ and ${\widehat {\bf X}}_{\rm act}^j$, respectively, based on ${\widetilde {\bf Y}}^j$, where the phase and permutation ambiguities are ignored, as in \emph{Algorithm~\ref{Alg_BiG-AMP}}.
		\label{Step_JCSE}
		\STATE // Phase Ambiguity Elimination:
		\vspace{0.5mm}
		\STATE ${\widehat \Sigma} = {\rm diag}\left(x_p / \left[{\widehat {\bf X}}_{\rm act}^j\right]_{:,1}\right)$, ${\widehat {\bf H}}_{\rm act}^j = {\widehat {\bf H}}_{\rm act}^j{\widehat \Sigma}^{-1}$, ${\widehat {\bf X}}_{\rm act}^j = {\widehat \Sigma}{\widehat {\bf X}}_{\rm act}^j$
		\label{Step_phaseCorrect}
		\STATE With ${\widehat {\bf X}}_{\rm act}^j$, perform $M$-PSK demodulation to obtain the estimated binary data matrix ${\widehat {\bf B}}_{\rm act}^j$.\!\!\!\!\!\!
		\label{Step_deMod}
		\STATE With ${\widehat {\bf B}}_{\rm act}^j$, perform CRC to acquire the corresponding checking result ${\bf c}^j \in {\mathbb C}^{{\widehat K}_a^j \times 1}$.
		\label{Step_CRC}
		\vspace{0.5mm}
		%\STATE ${\widehat {\bf H}_{\tilde n}}^j = {\widehat {\bf H}_{\tilde n}}^{j-1}$, ${\widehat {\bf X}}_{\tilde n}^j = {\widehat {\bf X}}_{\tilde n}^{j-1}$, ${\widehat {\bf B}}_{\tilde n}^j = {\widehat {\bf B}}_{\tilde n}^{j-1}$
		\STATE // Permutation Ambiguity Elimination:
		\vspace{0.5mm}
		\FOR {$k = 1, \cdots, {\widehat K}_a^j$}
		\label{Step_permuteCorrectStart}
		\STATE ${\widehat k} = {\rm bi2dec}\left(\left[{\widehat {\bf B}}_{\rm act}^j\right]_{k,{\cal B}_i}\right)$, ${\cal B}_i = \left\{1, \cdots, B_i\right\}$
		\label{Step_deviceIden}
		\IF {$0 \le {\widehat k} \le K$ $\&\&$ ${\bf c}^j\left(k\right) == 0$}
		\vspace{0.5mm}
		\STATE ${\widehat {\cal A}}^j = {\widehat {\cal A}}^{j-1} \cup {\widehat k}$, $\left[{\widehat {\bf B}}^j\right]_{{\widehat k},:} = \left[{\widehat {\bf B}}_{\rm act}^j\right]_{k,:}$
		\vspace{0.5mm}
		\label{Step_estUpdate1}
		\STATE $\left[{\widehat {\bf H}}^j\right]_{:,{\widehat k}} = \left[{\widehat {\bf H}}_{\rm act}^j\right]_{:,k}$, $\left[{\widehat {\bf X}}^j\right]_{{\widehat k},:} = \left[{\widehat {\bf X}}_{\rm act}^j\right]_{k,:}$
		\label{Step_estUpdate2}
		\ENDIF
		\label{Step_permuteCorrectEnd}
		\ENDFOR
		%\STATE // Soft Pilot-Based CSI Refinement
		%\STATE Re-modulate the correctly detected data bits $[{\widehat {\bf B}}^j]_{{\widehat {\cal A}}^j,:}$ as the soft pilot matrix ${\widehat {\bf X}}_p$.
		%\STATE Refine the CSI estimates as in (\ref{Eq_softEst}).
		\ENDFOR
		\RETURN ${\widehat {\cal A}}^j$, ${\widehat {\bf H}}^j$, ${\widehat {\bf X}}^j$, ${\widehat {\bf B}}^j$
	\end{algorithmic}
\end{breakablealgorithm}

For a specific active device, it has been revealed in \cite{Ref_ICASSP_Ngo'12} that the phase shifts of phase ambiguity are identical for all transmitted symbols.
Therefore, the phase ambiguity can be eliminated by computing the corresponding phase shift as in \emph{line \ref{Step_phaseCorrect}} of \emph{Algorithm \ref{Alg_sicSemiBlind}}, where $x_p$ is the common scalar pilot symbol inserted in the access signals for all the active devices.
Given the estimated signal matrix with phase correction, i.e., ${\widehat {\bf X}}_{\rm act}^j$, \emph{line \ref{Step_deMod}} further executes $M$-PSK demodulation to obtain the estimated binary data matrix ${\widehat {\bf B}}_{\rm act}^j$.
Moreover, leveraging the validity checking procedure and the inserted device ID bits, the permutation ambiguity is further resolved, as in \emph{lines \ref{Step_permuteCorrectStart}-\ref{Step_permuteCorrectEnd}} of \emph{Algorithm~\ref{Alg_sicSemiBlind}}.
Specifically, the CRC is firstly adopted to validate the correctness of the detected ID bits, as
\begin{equation} \label{Eq_crc}
	\begin{aligned}
		{\bf c}^j\left(k\right) = {\hat f}\left(\left[{\widehat {\bf B}}_{\rm act}^j\right]_{k,{\cal B}_i} \div {\bf p}_c\right), \forall k \in \left[{\widehat K_a^j}\right], 
	\end{aligned}
\end{equation}
where ${\cal B}_i$ is the index set of device ID bits, ${\hat f}\left(\cdot\right) = 1$ if the remainder of the binary division is non-zero and ${\hat f}\left(\cdot\right) = 0$ otherwise.
For the $k$th detected active device, if all the device ID bits are correctly recovered, i.e., ${\bf c}^j\left(k\right) = 0$, the estimated device ID can be used to identify the corresponding active device, which is added to the estimated active device set ${\widehat {\cal A}}^j$, and the corresponding estimates are updated, as in \emph{lines \ref{Step_estUpdate1} and \ref{Step_estUpdate2}}.

Obviously, the phase and permutation ambiguities can be effectively resolved based on the embedded RI, i.e., the device ID bits, the CRC bits, and the scalar pilot symbol.
Meanwhile, the amount of RI scales logarithmically with the number of devices, as detailed in {\em Section \ref{Sec_txDesign}}, which leads to a very small time resource consumption.
Moreover, due to the fact that the reliable BiG-AMP-based JCSE makes the error propagation controllable, the proposed SIC-based semi-blind detection scheme can further enhance the detection reliability by mitigating the inter-device interference iteratively.    

%Besides, the correctly detected data bits $[{\widehat {\bf B}}^j]_{{\widehat {\cal A}}^j,:}$ can be re-encoded and re-modulated as the soft pilots ${\widehat {\bf X}}_p$, which are used to enhance the channel estimation accuracy, as
%\begin{equation} \label{Eq_softEst}
%\begin{aligned}
%\left[{\widehat {\bf H}}^j\right]_{{\widehat {\cal A}}^j,:} = {\bf Y}{\bf X}_p^{\dag}.
%\end{aligned}
%\end{equation}
%With the reduced estimation error ${\widehat {\bf N}}$, the error propagation of SIC would be significantly mitigated.

%\begin{figure}[!t]
%     \centering
%     \includegraphics[width=1\columnwidth, keepaspectratio]
%     {Fig6/Fig6.eps}
%     \caption{\small{The proposed unified semi-blind detection framework for sourced and unsourced RA paradigms.}}
%     \label{Fig6}
%     \vspace{-2mm}
%\end{figure}

\subsection{Extension to Unsourced RA}

\begin{figure}[!t]
	\centering
	\includegraphics[width=0.85\columnwidth, keepaspectratio]
	{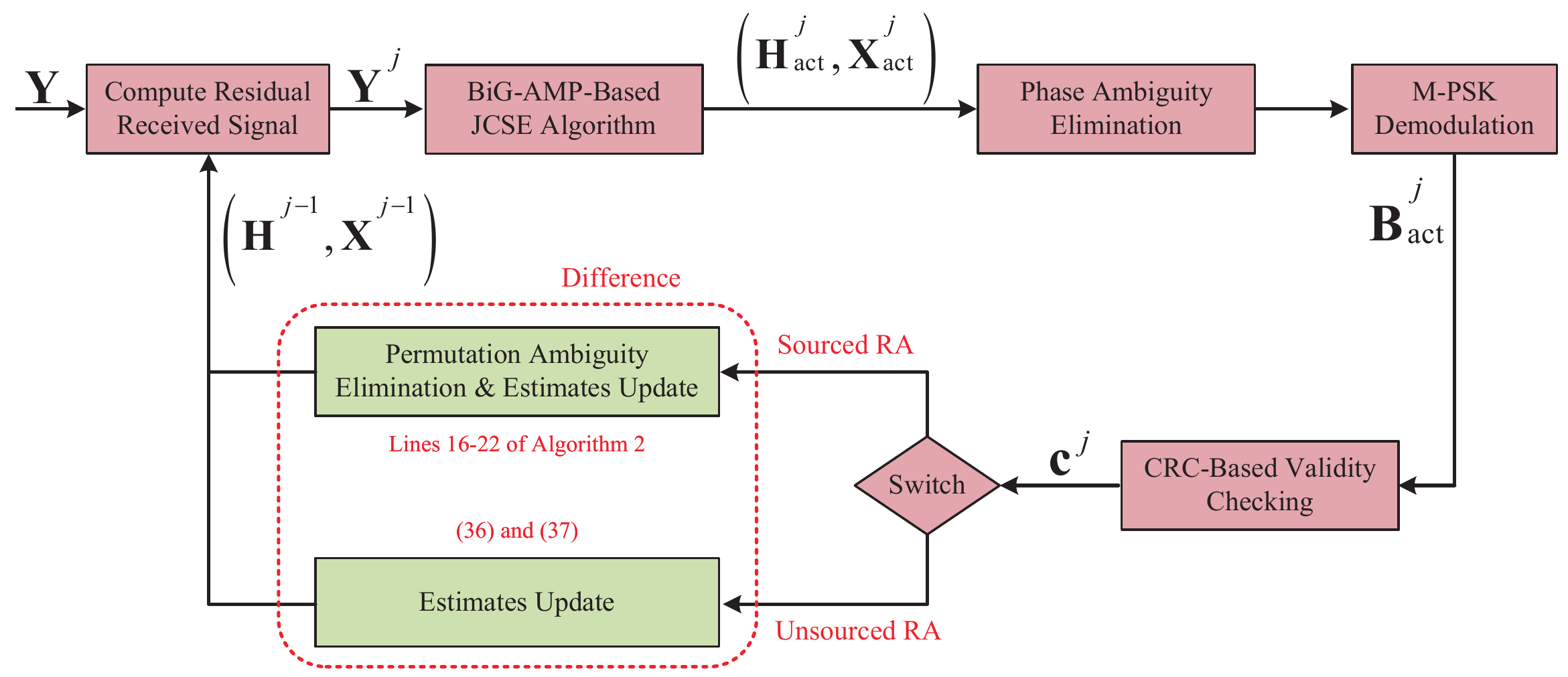}
	\caption{\small{Block diagram of the proposed unified SIC-based semi-blind detection scheme at the BS.}}
	\label{Fig_rxDesign}
	\vspace{-3mm}
\end{figure}

As analyzed in \emph{Section \ref{Sec_txDesign}}, the major difference between sourced and unsourced RA lies in that the permutation ambiguity in the unsourced RA does not have to be resolved.
Therefore, the proposed SIC-based semi-blind scheme can be directly applied to the unsourced RA by just making some minor modifications at the software level.
Specifically, since the CRC is utilized to validate the correctness of the detected payload data bits rather than the detected ID bits, the validity checking procedure in \emph{lines \ref{Step_CRC}} is modified to   
\begin{equation} \label{Eq_crc2}
	\begin{aligned}
		{\bf c}^j\left(k\right) = {\hat f}\left(\left[{\widehat {\bf B}}_{\rm act}^j\right]_{k,{\cal B}_d} \div {\bf p}_c\right), \forall k \in \left[{\widehat K_a^j}\right], 
	\end{aligned}
\end{equation}
where ${\cal B}_d$ is the index set of payload data bits.
Moreover, the step for device identification, i.e., \emph{line \ref{Step_deviceIden}}, is removed.
Finally, the update rules in \emph{lines \ref{Step_estUpdate1}} and \emph{\ref{Step_estUpdate2}} are modified as
\begin{equation}
{\widehat {\cal A}}^j = {\widehat {\cal A}}^{j-1} \cup \left({\widehat K}_a^{j-1}+k\right), {\rm if} \hspace{1mm} {\bf c}^j\left(k\right) == 0,
\end{equation}
\begin{equation}
\left[{\widehat {\bf H}}^j\right]_{:,{\widetilde {\cal A}}^j} = {\widehat {\bf H}_{\rm act}}^j,
\left[{\widehat {\bf X}}^j\right]_{:,{\widetilde {\cal A}}^j} = {\widehat {\bf X}}_{\rm act}^j,
\left[{\widehat {\bf B}}^j\right]_{:,{\widetilde {\cal A}}^j} = {\widehat {\bf B}}_{\rm act}^j,
\end{equation}
where ${\widehat {\cal A}}^j$ represents the active device set with correctly detected data bits and unknown identity, ${\widehat K}_a^{j-1}$ denotes the number of devices that have been detected in the $(j-1)$th SIC iteration, and ${\widetilde {\cal A}}^j = \left\{k|{\widehat K}_a^{j-1}+1 \le k \le {\widehat K}_a^{j-1}+{\widehat K}_a^j\right\}$.
In this context, we propose a unified semi-blind data detection scheme at the BS, as shown in Fig. \ref{Fig_rxDesign}.
Here, both RA paradigms share the same hardware modules and a software-defined switch is utilized to determine which RA mode is enabled.
This facilitates more flexible network deployment and reduces the cost of network re-configuration.

{\color{red} \subsection{Computational Complexity Analysis} \label{Section V-E}

For the practical implementation, the processing latency mainly depends on the computational complexity of the adopted receive algorithm.
In the non-orthogonal pilot-based coherent detection for sourced RA, the complexity of AMP-based joint ADD and CE is calculated by the function as
\begin{equation} \label{cmplt_amp}
	\begin{aligned}
		C_{\rm amp}\left(N, K, L_p, U_a\right) = U_a\left(4NKL_p + 3KL_p + 16NL_p + 20NK\right),
	\end{aligned}
\end{equation}
and the complexity of LS-based coherent data detection is calculated as
\begin{equation} \label{cmplt_ls}
	\begin{aligned}
		C_{\rm ls}\left(N, K_a, L_d\right) = K_a^3 + NK_a^2 + NK_aL_d,
	\end{aligned}
\end{equation}
where $U_a$ is the number of AMP iterations.
Therefore, the overall computational complexity is in the order of ${\cal O} \left[C_{\rm amp}\left(N, K, L_p, U_a\right) + C_{\rm ls}\left(N, K_a, L_d\right)\right]$.
While in the common codebook-based non-coherent detection for unsourced RA, the computational complexity mainly stems from the AMP-based codeword detection, which is in the order of ${\cal O} \left[C_{\rm amp}\left(N, 2^B, L, U_a\right)\right]$.

The computational complexity of the proposed unified semi-blind detection framework is mainly composed of three parts.
Specifically, the complexity of SVD-based rank selection is calculated as
\begin{equation} \label{cmplt_svd}
	\begin{aligned}
		\hspace{-10mm}	C_{\rm svd}\left(N, L\right) = 2NL^2 + L^3 + L + NL, 
	\end{aligned}
\end{equation}
the complexity of RI-aided initialization is calculated as
\begin{equation} \label{cmplt_init}
	\begin{aligned}
		C_{\rm init}\left(N, K, K_a, L, L_r, U_a\right) = C_{\rm amp}\left(N, K, L_r, U_a\right) + C_{\rm ls}\left(N, K_a, L-L_r\right), 
	\end{aligned}
\end{equation} 
and the complexity of BiG-AMP-based JCSE is calculated as
\begin{equation} \label{cmplt_jcse}
	\begin{aligned}
		\hspace{-20mm}	C_{\rm jcse}\left(N, K_a, L, U\right) = U\left(NK_a + K_aL + NL\right). 
	\end{aligned}
\end{equation}
Here, $L_r$ is the number of consumed symbol durations for transmitting the RI, which is expressed as
\begin{equation}\label{Eq_lenRef}
	L_r = \left\lceil\frac{B_i + B_c} {{\rm log}_2\left(M\right)}\right\rceil + 1.
\end{equation}
Further considering the SIC procedure, the overall computational complexity of the proposed semi-blind detection framework is in the order of ${\cal O}\left[C_{\rm sic}\left(N, K, K_a, L, L_r, U_a\right)\right]$ with
\begin{equation} \label{cmplt_basic}
	\begin{aligned}
		C_{\rm sic}\left(N, K, K_a, L, L_r, U_a, U\right) &= \sum_{j = 1}^J C_{\rm svd}\left(N, L\right) + C_{\rm init}\left(N, K, {\widehat K_a^j}, L, L_r, U_a\right) \\ 
		&\hspace{10mm} + \hspace{1.5mm} C_{\rm jcse}\left(N, {\widehat K_a^j}, L, U\right) + C_{\rm res}\left(N, \left|{\widehat {\cal A}^{j-1}}\right|_c, L\right), 
	\end{aligned}
\end{equation}
where $C_{\rm res}(N, |{\widehat {\cal A}^{j-1}}|_c, L) = N|{\widehat {\cal A}^{j-1}}|_cL$ is the complexity for computing the residual received signal, ${\widehat K_a^j}$ is the number of active devices to be estimated in the $j$th SIC iteration, $|{\widehat {\cal A}^{j-1}}|_c$ is the detected active devices in the $(j-1)$th SIC iteration, and $J$ is the number of SIC iterations.
Since the RI-aided initialization strategy is only applicable for sourced RA, the complexity of RI-aided initialization, i.e., $C_{\rm init}\left(N, K, K_a, L, L_r, U_a\right)$, should be removed from (\ref{cmplt_basic}) when unsourced RA is considered.}

{\color{red} \section{ Sourced RA and Unsourced RA Coexistence Schemes} \label{Section VI}

For simplicity, the previous descriptions on the proposed unified transceiver design assume that the BS provides only one of the sourced and unsourced RA services during a given time interval, and the RA mode may switch between sourced and unsourced RA in different time intervals.
Meanwhile, the enabled RA mode is assumed to be known in advance at both the devices and the BS.
These assumptions fail to consider the more general sourced RA and unsourced RA (SRA-URA) coexistence scenarios having unknown device access requirements.
To this end, we further develop two SRA-URA coexistence schemes, where the aforementioned unified transceiver design can be directly applied by making minor software-level updates.

\subsection{Orthogonal SRA-URA Coexistence Scheme} \label{Section VI-A}

Considering the devices with periodic uplink traffic and predictable access requirements, such as the sensors that need to report their data periodically, we first propose an orthogonal SRA-URA coexistence scheme.
Specifically, for a specific time interval, the potential devices are divided into two groups, i.e., sourced and unsourced RA device groups, according to their practical access requirements.
Meanwhile, the time-frequency resources reserved for grant-free RA are divided into multiple orthogonal resource blocks (RBs), which are then allocated to the two device groups for avoiding inter-group interferences.
Here, due to the predictable uplink traffic and access requirements, the associations between the device groups and the orthogonal RBs are pre-configured.
In this context, the received signals of sourced and unsourced RA are distinguishable at the BS.
Therefore, the proposed unified transceiver design in \textit{Sections IV} and \textit{V} can be independently applied to all RBs for semi-blind data detection, and the RA mode can flexibly switch between sourced and unsourced RA according to the served device type in different RBs.
The proposed orthogonal SRA-URA coexistence scheme is inspired by the traditional orthogonal frequency division multiple access (OFDMA) developed in the fourth-generation (4G) Long-Term Evolution (LTE), where the interferences among all the active devices are avoided through orthogonal resource allocation.
The key difference lies in that only the access signals from different device groups are orthogonal, while the signals from the same device group are still overlapped on the same RB.

\subsection{Non-Orthogonal SRA-URA Coexistence Scheme} \label{Section VI-B}

In practice, since a considerably number of devices may randomly access the network and change their access requirements, the application scenario of the aforementioned orthogonal SRA-URA coexistence scheme is still very limited.
To overcome this limitation, we further propose a non-orthogonal SRA-URA coexistence scheme, where the access signals of both types of devices, i.e., sourced and unsourced RA devices, are directly transmitted exploiting the same time-frequency resources without uplink resource pre-allocation or scheduling.
In this case, the received signals of sourced and unsourced RA are overlapped at the BS and unable to be separated.
By exploiting the common receive modules of sourced and unsourced RA paradigms, i.e., \textit{lines 1-14} of \textit{Algorithm 2}, the overlapped received signals can be jointly processed to obtain the estimated data packets of active devices, while their adopted RA modes are still unavailable.
To tackle this issue, we propose to add a one-bit mode indicator at the beginning of the data packet, which serves as the reference information and indicates the adopted RA mode.
Here, the mode indicator takes 1 for sourced RA and 0 for unsourced RA. 
Since there are two types of RA modes, only one-bit reference information is sufficient to identify which RA mode is adopted by the active devices.
At this point, with the estimated data packets and corresponding mode indicators, the remaining steps of the proposed semi-blind detection scheme can be executed to acquire the final estimates of the payload data bits. 
Specifically, for a specific detected active device, if its RA mode is judged to be sourced RA, the corresponding estimated data packet is processed by \textit{lines 16-22} of \textit{Algorithm 2} for permutation ambiguity elimination and estimates update; otherwise, the estimated data packet is processed by (36) and (37) for estimates update, as detailed in \textit{Section V} and illustrated in Fig. 5.
It is clear that, based on the proposed unified transceiver design presented in \textit{Sections IV} and \textit{V}, the aforementioned non-orthogonal SRA-URA coexistence scheme can be realized by making minor software-level updates.}

\section{URLLC Enhancements}
\label{URLLC}

{\color{red} According to previous discussions, the proposed unified semi-blind detection framework facilitates massive URLLC via simplifying the access scheduling, improving the payload efficiency, reducing the computational complexity, and enhancing the JCSE reliability.
However, these are still not enough to satisfy the stringent latency and reliability requirements in the context of massive devices, e.g., over $99.99999\%$ reliability within 1 ms user plane latency for 32 bytes~\cite{Ref_IoTJ_Ding'22}.
Indeed, it is generally difficult for a single grant-free MRA technique to satisfy these requirements and several key enabling techniques should be further integrated to achieve the goal.}

\subsection{Multi-Carrier Deployment}

\begin{figure}[!t]
	\centering
	\includegraphics[width=0.7\columnwidth, keepaspectratio]
	{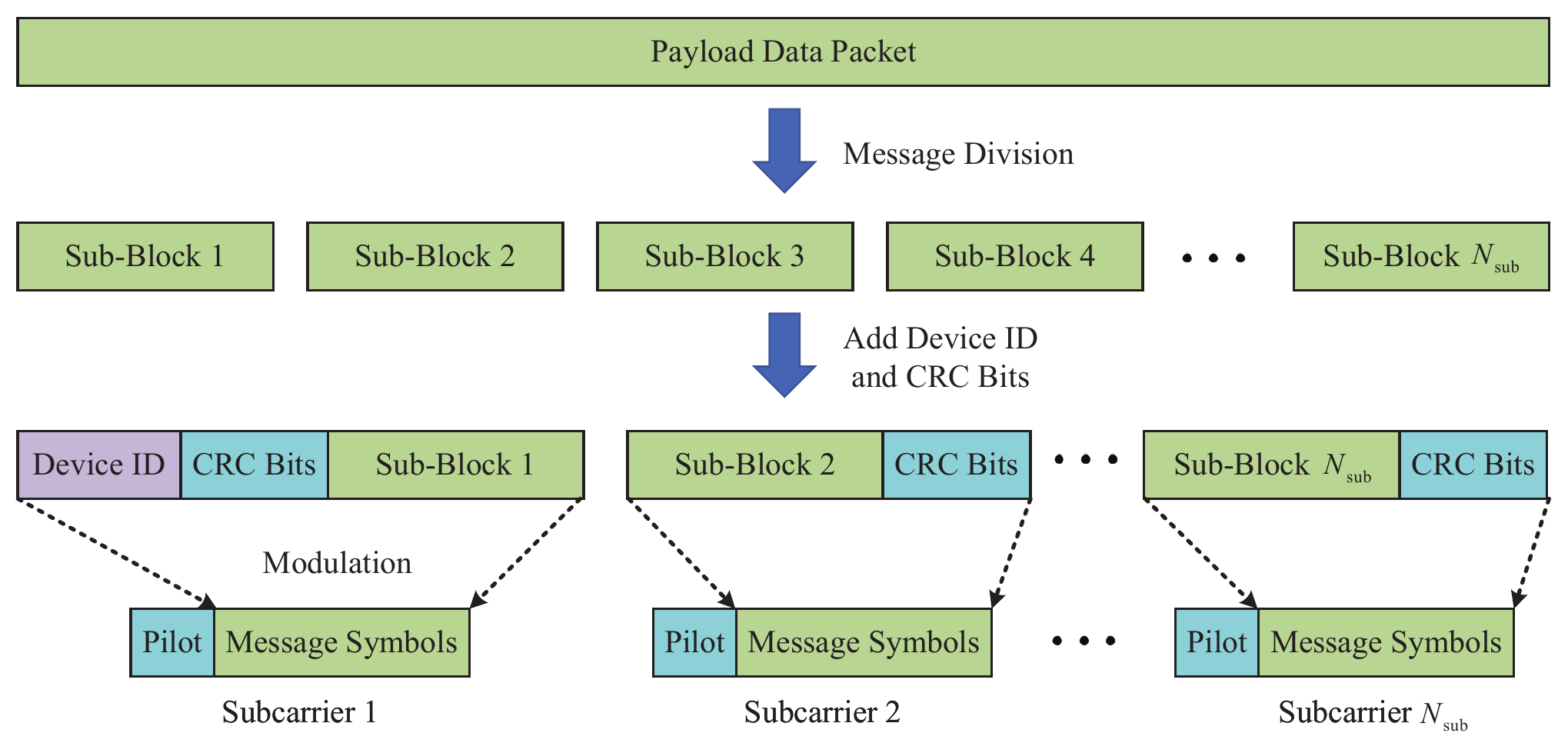}
	\caption{\small{Multi-carrier deployment of the proposed semi-blind detection framework.}}
	\label{Fig_mcDeploy}
	\vspace{-3mm}
\end{figure}

The basic version of the proposed semi-blind detection framework considers the single-carrier transmission.
In practical orthogonal frequency division multiplexing (OFDM) systems, it can be directly deployed by selecting one of the subchannels for grant-free MRA, while the remaining subchannels are reserved for other purposes, such as control signaling exchanges.
Meanwhile, it can be easily extended to multi-carrier deployment, where the payload data bits are delivered in parallel at multiple subcarriers for further reduced transmission latency.
Specifically, for each active device, its data packet is uniformly split into $N_{\rm sc}$ sub-blocks, where $N_{\rm sc}$ is the number of occupied subcarriers.
Then, the device ID bits, the CRC bits, and the scalar pilot symbol are inserted in each sub-blocks to eliminate the ambiguities, as detailed in {\em Section~\ref{Sec_txDesign}}.
On this basis, the single-carrier version of the proposed SIC-based semi-blind detection scheme is employed to detect the sub-blocks carried by different subcarriers.
Finally, the original data packet is acquired by stitching the detected sub-blocks together according to the device ID. 
However, it is not efficient to insert the device ID bits in all the sub-blocks, which significantly degrades the payload efficiency.
The authors in~\cite{Ref_TSP_Gao'15} have revealed that the subchannels across different subcarriers generally have a common sparsity pattern in the angular domain.
Meanwhile, the AoAs of different devices are usually distinguishable.
Considering this characteristic, we propose a more efficient deployment for the improved payload efficiency, where only one sub-block is selected to carry the device ID bits to eliminate the permutation ambiguity of the corresponding subchannels, as depicted in Fig. \ref{Fig_mcDeploy}.
While the permutation ambiguities of the remaining subchannels are resolved by leveraging the fact that the subchannels having a common angular-domain sparsity pattern belong to the same active device.
This can be realized by various clustering algorithms, such as $K$-means algorithm in \cite{Ref_Tcom_Xie'22}.
{\color{red} Based on the above descriptions, the complexity of the multi-carrier deployment of the proposed detection framework is calculated as
\begin{equation} \label{cmplt_mc}
	\begin{aligned}
		C_{\rm mc}\left(N, K, K_a, L, L_r, U_a, U, N_{\rm sc}\right) &= C_{\rm sic}\left(N, K, K_a, L_1, L_r, U_a, U\right) \\ & \hspace{5mm} + \left(N_{\rm sc}-1\right)\left[C_{\rm svd}\left(N, L_2\right) + C_{\rm jcse}\left(N, K_a, L_2, U\right)\right], 
	\end{aligned}
\end{equation}
where $L_{\rm sc} = (L-L_r)/N_{\rm sc}$ is the payload signal length at each subcarrier, $L_1 = L_{\rm sc} + L_r$ is the overall signal length of the first subcarrier, $L_2 = L_{\rm sc} + L_r - L_i$ is the overall signal length of the remaining subcarriers, and $L_i$ is the signal length for transmitting the device ID bits.}

\subsection{Flexible Frame Structure}

\begin{figure}[!t]
	\centering
	\includegraphics[width=0.7\columnwidth, keepaspectratio]
	{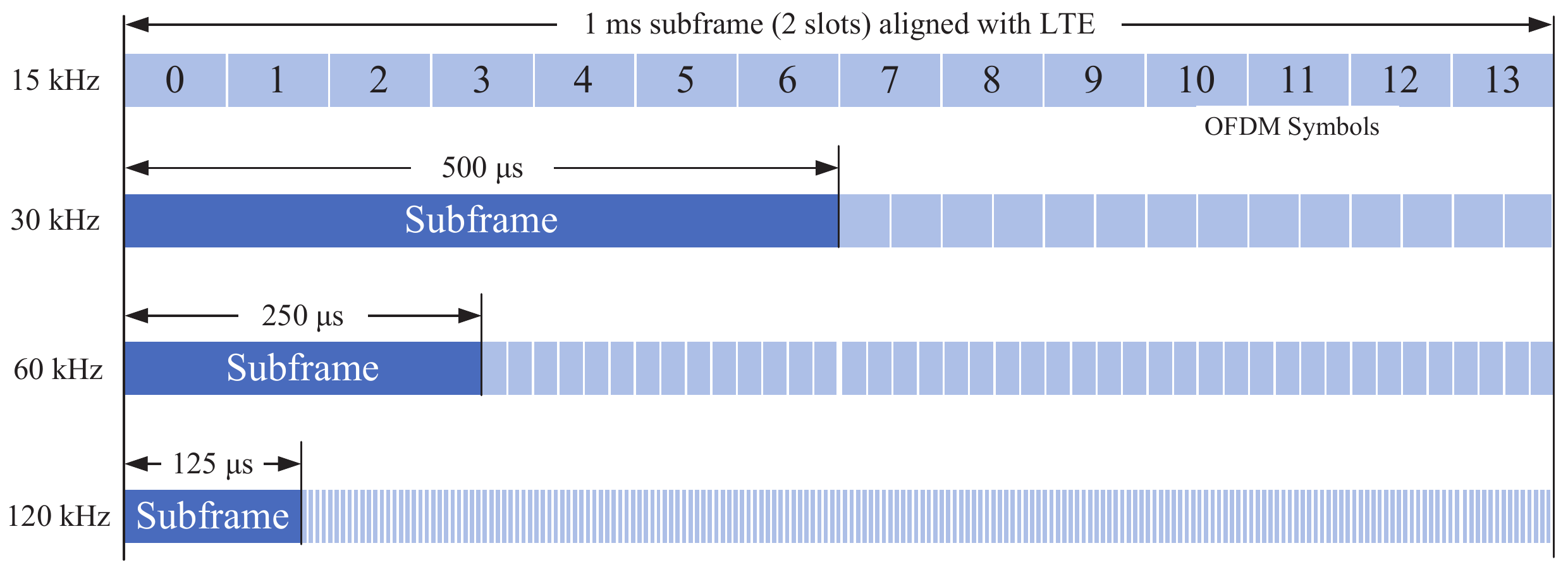}
	\caption{{\color{red} \small{The flexible frame structure introduced in 5G NR, where different SCSs affects the slot duration and TTI.}}}
	\label{Fig_FlxFrameStr}
	\vspace{-3mm}
\end{figure}

In addition to payload efficiency, the transmission time interval (TTI) also plays an important role in contributing to the transmission latency \cite{Ref_IoTJ_Ding'22}.
Hence, reducing TTI is another key to satisfying the ultra-low user plane latency of massive URLLC.
{\color{red} In the 4G LTE, the subcarrier spacing (SCS) is fixed at 15 kHz and  each TTI contains 14 OFDM symbols, leading to a TTI (equals to two slots) of 1ms.}
%In the forth-generation (4G) Long-Term Evolution (LTE), the subcarrier spacing (SCS) is fixed at 15 kHz and each TTI contains 14 OFDM symbols, leading to a TTI (equals to a slot)  of 1ms.
This is only the transmission time on the air interface.
The overall user plane latency would be much larger than 1 ms by further considering other delay components.
Therefore, the 5G New Radio (NR) has introduced a more flexible frame structure, where the TTI can be shortened by using the scalable SCS \cite{Ref_ISWCS_Li'18}.
Specifically, each frame with 10 ms consists of 10 subframes and the number of slots within a certain subframe depends on the SCS, as illustrated in Fig.~\ref{Fig_FlxFrameStr}.
Furthermore, each slot is composed of 14 OFDM symbols.
By using different SCSs in 5G NR, different slot durations and TTIs are configurable.
For example, 15 kHz SCS with 14 symbols spanning the entire 1 ms subframe corresponds to the LTE's configuration.
While at 240 kHz SCS, 14 symbols are squeezed into a mini-slot with 62.5 $\mu{\rm s}$, thus significantly reducing the transmission latency. 

\subsection{Concurrent Access Mechanism}

By reducing the transmission latency, the aforementioned multi-carrier deployment and flexible frame structure also create more retransmission opportunities within a target latency.
As a result, various hybrid automatic repeat request (HARQ) transmission schemes, including reactive HARQ, $K$-repetition HARQ, and proactive HARQ, can be incorporated into grant-free MRA for further enhanced reliability \cite{Ref_IoTJ_Ding'22}.
However, the number of retransmission times is still very limited due to the stringent latency requirement.
To overcome this limitation, we further propose a concurrent access mechanism in this paper, which resorts to the relatively richer frequency resource for diversity gain.
Specifically, the whole bandwidth is divided into multiple independent sub-bands.
Moreover, the same payload data is repeatedly delivered in these sub-bands, where the proposed semi-blind detection framework is employed for each sub-band.
In this context, the detection reliability could be dramatically improved due to the frequency diversity.
More specifically, although an active device may be missed in a specific sub-band, it can be successfully detected in other sub-bands with a high probability. 
{\color{red} The overall computational complexity of the URLLC-enhanced semi-blind detection framework is in the order of  ${\cal O}\left[N_bC_{\rm mc}\left(N, K, K_a, L, L_r, U_a, U, N_{\rm sc}\right)\right]$, where $N_b$ is the number of sub-bands.}

\subsection{Adaptive Transmit Power Control}

In the proposed semi-blind detection framework, the corresponding whole payload data packet would be lost if an active device is not successfully detected, i.e., miss detection.
On the other hand, all the active devices have an identical transmit power.
Due to the severe path loss, the active devices located in the cell edge generally suffer from a far smaller received SNR than those in the cell center.
This leads to a high miss detection probability of active devices in the cell edge and becomes a major limiting factor for improving the data detection reliability \cite{Ref_WCM_Ke'21}.
To tackle this issue, we propose an adaptive transmit power control (ATPC), where the transmit power of the $k$th device is given as $P_k = {P_{\rm max}d_k^{\tilde \alpha}}/{d_{\rm max}^{\tilde \alpha}}$.
Here, $P_{\rm max}$ is the maximum transmit power, $d_k$ is the distance between the $k$th device and the BS, ${\tilde \alpha}$ is the path loss decay exponent, and $d_{\rm max}$ is the cell radius.
In this context, all the active devices will have a similar received SNR at the BS, which significantly improves the data detection reliability by reducing the miss detection probability.

\section{Numerical Results}
\label{Sec_simResults}

\begin{table}[t]
	\caption{Simulation Parameters \cite{Ref_TSP_Ke'20}}
	\label{Tab_simPara}
	\centering
	\begin{tabular}{@{}cc@{}}
		\toprule
		Parameter  &  Value \\ 
		\midrule
		\vspace{0.2mm}
		Number of potential devices $K$  &  500 \\
		\vspace{0.2mm}
		Number of BS antennas $N$  &  512 \\
		\vspace{0.2mm}
		Number of payload data bits $B_d$  &  100 \\
		\vspace{0.2mm}
		Number of CRC bits $B_c$  &  8 \\
		\vspace{0.2mm}
		Generator polynomial of CRC ${\bf p}_c$  &  $x^8 + x^7 + x^6 + x^4 + x^2 + 1$ \\ 
		\vspace{0.2mm}
		Modulation order $M$  &  2 \\
		\vspace{0.2mm}
		Carrier frequency  &  3.9 GHz \\
		\vspace{0.2mm}
		System bandwidth  &  400 MHz \\
		\vspace{0.2mm}
		Number of MPCs $P$   & ${\cal U}\left(P; 31, 61\right)$ \\
		\vspace{0.2mm}
		Angular spread in degree  &  $10^{\circ}$ \\
		\vspace{0.2mm}
		Complex gain of the MPCs $\beta_{k,p}$  &  ${\cal CN}\left(\beta_{k,p}; 0, 1\right)$ \\
		\vspace{0.2mm}
		Maximum transmit power $P_{\rm max}$  &  35 dBm \\
		\vspace{0.2mm}
		Background noise power  &  -174 dBm/Hz \\
		\vspace{0.2mm}
		Number of SIC iterations $J$  &  3 \\
		\vspace{0.2mm}
		Number of BiG-AMP iterations $U$  &  500 \\
		\vspace{0.2mm}
		Termination threshold $\epsilon_{\rm amp}$  &  $10^{-5}$ \\
		\vspace{0.2mm}
		Termination threshold $\epsilon_{\rm sic}$  &  $10^{-5}$ \\
		\vspace{0.2mm}
		Device-to-BS distance $d_k$ in km  &  ${\cal U}\left(d_k; 0.1, 1\right)$ \\
		\vspace{0.2mm}
		Path loss in dB at the distance $d_k$   &  $128.1 + 37.6{\rm log}_{10}\left(d_k\right)$  \\
		\bottomrule
	\end{tabular}
	\vspace{-2.5mm}
\end{table}

This section conducts exhaustive Monte-Carlo simulations to assess the performance of the proposed unified semi-blind detection framework.
We consider a grant-free massive URLLC scenario in massive MIMO systems, where a BS equipped with an $N$-antenna ULA is employed to serve $K$ single-antenna devices.
The devices are uniformly distributed in the BS's coverage and only $K_a$ out of the total $K$ devices are active within any given time interval.
The massive MIMO channels are generated as in (\ref{Eq_chModel}).
Moreover, considering the perfect synchronization between different devices, we further assume the device activity and the massive MIMO channels remain unchanged during the considered transmission duration.
The assumed simulation parameters are provided in \emph{Table~\ref{Tab_simPara}} unless otherwise specified.
According to the practical access requirements, the system can flexibly switch to either sourced or unsourced RA mode, where the proposed unified transceiver design detailed in \emph{Sections \ref{Sec_txDesign} and \ref{Sec_rxDesign}} is adopted.
Here, we first focus on the basic version of the proposed semi-blind detection framework, then the effectiveness of the URLLC-enhanced version is further verified. 
{\color{red} All simulation results are obtained by averaging over 10000 independent channel realizations.}

\subsection{Performance of Sourced RA}
\label{Sub_pmSourcedRA}

For sourced RA, the state-of-the-art non-orthogonal pilot-based coherent detection framework detailed in \emph{Section \ref{Sub_coheDet}} is compared as the benchmark.
Particularly, based on the angular-domain received pilot signal, the advanced generalized MMV-AMP algorithm is employed for joint ADD and CE, as in \cite{Ref_TSP_Ke'20}.
%In the proposed semi-blind detection framework, the number of consumed symbol durations for transmitting the device ID bits, the CRC bits, and the scalar pilot symbol is calculated as
%\begin{equation}\label{Eq_lenRef}
%	L_r = \left\lceil\frac{B_i + B_c} {{\rm log}_2\left(M\right)}\right\rceil + 1.
%\end{equation}
The length of non-orthogonal pilot sequence in the baseline scheme is set to $L_p = L_r$ for comparison fairness.
For performance evaluation, we consider the activity error rate (AER) of ADD and the bit error rate (BER) of data detection, which are defined as
\begin{align}
	{\rm AER} &= \frac{\left|{\cal A} - {\widehat {\cal A}}\right|_c + \left|{\widehat {\cal A}} - {\cal A}\right|_c} {K}, \label{Eq_AER} \\
{\rm BER} &=  \frac{\left\|\left[{\bf B}\right]_{{\cal A} \cap {\widehat {\cal A}},{\cal B}_d} - [{\widehat {\bf B}}]_{{\cal A} \cap {\widehat {\cal A}},{\cal B}_d}\right\|_0 + B_d\left|{\cal A} - {\widehat {\cal A}}\right|_c} {K_aB_d}, \label{Eq_BER}
\end{align}
%\begin{align}
%	{\rm AER} &= \frac{\left|{\cal A} - {\widehat {\cal A}}\right|_c + \left|{\widehat {\cal A}} - {\cal A}\right|_c} {K}, \label{Eq_AER} \\ 
%	{\rm BER} &=  \frac{\left\|\left[{\bf B}\right]_{{\cal A} \cap {\widehat {\cal A}},{\cal B}_d} - [{\widehat {\bf B}}]_{{\cal A} \cap {\widehat {\cal A}},{\cal B}_d}\right\|_0 + B_d\left|{\cal A} - {\widehat {\cal A}}\right|_c} {K_aB_d}, \label{Eq_BER}
%\end{align}
respectively.
Here, the set ${\cal B}_d = \left\{B_i+B_c+1, \cdots, B\right\}$ with $\left|{\cal B}_d\right|_c = B_d$ denotes the column indexes corresponding to the payload data bits in the binary data matrix ${\bf B}$.
The AER takes both miss detection and false alarm into account, as expressed in the numerator of (\ref{Eq_AER}).
The BER also consists of two parts: the number of error bits due to the failure of symbol detection and the number of bits that are lost due to the miss detection, cf. the numerator of (\ref{Eq_BER}). 

%\begin{figure*}[!t]
%	\captionsetup{font={footnotesize}, name = {Fig.}, labelsep = period}
%	\centering
%	\setcounter{subfigure}{0}
%	\subfigure[]{\includegraphics[width=3 in]{Fig8/Fig_Base_AER.eps}}\hspace{5mm}
%	\subfigure[]{\includegraphics[width=3 in]{Fig8/Fig_AER_MDFA.eps}}
%	\vspace{-5mm}
%	\caption{\color{red} \small{Activity detection performance comparison of the conventional non-orthogonal pilot-based coherent detection framework and the proposed semi-blind detection framework: (a) Overall AER for different numbers of BS antennas $N$; (b) Separate AER for $N = 512$.}}
%	\label{Fig_AERBasic}
%	\vspace{-3mm}
%\end{figure*}

\begin{figure*}[t]
	\captionsetup{font={footnotesize}, name = {Fig.}, labelsep = period}
	\centering
	\setcounter{subfigure}{0}
	\subfigure[]{\includegraphics[width=2.1 in]{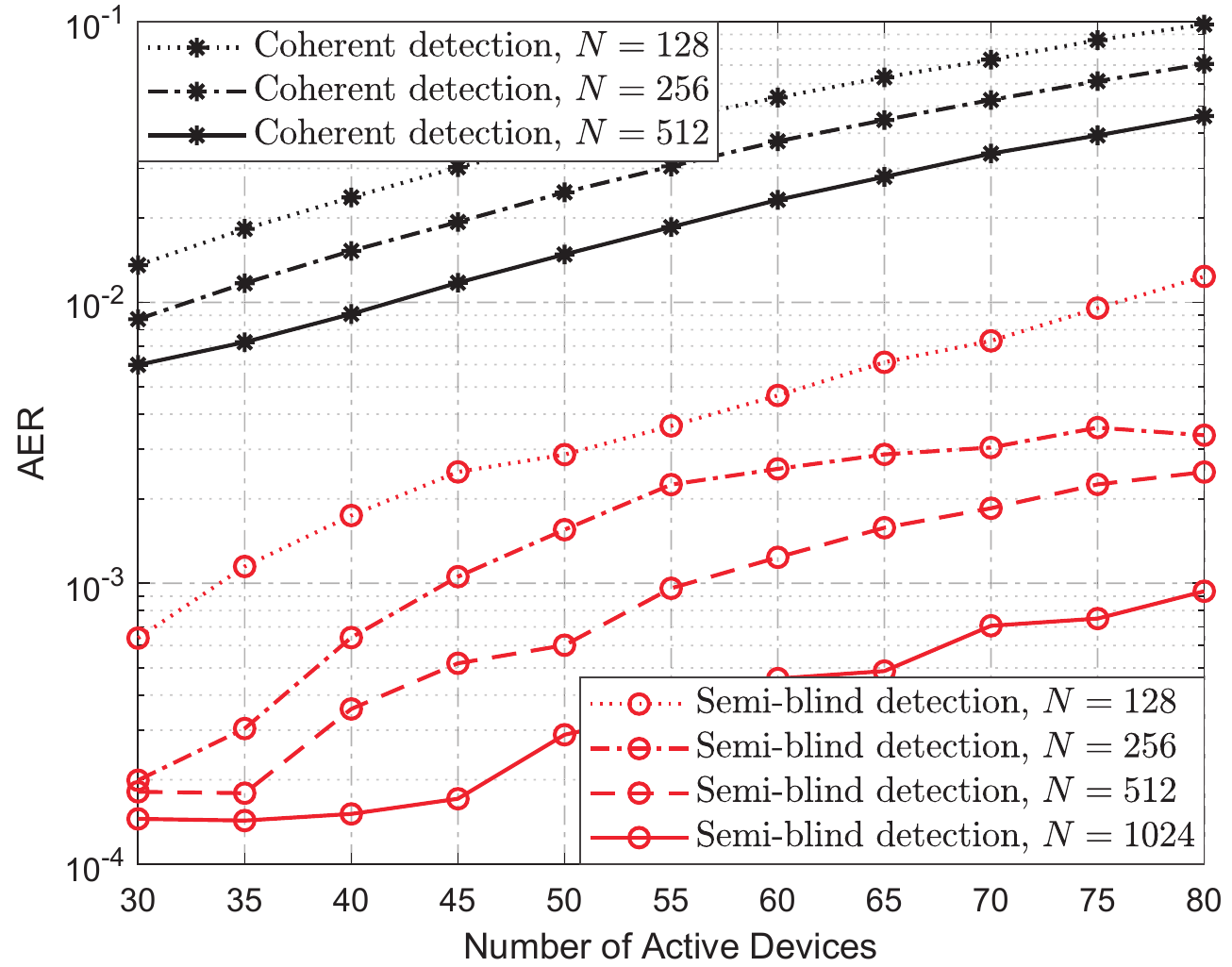}}\hspace{1mm}
	\subfigure[]{\includegraphics[width=2.1 in]{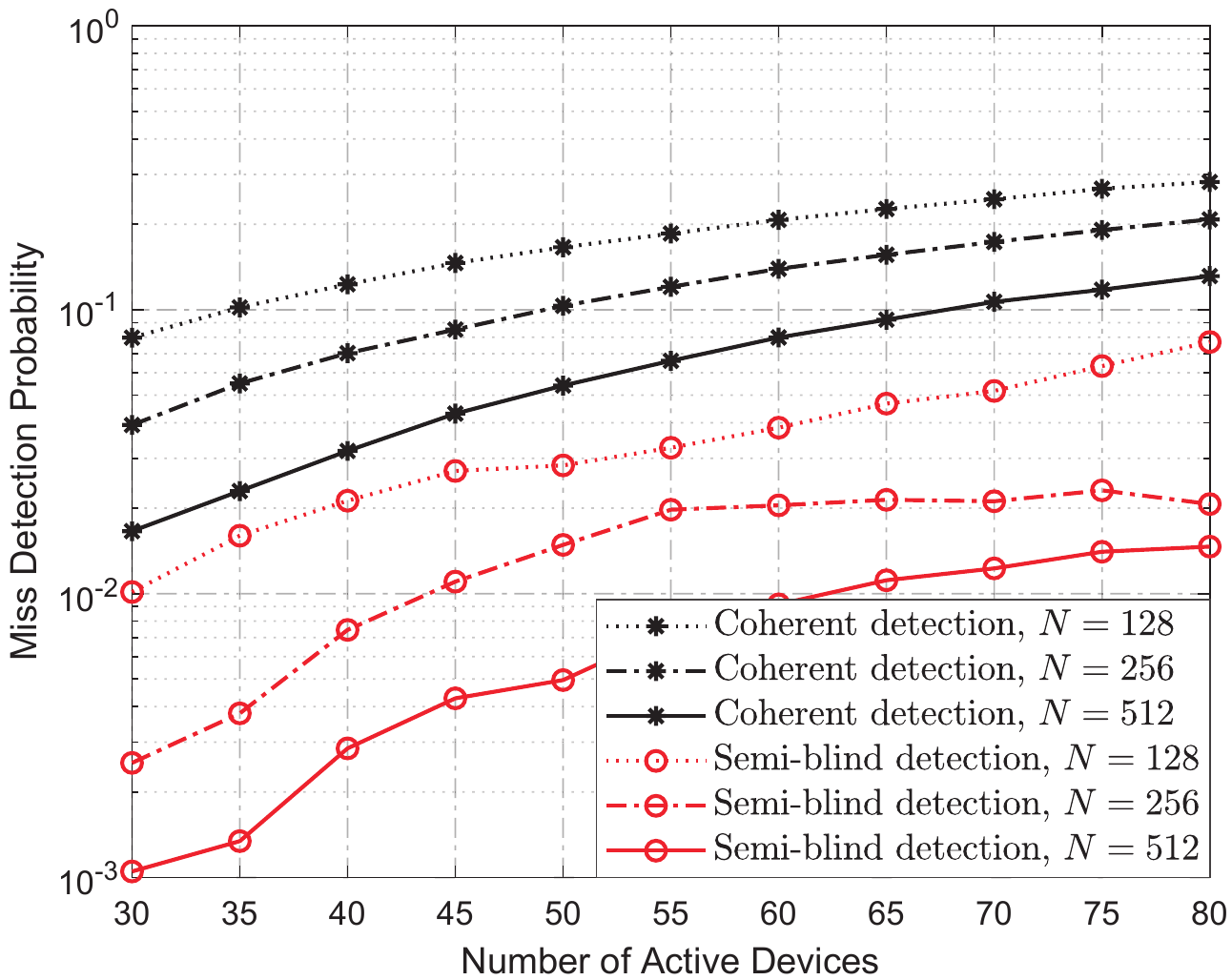}}\hspace{1mm}
	\subfigure[]{\includegraphics[width=2.1 in]{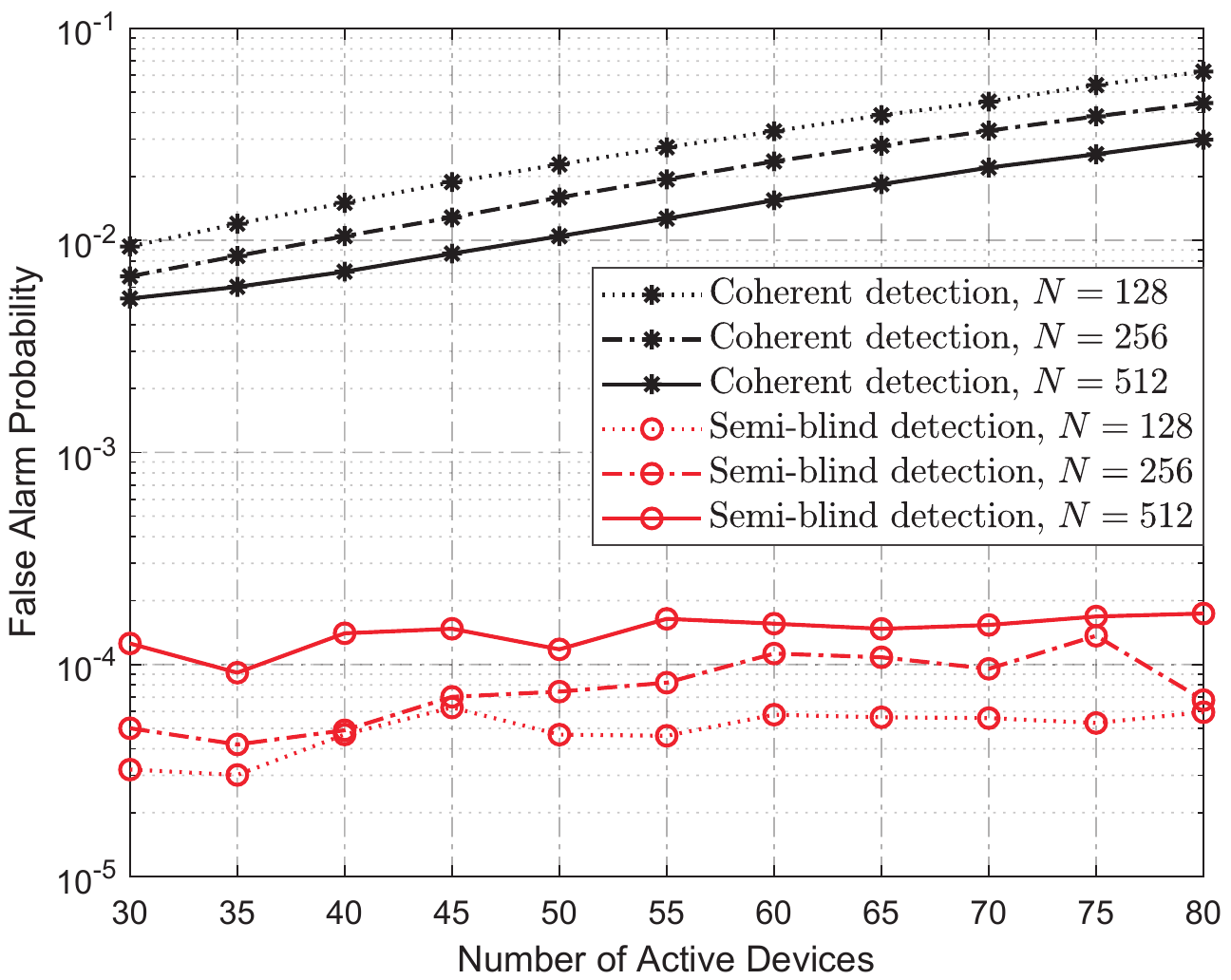}}
	\vspace{-5mm}
	\caption{\color{red} {\small{Activity detection performance comparison of the traditional non-orthogonal pilot-based coherent detection framework and the proposed semi-blind detection framework under different numbers of BS antennas $N$: (a) AER performance; (b) Miss detection probability; (c) False alarm probability.}}}
	\label{Fig_AERBasic}
	\vspace{-3mm}
\end{figure*}

%\begin{figure*}[h]
%	\captionsetup{font={footnotesize}, name = {Fig.}, labelsep = period}
%	\centering
%	\setcounter{subfigure}{0}
%	\subfigure[]{\includegraphics[width=3 in]{Fig9/Fig_Base_BER.eps}}\hspace{5mm}
%	\subfigure[]{\includegraphics[width=3 in]{Fig9/Fig_BER_MDDE.eps}}
%	\vspace{-5mm}
%	\caption{\color{red} {\small{Data detection performance comparison of the conventional non-orthogonal pilot-based coherent detection framework and the proposed semi-blind detection framework: (a) Overall BER for different numbers of BS antennas $N$; (b) Separate BER for $N = 512$.}}}
%	\label{Fig_Base_BER}
%	\vspace{-3mm}
%\end{figure*}

\begin{figure*}[t]
	\captionsetup{font={footnotesize}, name = {Fig.}, labelsep = period}
	\centering
	\setcounter{subfigure}{0}
	\subfigure[]{\includegraphics[width=3 in]{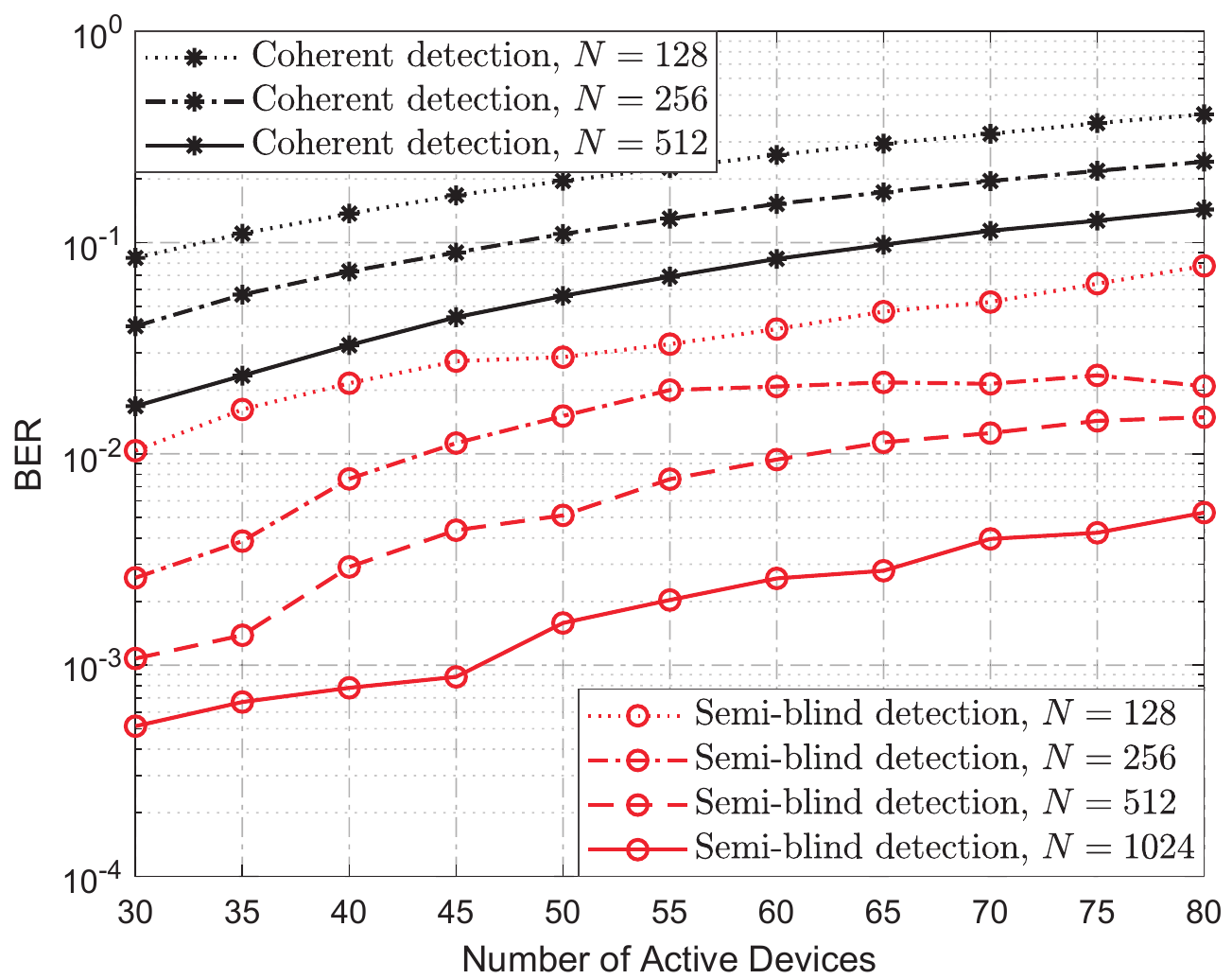}}\hspace{1mm}
	\subfigure[]{\includegraphics[width=3 in]{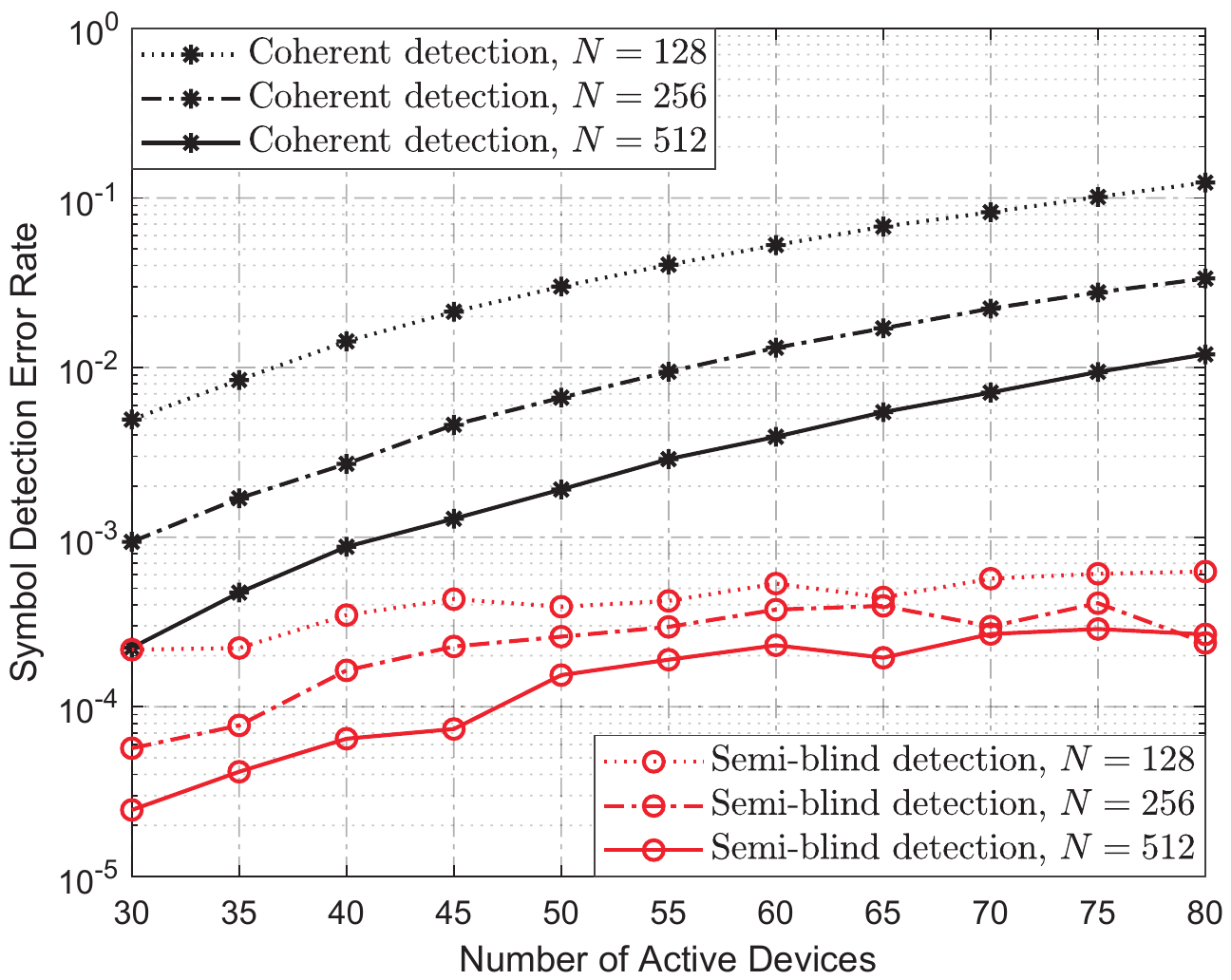}}
	\vspace{-5mm}
	\caption{\color{red} {\small{Data detection performance comparison of the traditional non-orthogonal pilot-based coherent detection framework and the proposed semi-blind detection framework under different numbers of BS antennas $N$: (a) Overall BER; (b) Symbol detection error rate.}}}
	\label{Fig_Base_BER}
	\vspace{-3mm}
\end{figure*}

%\begin{figure}[!t]
%	\centering
%	\includegraphics[width=0.5\columnwidth, keepaspectratio]
%	{Fig6/BER.eps}
%	\caption{\small{BER performance comparison of the proposed semi-blind detection framework and the conventional non-orthogonal pilot-based coherent detection framework under different numbers of BS antennas $N$.}}
%	\label{Fig_BERBasic}
%	\vspace{-3mm}
%\end{figure}
{\color{red} In Figs. \ref{Fig_AERBasic} and \ref{Fig_Base_BER}, we first compare the sourced RA performances of the conventional non-orthogonal pilot-based coherent detection framework and the proposed semi-blind detection framework.
To validate the most fundamental superiority of the proposed detection framework, the SIC procedure is disabled to exclude the performance gain provided by SIC.
Meanwhile, both detection frameworks occupy the same number of time-frequency resources for comparison fairness.
Meanwhile, both detection frameworks occupy the same number of time-frequency resources for comparison fairness.}
%In Figs. \ref{Fig_AERBasic} and \ref{Fig_BERBasic}, we first compare the sourced RA performance of the proposed semi-blind detection framework and the conventional non-orthogonal pilot-based coherent detection framework, where the SIC procedure is disabled.
%Here, both detection frameworks occupy the same number of time-frequency resources for comparison fairness.
{\color{red} As can be observed, the AER and BER performances of all the considered schemes degrade as the number of active devices increases.
This is because a larger number of active devices indicates severer inter-device interferences and a larger number of unknown variables to be estimated.}
{\color{red} Note that only the devices whose estimated data packet passes the CRC will be identified as the active devices.
Therefore, in the proposed semi-blind detection framework, the miss detection probability is generally larger than the false alarm probability.}
{\color{red} Meanwhile, the proposed semi-blind detection framework achieves much better AER and BER performance than the baseline scheme.
This verifies the superiority of the proposed detection framework in combating inter-device interferences when the same number of physical resources is consumed for grant-free MRA.}
{\color{red} As for the baseline scheme, the length of non-orthogonal pilot sequence is too short to realize satisfactory ADD and CE performance, which leads to an inaccurate signal matrix estimate in (\ref{Eq_dataDet}) and becomes the major limiting factor of the reliable data detection.
The performance of the baseline scheme can be improved by increasing the pilot length, but the payload efficiency would be significantly degraded, especially for massive URLLC with short data packets.
While for the proposed semi-blind detection framework, the channel and signal matrices are jointly estimated via the advanced BiG-AMP algorithm, which does not rely heavily on the length of pilot sequence, thus reaps a better performance.}
Besides, the sourced RA performance of both detection frameworks becomes better as the number of BS antennas increases.
This is because a larger number of BS antennas indicates the enhanced angular-domain sparsity of the MRA channel matrix, which leads to the improved ADD and CE performance in the pilot phase.
In this context, a more accurate initialization for JCSE is available and the BiG-AMP algorithm will converge to the global optimum with a higher probability and a faster speed.
On the other hand, more spatial observations are available by equipping more antennas at the BS, which further improves the JCSE performance.
Indeed, the BS must have a sufficient number of measurement samples, i.e., $N \gg K_a$, to avoid over-fitting, thus allowing more flexible and richer channel matrix estimates \cite{Ref_JMLR_Spielman'12, Ref_Tcom_Zhang'18}.
Therefore, we conclude that the massive MIMO shows great benefits in grant-free MRA and makes the semi-blind detection framework practical.

\begin{figure*}[t]
	\captionsetup{font=small, name = {Fig.}, labelsep = period}
	\centering
	\subfigure[]{\includegraphics[width=0.47\columnwidth]{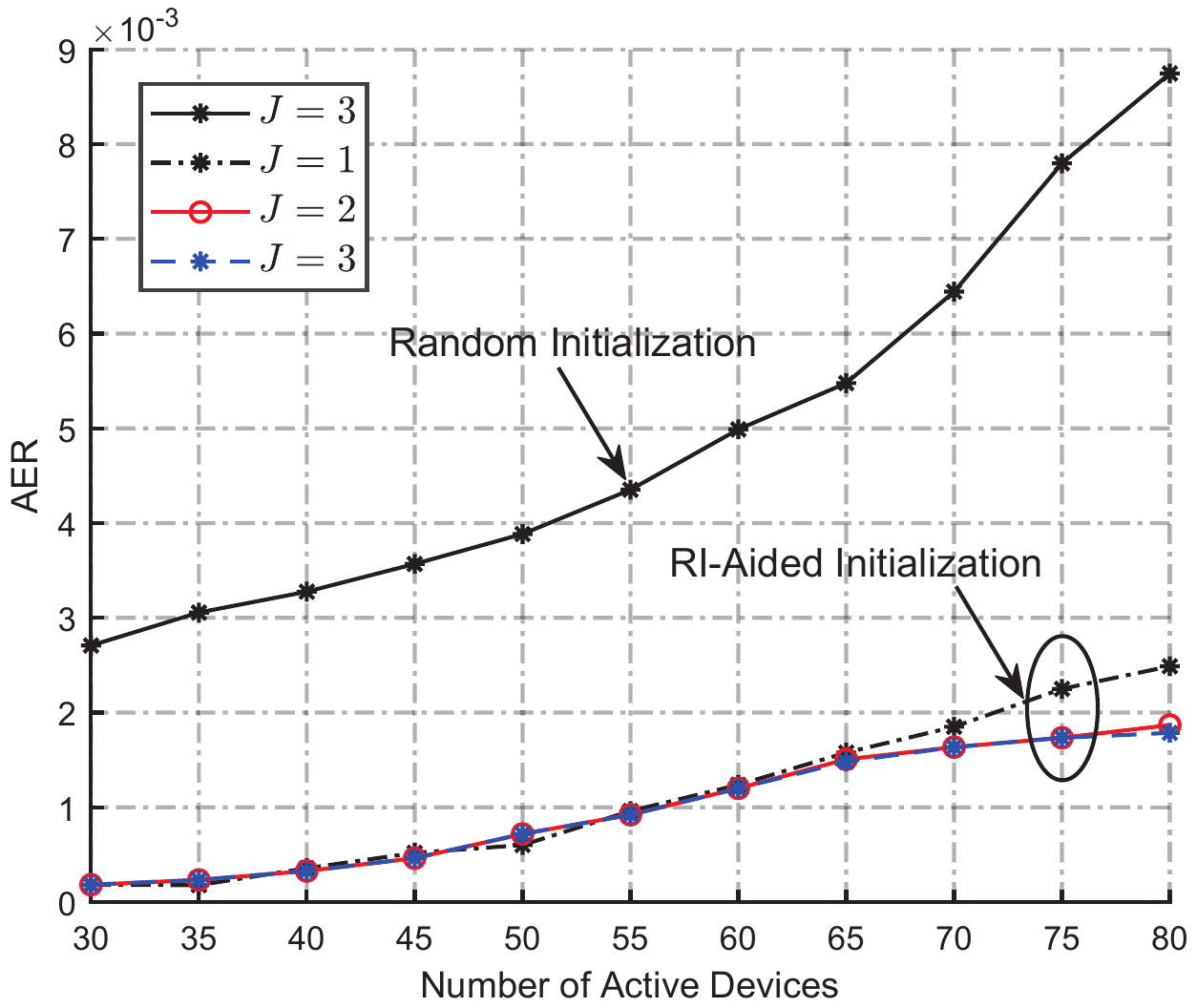}}
	\subfigure[]{\includegraphics[width=0.49\columnwidth]{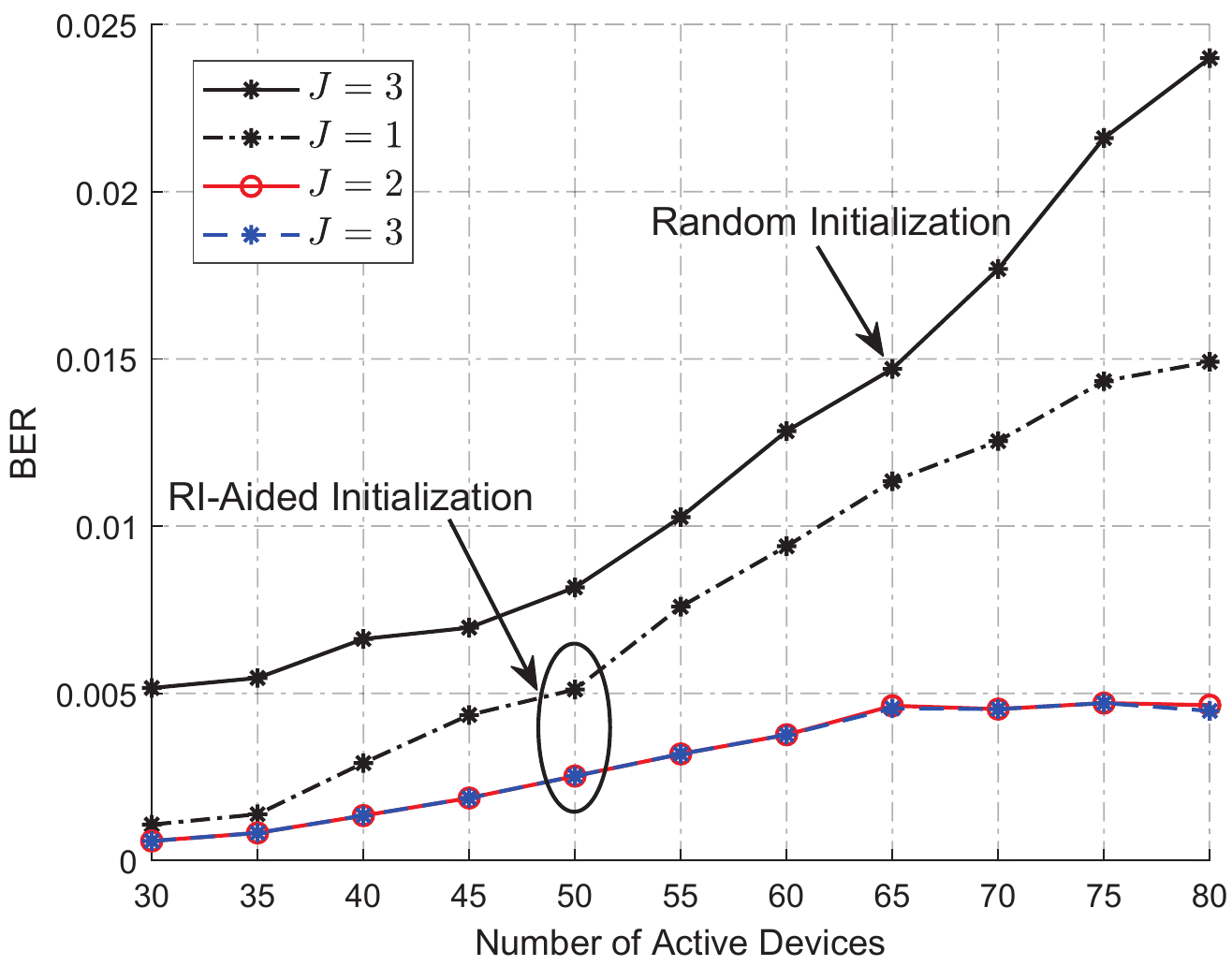}}
	\caption{\small{Sourced RA performance of the proposed semi-blind detection framework under different numbers of SIC iterations $J$ and initialization strategies.: (a) AER performance; (b) BER performance.}}
	\label{Fig_SIC}
	\vspace{-3mm}
\end{figure*}

\begin{figure*}[t]
	\captionsetup{font={footnotesize}, name = {Fig.}, labelsep = period}
	\centering
	\setcounter{subfigure}{0}
	\subfigure[]{\includegraphics[width=3 in]{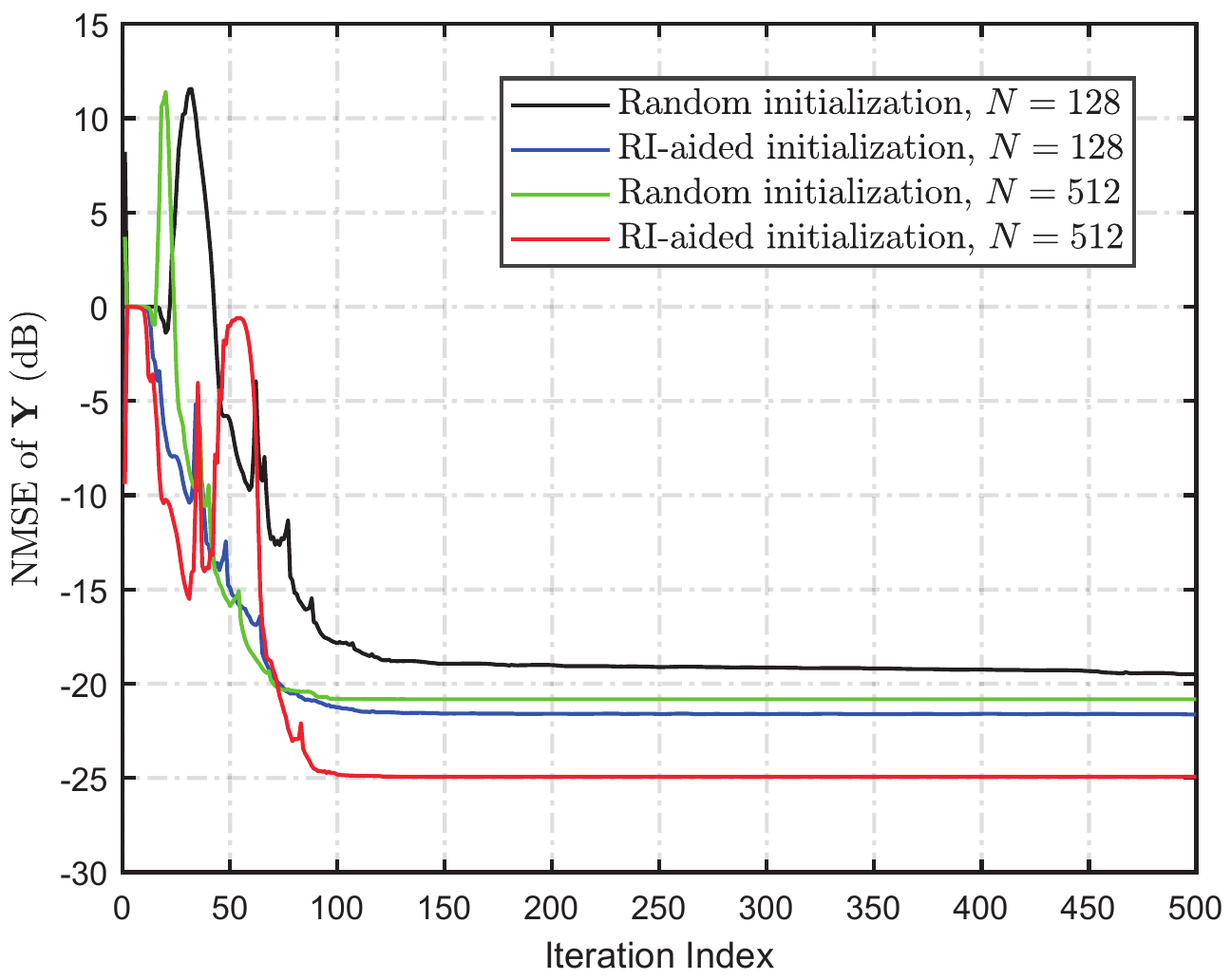}}\hspace{5mm}
	\subfigure[]{\includegraphics[width=2.93 in]{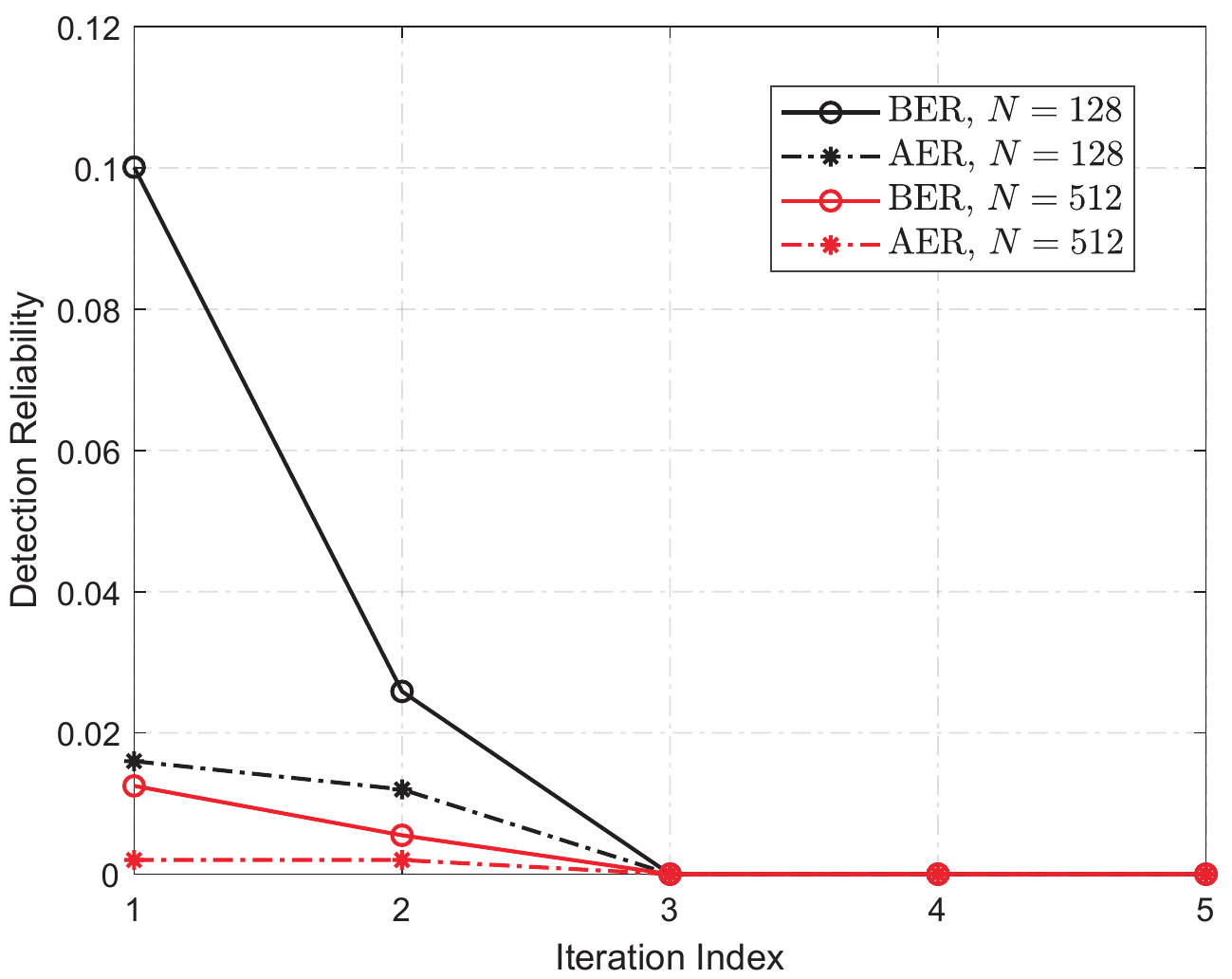}}
	\vspace{-5mm}
	\caption{\color{red} {\small{Convergence of the proposed BiG-AMP-based JCSE algorithm and the proposed SIC-based semi-blind detection scheme, where $K_a = 80$ is considered: (a) BiG-AMP iteration; (b) SIC iteration.}}}
	\label{Fig_Converge}
	\vspace{-3mm}
\end{figure*}

In Fig. \ref{Fig_SIC}, we further investigate the effectiveness of the proposed SIC-based semi-blind detection scheme and the RI-aided initialization strategy. 
Note that the SIC procedure is disabled when the number of SIC iterations is set to $J = 1$.
It is clear that the proposed semi-blind detection framework reaps better AER and BER performance as the number of SIC iterations increases, and only $J = 3$ iterations are sufficient to converge.
This is due to the fact that the reliable BiG-AMP-based JCSE makes the error propagation of SIC controllable.
Meanwhile, as the SIC iterations proceed, the signal components associated with the reliably detected active devices, i.e., whose detected device ID bits passes the CRC, are removed from the received signal, which mitigates the inter-device interference in the following SIC iterations.
On this basis, the activity and the payload data of residual active devices can be detected with improved reliability.
Moreover, the proposed RI-aided initialization strategy outperforms the traditional random initialization strategy.
This demonstrates that the traditional BiG-AMP algorithm using the traditional random initialization is easy to stuck in the local extremum, while the proposed RI-aided initialization can guarantee the algorithm to converge to the global optimum. 

\begin{figure*}[t]
	\captionsetup{font={footnotesize}, name = {Fig.}, labelsep = period}
	\centering
	\setcounter{subfigure}{0}
	\subfigure[]{\includegraphics[width=3 in]{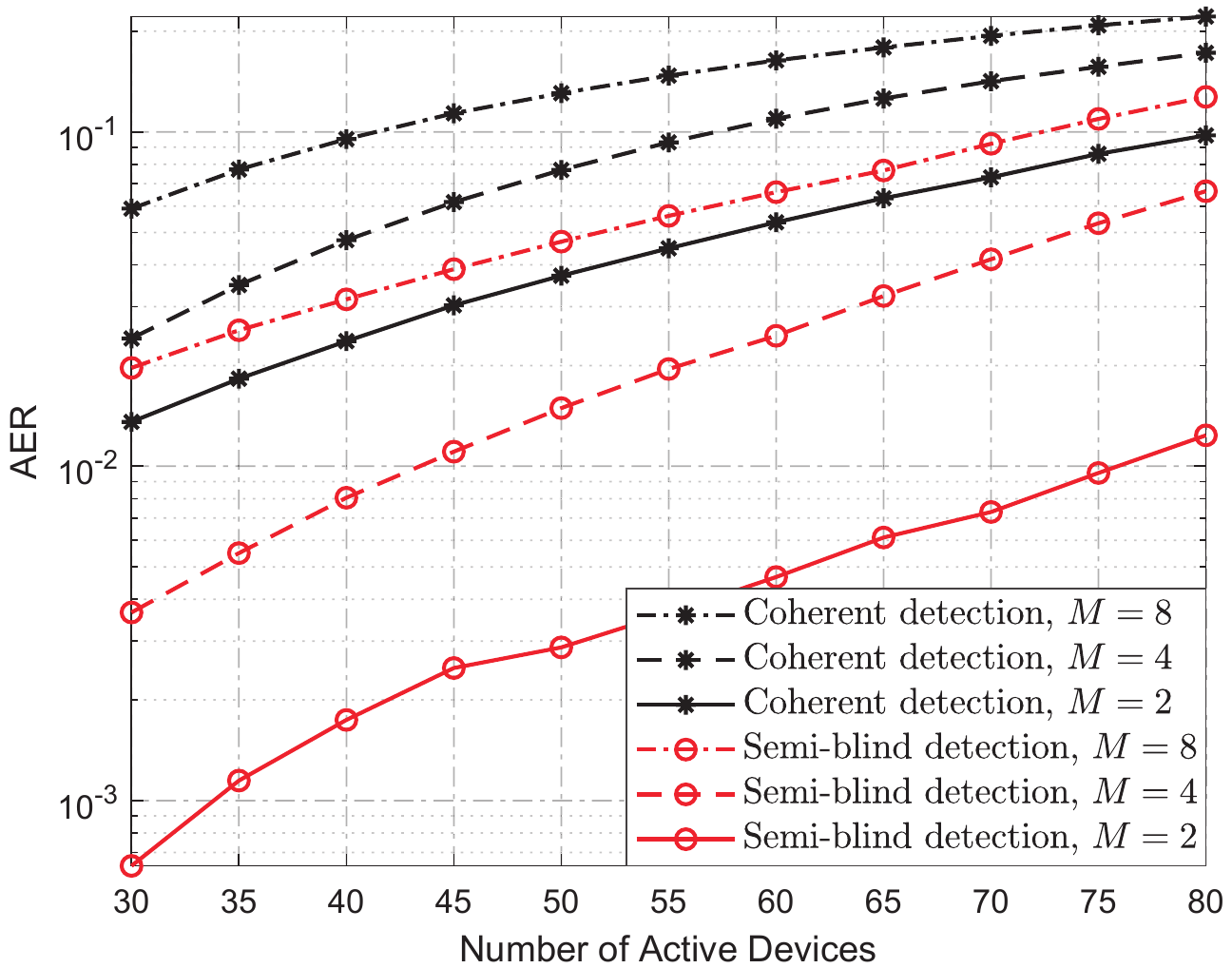}}\hspace{5mm}
	\subfigure[]{\includegraphics[width=3 in]{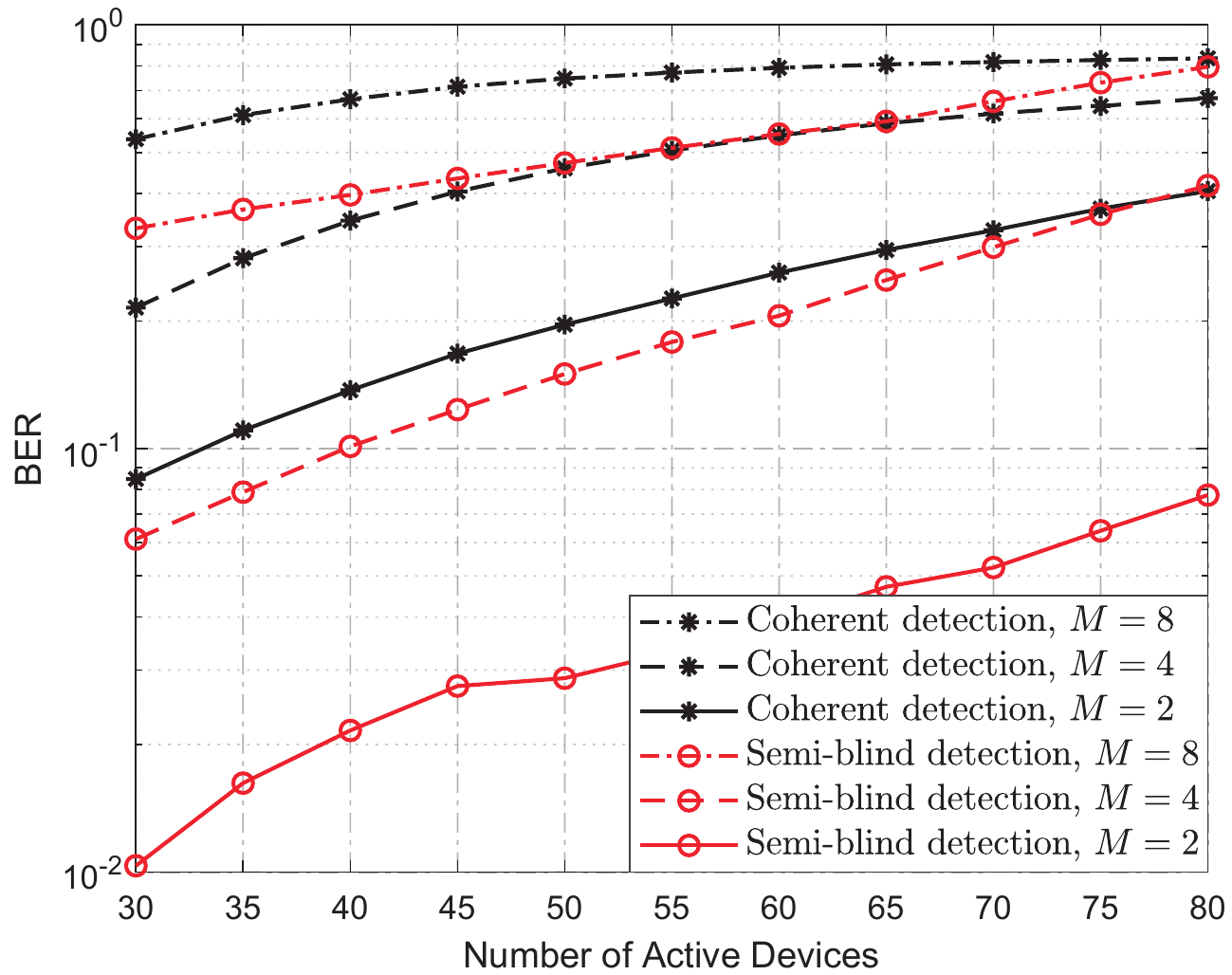}}
	\vspace{-5mm}
	\caption{\color{red} {\small{Sourced RA performance comparison of the traditional coherent detection framework and the proposed semi-blind detection framework under different modulation orders $M$: (a) AER performance; (b) BER performance.}}}
	\label{Fig_mOrder}
\end{figure*}

{\color{red} Fig. \ref{Fig_Converge} validates the convergence of the proposed BiG-AMP-based JCSE algorithm and the proposed SIC-based semi-blind detection scheme, where only $U = 100$ and $J = 3$ are sufficient to converge.}
{\color{red} Fig. \ref{Fig_mOrder} compares the sourced RA performances of the traditional non-orthogonal pilot-based coherent detection framework and the proposed semi-blind detection framework, where $N = 128$ and different modulation orders are considered.
It is clear that the AER and BER performances of both detection frameworks degrade when a higher modulation order is adopted.
This is because a higher modulation order leads to smaller Euclidean distances among the constellations and a worse RI-aided initialization.
However, the proposed detection framework still achieves a much better performance than the traditional coherent detection framework dedicated to sourced RA.}

%\begin{figure}[!t]
%	\centering
%	\includegraphics[width=0.5\columnwidth, keepaspectratio]
%	{Fig10/AER.eps}
%	\caption{\small{AER performance of the proposed semi-blind detection framework with different numbers of CRC bits $B_c$.}}
%	\label{Fig10}
%	\vspace{-3mm}
%\end{figure}
%
%\begin{figure}[!t]
%	\centering
%	\includegraphics[width=0.5\columnwidth, keepaspectratio]
%	{Fig11/BER.eps}
%	\caption{\small{BER performance of the proposed semi-blind detection framework with different numbers of CRC bits $B_c$.}}
%	\label{Fig11}
%	\vspace{-3mm}
%\end{figure}
\begin{figure*}[t]
	\captionsetup{font={footnotesize}, name = {Fig.}, labelsep = period}
	\centering
	\setcounter{subfigure}{0}
	\subfigure[]{\includegraphics[width=3 in]{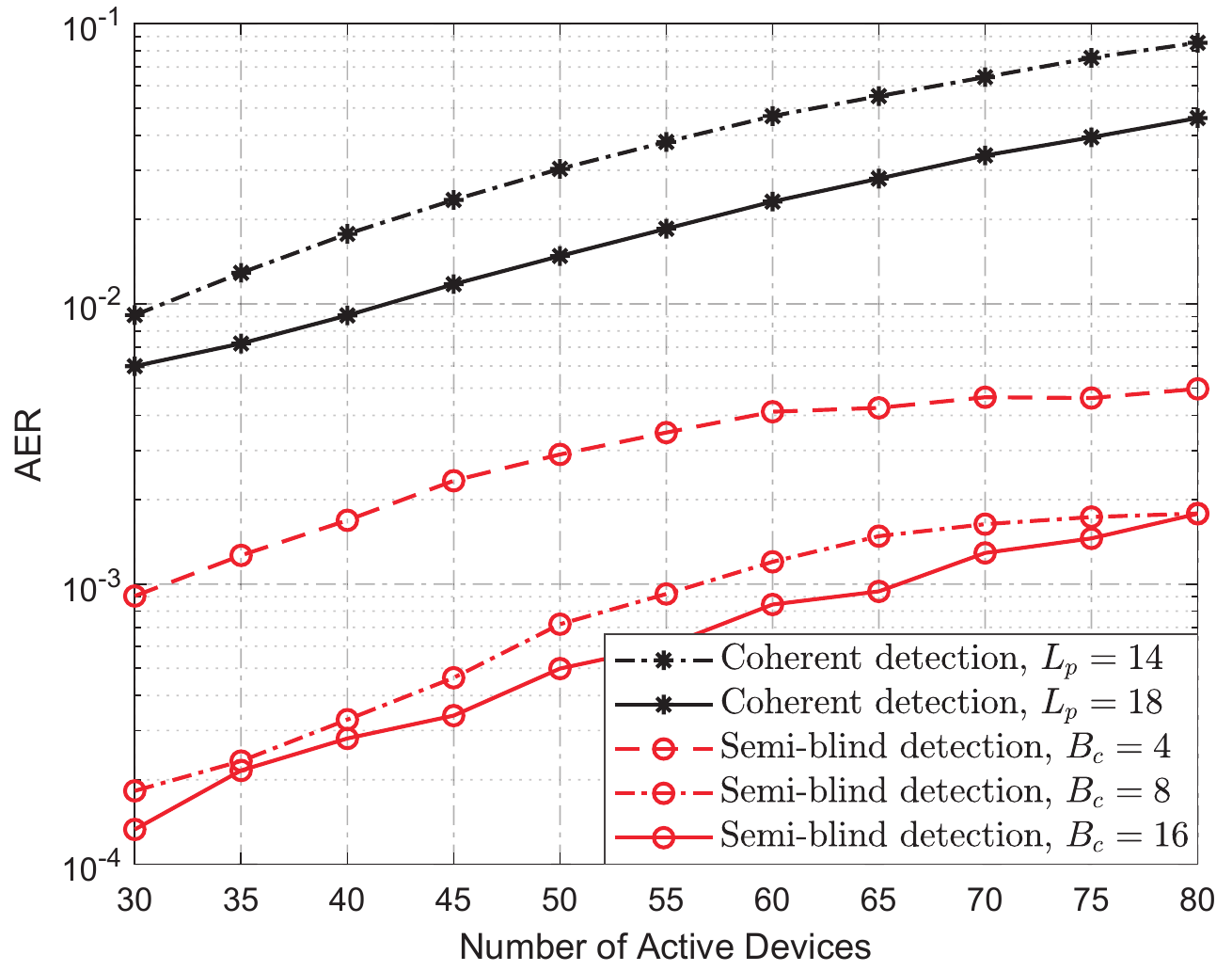}}\hspace{5mm}
	\subfigure[]{\includegraphics[width=3 in]{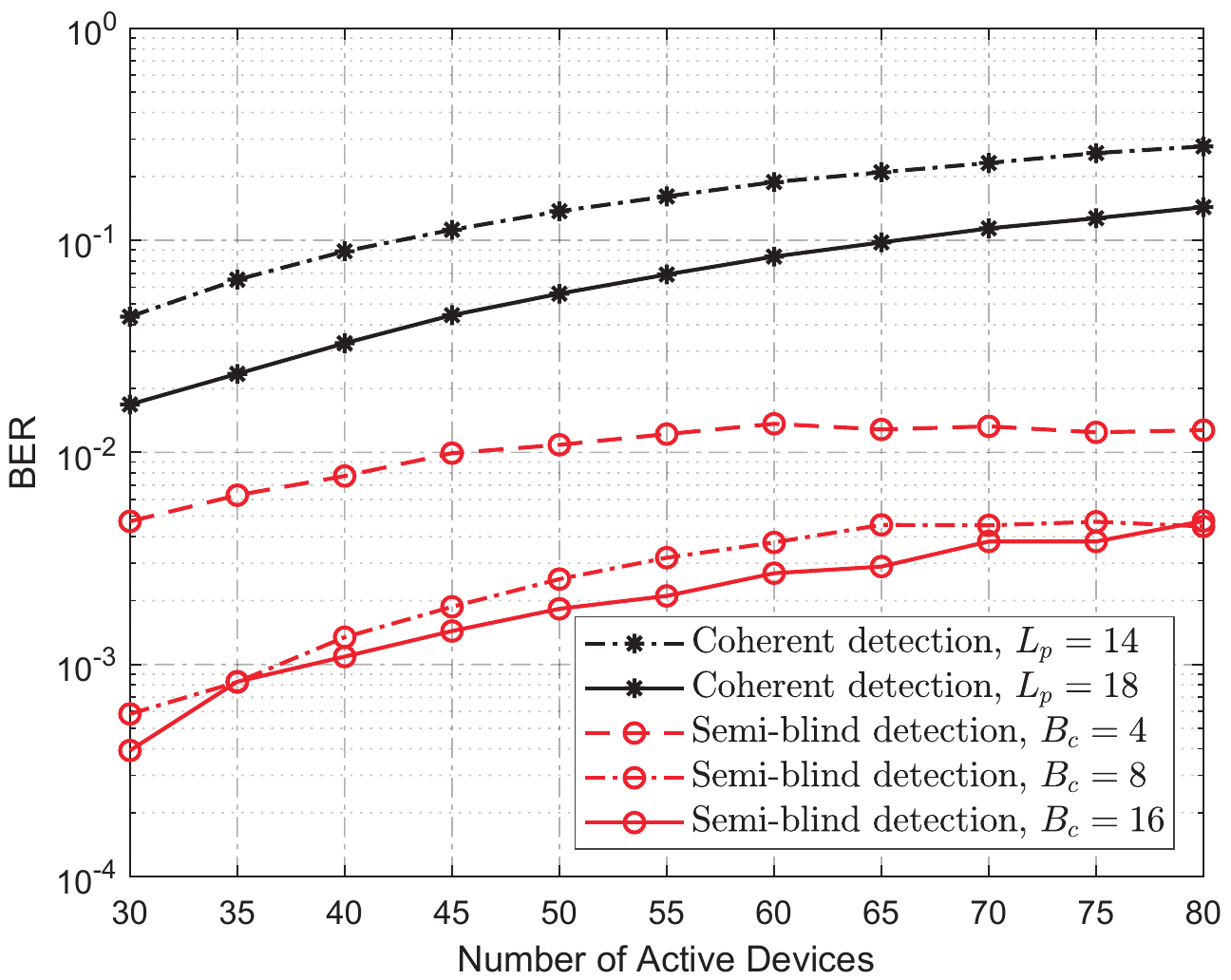}}
	\vspace{-5mm}
	\caption{\small{Sourced RA performance of the proposed semi-blind detection framework for different numbers of CRC bits $B_c$: (a) AER performance; (b) BER performance.}}
	\label{Fig_CRC}
\end{figure*}

In Fig. \ref{Fig_CRC}, the AER and BER performances of the proposed semi-blind detection framework for different numbers of CRC bits $B_c$ are also studied.
In the simulations, $B_c \in \left[4, 8, 16\right]$ are investigated, where the corresponding $L_p = L_r$ with $L_r \in \left[14, 18\right]$ are considered in the baseline scheme for comparison fairness.
The relationship between $B_c$ and $L_r$ is provided in (\ref{Eq_lenRef}).
The generator polynomials of 4-bit and 16-bit CRC codes are given as $x^4 + x^3 + x^2 + x + 1$ and $x^{16} + x^{15} + x^2 + 1$, respectively.
{\color{red} Meanwhile, the payload efficiency of the proposed semi-blind detection framework is defined as
\begin{equation} \label{Eq_PE_}
	{\rm PE}_{\rm sbd} = \frac{B_d}{\lceil B/{\rm log}_2\left(M\right)\rceil + 1},
\end{equation}
that is, the number of payload data bits to the number of consumed symbol durations for transmission. 	
As shown in the Fig. 13 and (48), a larger $B_c$ improves both AER and BER performance, but at the expense of payload efficiency.}
%As shown in the figures, the number of the attached CRC bits influences the error detection performance.
%A larger $B_c$ will reduce the miss detection and false alarm probabilities of error detection, but at the expense of payload efficiency.
Meanwhile, the performance gain is limited when $B_c > 8$.
With the limited number of CRC bits, the generator polynomial can be optimized to fulfill a specified error detection performance \cite{Ref_Tcom_Baicheva'08}. 

\begin{figure*}[t]
	\captionsetup{font={footnotesize}, name = {Fig.}, labelsep = period}
	\centering
	\setcounter{subfigure}{0} 
	\subfigure[]{\includegraphics[width=3 in]{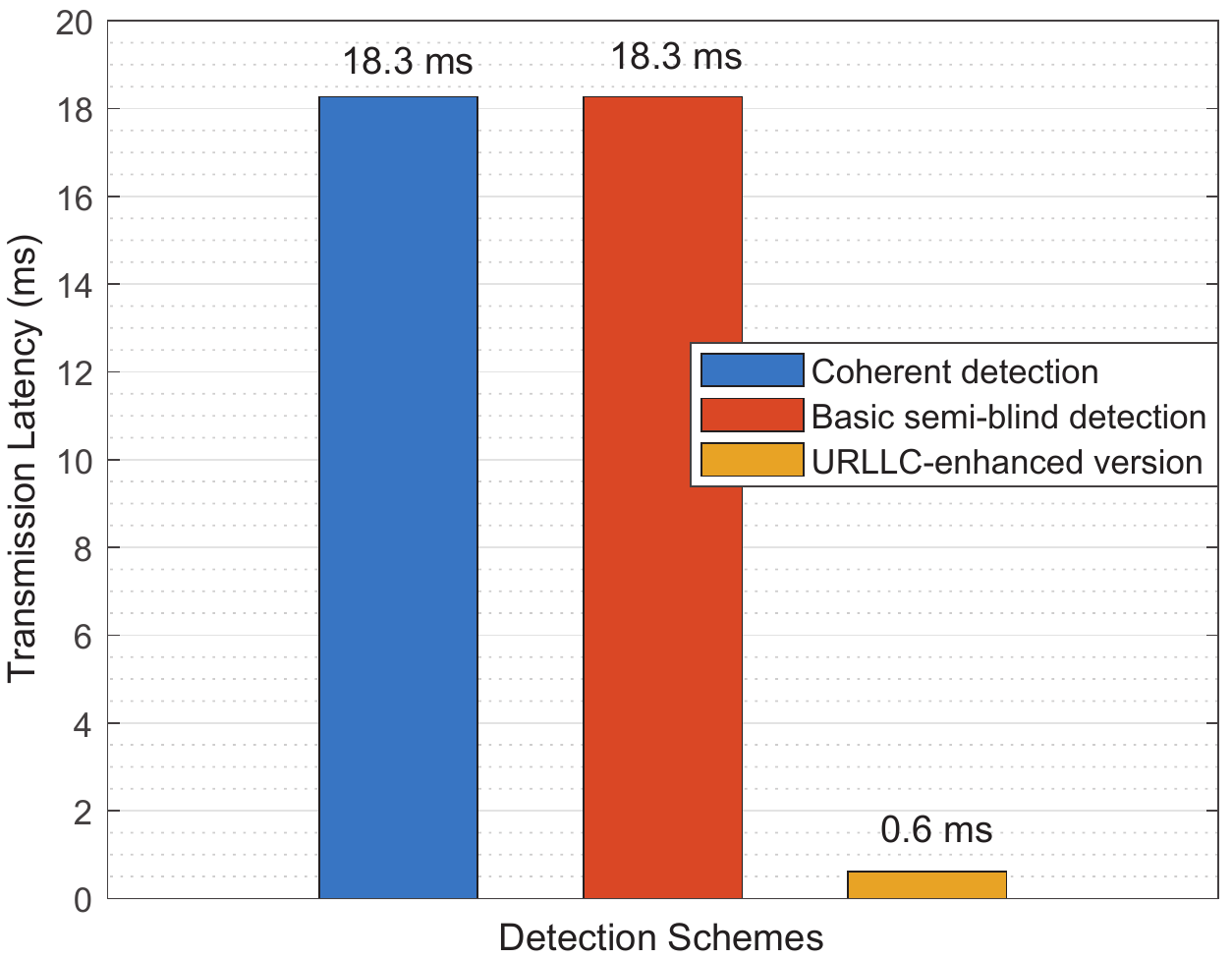}}\hspace{5mm}
	\subfigure[]{\includegraphics[width=3 in]{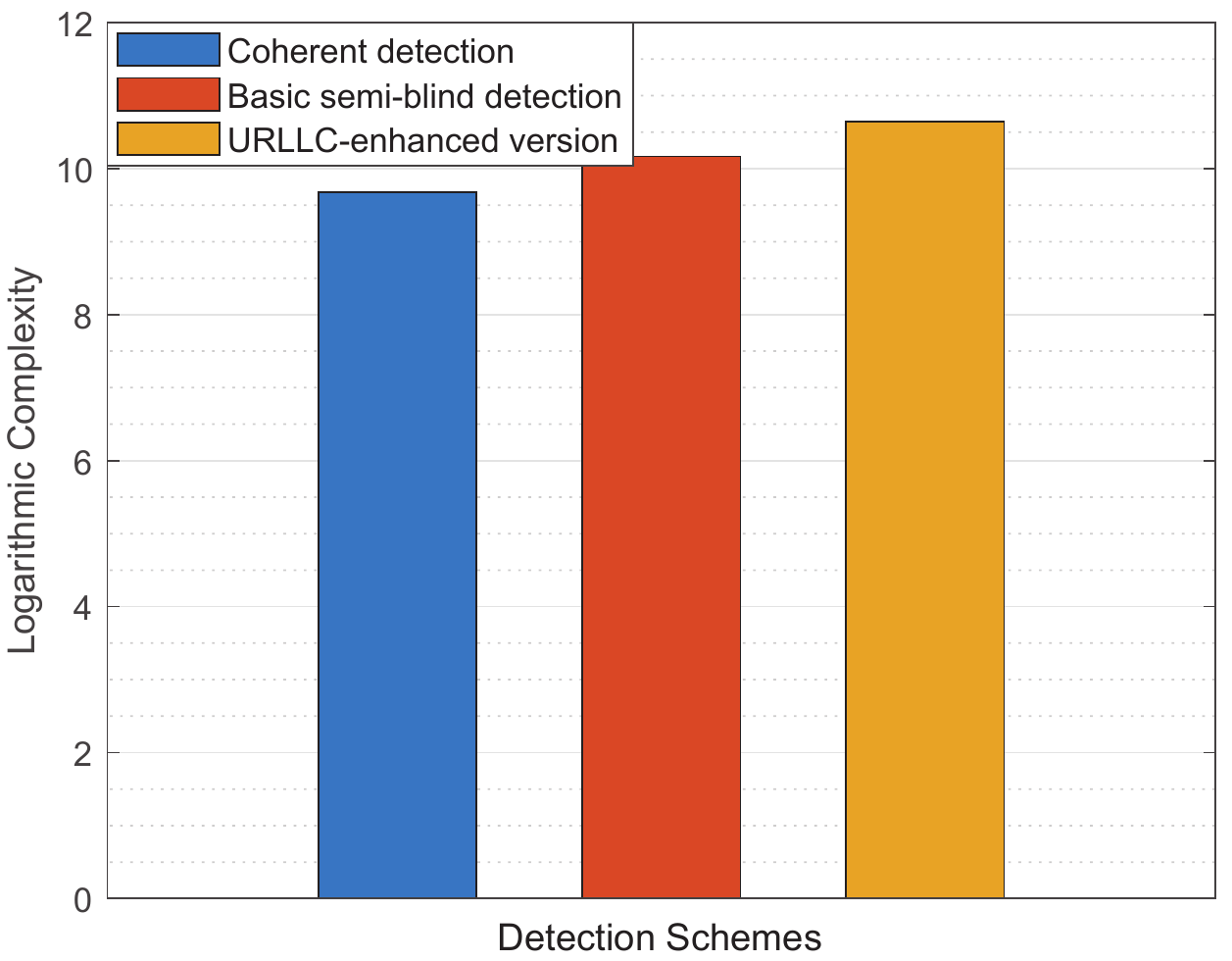}}
	\vspace{-5mm}
	\caption{\color{red} {\small{The user-plane latency comparison of the traditional coherent detection framework, the basic semi-blind detection framework, and the URLLC-enhanced semi-blind detection framework, where $N = 512$, $K_a = 50$, and $B_d = 256$ are considered: (a) Transmission latency; (b) Processing latency.}}}
	\label{Fig_Latency_SRA}
	\vspace{-3mm}
\end{figure*}

%\begin{figure}[!t]
%	\centering
%	\includegraphics[width=0.5\columnwidth, keepaspectratio]
%	{Fig9/BER_URLLC.eps}
%	\caption{\small{Detection reliability comparison of the basic version and the URLLC-enhanced version of the proposed semi-blind detection framework.}}
%	\label{Fig_BERURLLC}
%	\vspace{-3mm}
%\end{figure}
\begin{figure}[!t]
	\captionsetup{font=small, name = {Fig.}, labelsep = period}
	\centering
	\subfigure[]{\includegraphics[width=0.48\columnwidth]{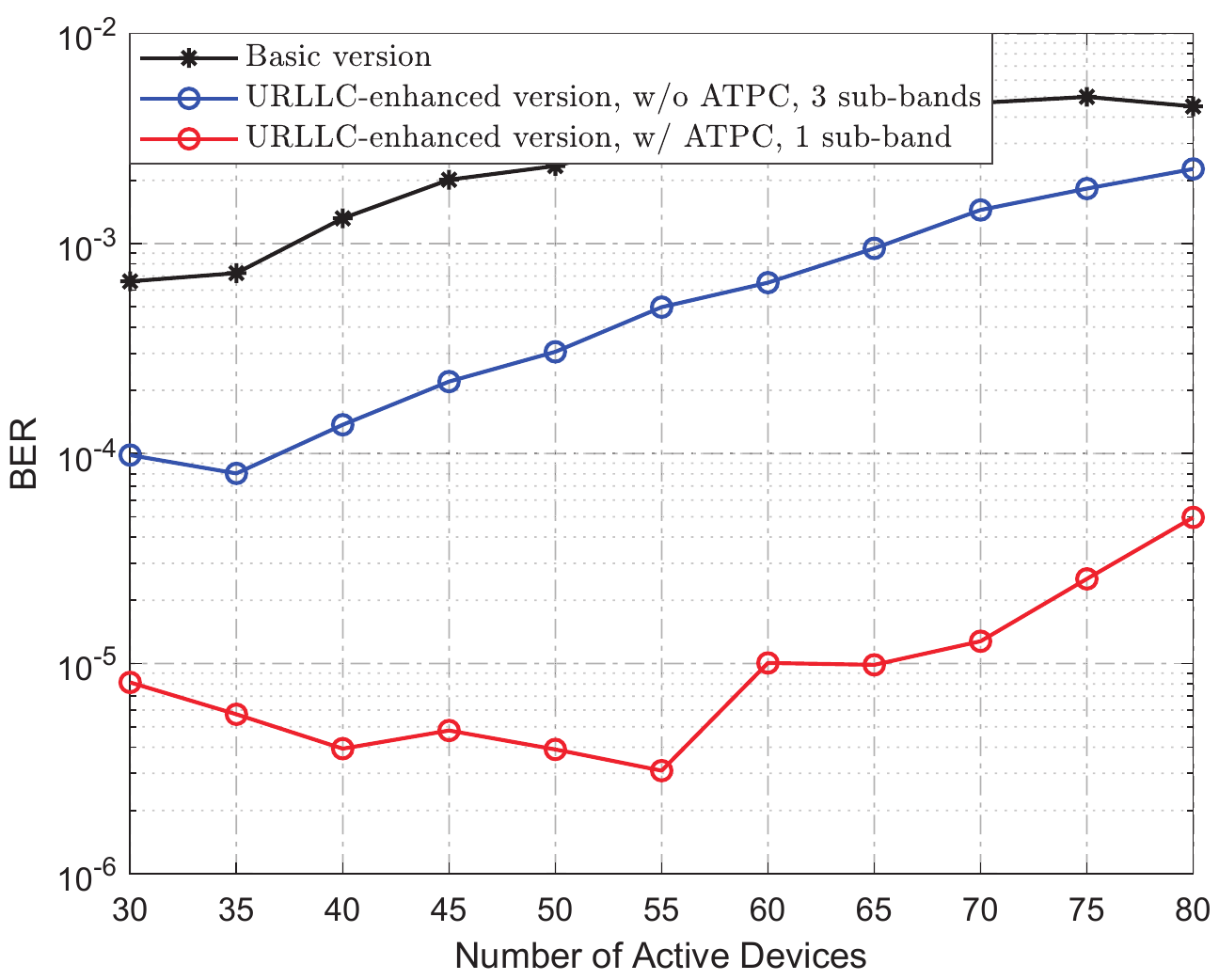}}
	\subfigure[]{\includegraphics[width=0.48\columnwidth]{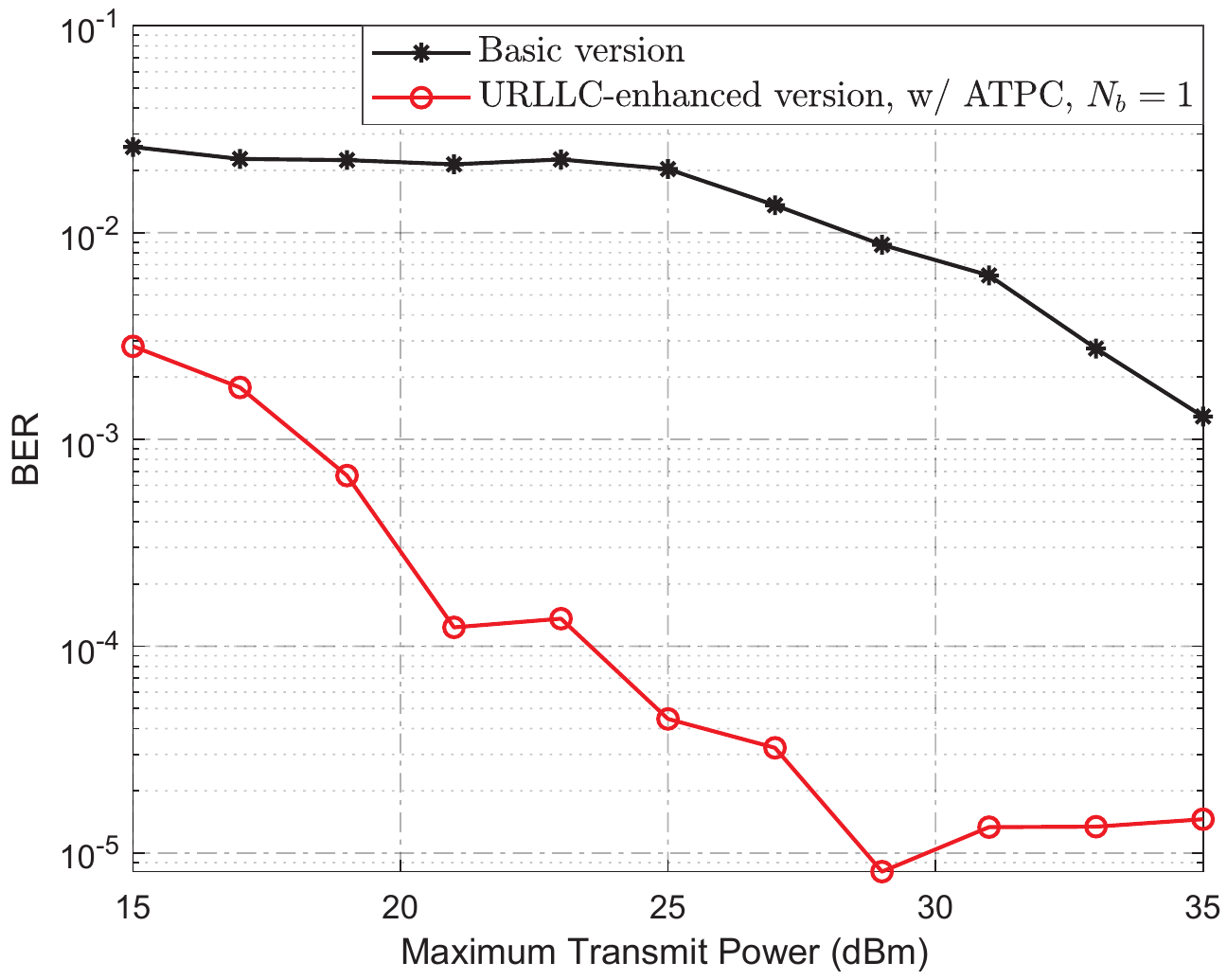}}
	\caption{{\color{red} \small{Data detection performance comparison of the basic version and the URLLC-enhanced version of the proposed semi-blind detection framework for $N = 512$ and $B_d = 256$: (a) BER under different numbers of active devices; (b) BER under different maximum transmit powers with $K_a = 50$.}}}
	\label{Fig_BERURLLC}
	\vspace{-3mm}
\end{figure}

The aforementioned simulation results focus the basic version of the proposed semi-blind detection framework, which is still not enough to satisfy the stringent latency and reliability requirements of massive URLLC.
Therefore, the enabling techniques introduced in {\em Section \ref{URLLC}} should be further integrated to obtain a URLLC-enhanced version of the proposed detection framework.
{\color{red} Fig. \ref{Fig_Latency_SRA}  compares the user-plane latency of the traditional non-orthogonal pilot-based coherent detection for sourced RA, the basic semi-blind detection framework, and the URLLC-enhanced semi-blind detection framework.
The user plane latency mainly consists of the transmission latency, propagation latency, and receive processing latency.
The propagation latency only depends on the distance between the device and the BS, which is at a maximum of 3.3 $\mu$s for a cell radius of 1 km.
Therefore, the propagation latency is identical for all the considered schemes.
The processing latency generally depends on the computational complexity of the receive algorithm and the computing power of the processing unit.
Since the available computing resources are identical for all the considered schemes, we mainly analyze the processing latency in terms of the computational complexity, which is quantified by the number of required complex multiplications.
As can be observed, by exploiting the multi-carrier deployment, the proposed URLLC-enhanced semi-blind detection framework achieves a significantly reduced transmission latency than its counterparts.
Moreover, the basic semi-blind detection framework has a slightly higher computational complexity than the non-orthogonal pilot-based coherent detection.
Meanwhile, for the URLLC-enhanced version, the complexity will further increase by incorporating the concurrent access mechanism.
However, it should be noted that the complexities of the all considered schemes are in the same order of magnitudes.
Considering the rapid development of high-performance processing units and the rich computing resources at the BS, the increased processing latency is expected to be very minor.}
{\color{red} Since the traditional coherent detection framework and the proposed basic semi-blind detection framework occupy the same number of time-frequency resources, they have an identical transmission latency. }
Hence, the 1 ms user-plane latency can be satisfied as long as the processing latency is less than 0.4 ms.
{\color{red} Fig.~\ref{Fig_BERURLLC} verifies the effectiveness of the proposed concurrent access mechanism and ATPC in improving data detection reliability.
It is observed that the URLLC-enhanced version effectively satisfy the 99.99999$\%$ reliability requirement when $K_a \le 65$.}
%verifies the effectiveness of the proposed concurrent access mechanism and ATPC in improving the data detection reliability.
%It is observed that the URLLC-enhanced version effectively approaches the $99.99999\%$ reliability requirement by reducing the miss detection probability.
Moreover, increasing the maximum transmit power further improves the detection reliability.

%Fig. shows that the proposed multi-carrier deployment and the adopted flexible frame structure significantly reduce the transmission latency.

%Since the propagation latency is fixed given the device-to-BS distance and it is difficult to analyze the processing latency in quantity, we only compare the transmission latency here.

%Note that the maximum propagation latency is about $3.3 \, {\mu}s$ for a cell with radius of 1 km.
%Meanwhile, since the computational complexity scales linearly with the problem dimensions and the BS has rich computing resources, the processing latency can be far smaller than 1 ms.

\subsection{Performance of Unsourced RA}
\label{Sub_pmUnsourcedRA}

\begin{figure}[!t]
	\centering
	\includegraphics[width=0.5\columnwidth, keepaspectratio]
	{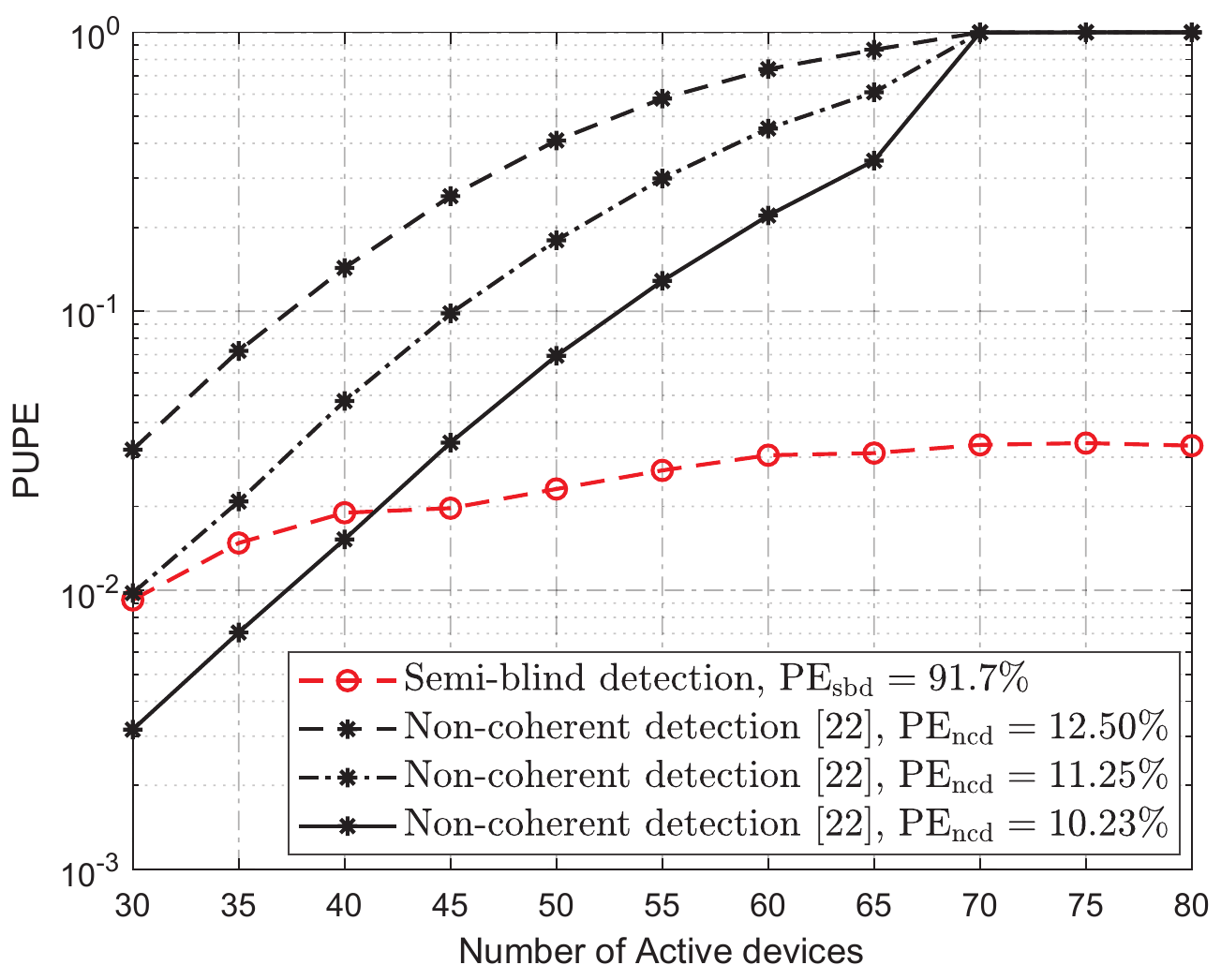}
	\caption{\small{PUPE performance comparison of the proposed semi-blind detection framework and the conventional common codebook-based non-coherent detection framework.}}
	\label{Fig_PUPE}
	\vspace{-3mm}
\end{figure}

\begin{figure*}[h]
	\captionsetup{font={footnotesize}, name = {Fig.}, labelsep = period}
	\centering
	\setcounter{subfigure}{0}
	\subfigure[]{\includegraphics[width=3 in]{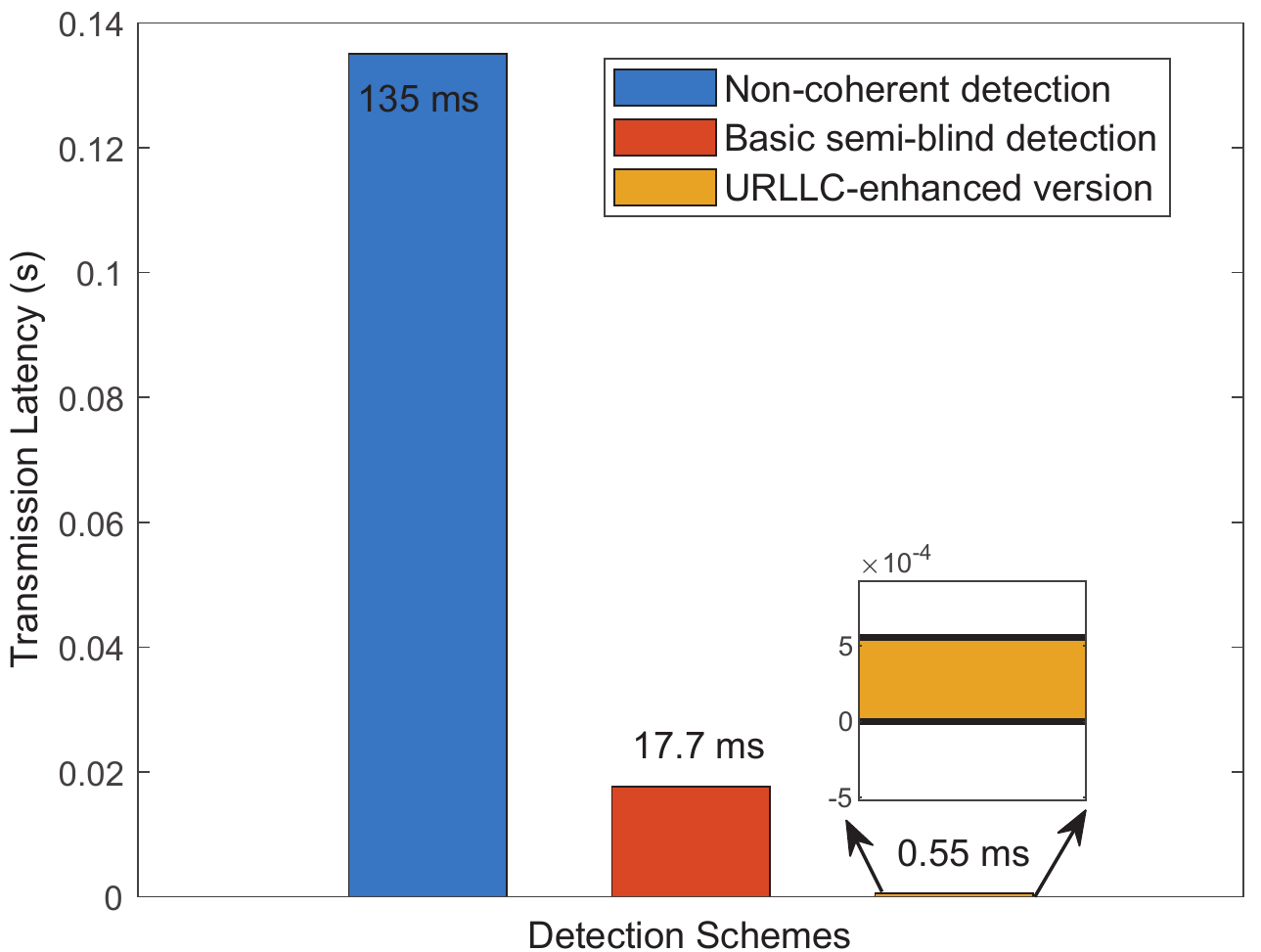}}\hspace{5mm}
	\subfigure[]{\includegraphics[width=2.93 in]{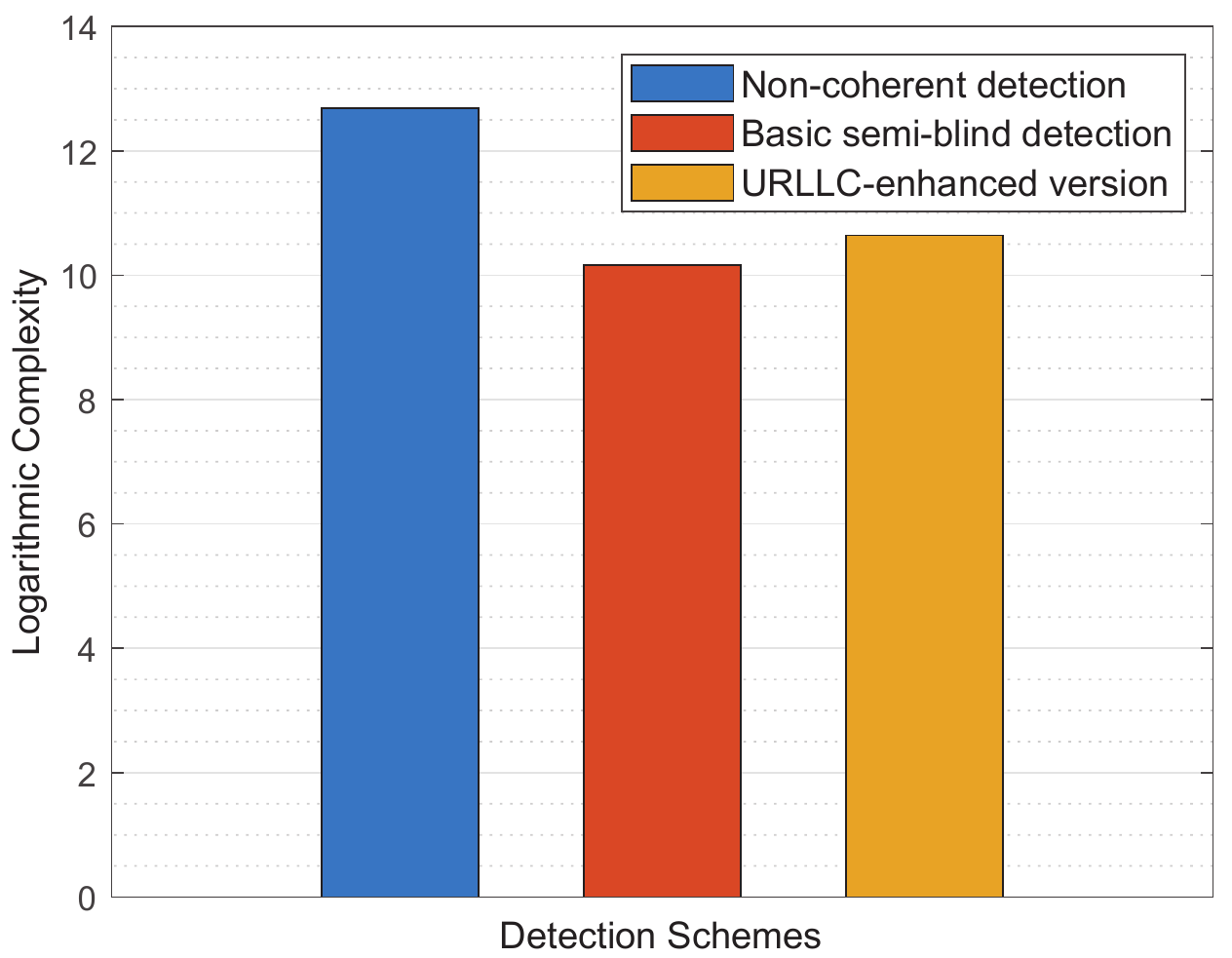}}
	\vspace{-5mm}
	\caption*{\color{red} {\small{Fig. 17. The user-plane latency comparison of the traditional non-coherent detection framework, the basic semi-blind detection framework, and the URLLC-enhanced semi-blind detection framework: (a) Transmission latency; (b) Processing latency.}}}
	\label{Fig17}
\vspace{-3mm}
\end{figure*}

For unsourced RA, we employ the most widely studied common codebook-based non-coherent detection framework as the benchmark, which was briefly introduced in \emph{Section \ref{Sub_nonCoherentDet}} and detailed in \cite{Ref_ISIT_Polyanskiy'17}.   
Moreover, the coupled CS-based coding scheme is adopted for reducing the computational complexity, where the concatenated coding is utilized to couple an outer tree code and an inner CS code \cite{Ref_TIT_Amalladinne'20}.
Specifically, for each active device, its payload data of $B = 180$ bits is non-uniformly divided into $Q = 32$ fragments and the length of the $q$th fragment is $b_q$ satisfying $\sum_{q=1}^{Q} b_q = B$. 
To realize the coupling of different fragments, $a_q$ redundant parity check bits are added to the end of each fragment to form a sub-block with a fixed-length of ${\widetilde B} = b_q + a_q = 12$.
Considering the typical simulation setup in \cite{Ref_TIT_Amalladinne'20}, the parity profile ${\bf a} = [a_1, a_2, \cdots, a_Q]$ is set to ${\bf a} = [0, 6, \cdots, 6, 12, 12, 12]$.
Subsequently, the inner encoder maps each sub-block to a codeword of a common codebook with size $L \times 2^{\widetilde B}$, which will be transmitted over $L$ successive symbol durations.
Due to the absence of device identification, i.e., the permutation ambiguity is ignored, the widely adopted RA performance metrics including AER and BER are not available.
In this paper, the unsourced RA performance is evaluated in terms of per-user probability of error (PUPE), defined as the average fraction of transmitted messages not contained in the detected message list, i.e., 
\begin{equation}\label{Eq_metricURA}
	{\rm PUPE} = \frac{K_a - \left|{\cal M} - {\widehat {\cal M}}\right|_c} {K_a},
\end{equation}
where ${\widehat {\cal M}} = \left\{ [{\widehat {\bf B}}]_{k,{\cal B}_d} | k \in [K_a]\right\}$ and ${\cal M} = \left\{ [{\bf B}]_{k,{\cal B}_d} | k \in {\cal A}\right\}$ are the detected message list and the transmitted message list, respectively.

Fig. \ref{Fig_PUPE} examines the unsourced RA performance of the proposed semi-blind detection framework and the traditional common codebook-based non-coherent detection framework, where the codeword lengths $L \in \left[45, 50, 55\right]$ are considered and the corresponding payload efficiencies are indicated.
The payload efficiency is defined as the number of transmitted payload data bits to the total number of consumed symbol durations, which turns out to be
\begin{equation}\label{Eq_PE1}
	{\rm PE}_{\rm ncd} = \frac{B_d} {QL},
\end{equation}
for the traditional common codebook-based non-coherent detection and 
\begin{equation}\label{Eq_PE2}
	{\rm PE}_{\rm sbd} = \frac{B_d} {\lceil B/{\rm log}_2\left(M\right) \rceil + 1},
\end{equation} 
for the proposed semi-blind detection.
As shown in Fig. \ref{Fig_PUPE}, the proposed semi-blind detection framework significantly outperforms the baseline scheme and offers a much better payload efficiency.
This is because, in the baseline scheme, a large proportion of time resources are consumed to transmit the redundant parity check bits to guarantee the reliable stitching of the message fragments.
Indeed, even under a very low payload efficiency, e.g., ${\rm PE}_{\rm ncd} = 10.23\%$, the corresponding number of measurements, i.e., the codeword length $L$, is too small to achieve the reliable CS-based inner decoding for the baseline scheme.
Although the performance can be further improved by increasing the codeword length, the payload efficiency would continue to degrade, as in (\ref{Eq_PE1}).
While for the proposed detection framework, only a small proportion of time resources are consumed to transmit a $B_c$-bit CRC code and a scalar pilot symbol, leading to a payload efficiency over $90\%$.
Therefore, compared to the baseline scheme, the proposed semi-blind detection framework has a better tradeoff between the transmission latency and the detection reliability.   
{\color{red} Fig. 17 compares the user-plane latencies of the traditional non-coherent detection framework, the basic semi-blind detection framework, and the URLLC-enhanced semi-blind detection framework.
Here, the transmission latency of the traditional non-coherent detection framework is computed as
\vspace{-1.5mm}
\begin{equation} \label{Eq_delayNCD}
	d_{\rm ncd} = \frac{QL}{N_{\rm scs}},
	\vspace{-1.5mm}
\end{equation}
where $N_{\rm scs}$ is the subcarrier spacing.
While for the proposed unified detection framework, the transmission latency for unsourced RA is given as 
\begin{equation} \label{Eq_delaySBD}
	d_{\rm sbd} = \frac{\lceil(B_d + B_c)/{\rm log}_2(M)\rceil + 1}{N_{\rm scs} N_{\rm sc}}.
	\vspace{-1.5mm}
\end{equation}
The subcarrier spacing is set to 240 kHz for URLLC-enhanced scheme and 15 kHz for other schemes.
Compared with the traditional non-coherent detection framework, the proposed unified semi-blind detection framework significantly reduces the transmission latency and computational complexity thus leading to a much smaller user-plane latency.}

\section{Conclusion}
\label{Sec_conclusion}

This paper has proposed a unified semi-blind detection framework for grant-free sourced and unsourced RA, which effectively facilitates the massive URLLC in massive MIMO systems.
Under this framework, the system can flexibly switch to either sourced or unsourced RA mode according to the practical heterogeneous access requirements, making the network configuration more efficient and economical.
By leveraging the large spatial degrees-of-freedom offered by the massive MIMO BS, we have developed a BiG-AMP-based JCSE algorithm to jointly infer the channel and signal matrices, where a rank selection approach and a RI-aided initialization strategy are incorporated for the reduction of computational complexity and the improvement of estimation reliability, respectively.
Moreover, a small amount of RI is embedded in the access signal to eliminate the inherent phase and permutation ambiguities and the SIC technique has been introduced for further enhanced detection reliability.
Besides, the four enabling techniques have also been integrated to satisfy the stringent latency and reliability requirements of massive URLLC.
Numerical results have revealed that the proposed semi-blind detection framework offers a much better scalability-latency-reliability tradeoff than its counterparts dedicated to either sourced or unsourced RA, and thus it is more attractive for supporting massive URLLC. 


\begin{thebibliography}{99}
	
\bibitem{Ref_IWCMC_Ke'22}
M. Ke, Z. Gao, S. Tan {\em et al.},
``Massive MIMO-enabled semi-blind detection for grant-free massive connectivity,"
in \emph{Proc IEEE Int. Wireless Commun. Mobile Comput. (IWCMC)}, Dubrovnik, Croatia, July 2022, pp. 38-43.

%\bibitem{Ref_CS&T_Fuqaha'15}
%A. Al-Fuqaha, M. Guizani, M. Mohammadi, M. Aledhari, and M. Ayyash,
%``Internet of Things: A survey on enabling technologies, protocols, and applications,"
%\emph{IEEE Commun. Surveys Tuts.}, vol. 17, no. 4, pp. 2347-2376, 4th Quart., 2015.

%\bibitem{Ref_IOTJ_Chettri'20}
%L. Chettri and R. Bera,
%``A comprehensive survey on Internet of Things (IoT) toward 5G wireless systems,"
%\emph{IEEE Internet Things J.}, vol. 7, no. 1, pp. 16-32, Jan. 2020.

\bibitem{Ref_IOTJ_Nguyen'22}
D. C. Nguyen, M. Ding, P. N. Pathirana {\em et al.},
``6G Internet-of-Things: A comprehensive survey,"
\emph{IEEE Internet Things J.}, vol. 9, no. 1, pp. 359-383, Jan. 2022.

\bibitem{Ref_Tcom_Popovski'19}
P. Popovski, C. Stefanovic, J. N. Jimmy {\em et al.},
``Wireless access in ultra-reliable low-latency communication (URLLC),"
{\em IEEE Trans. Commun.}, vol. 67, no. 8, pp. 5783-5801, Aug. 2019.

\bibitem{Ref_Network_Saad'20}
W. Saad, M. Bennis, and M. Chen,
``A vision of 6G wireless systems: Applications, trends, technologies, and open research problems,"
{\em IEEE Network}, vol. 34, no. 3, pp. 134-142, Oct. 2019.

\bibitem{Ref_Access_Popovski'18}
P. Popovski, K. F. Trillingsgaard, O. Simeone, and G. Durisi,
``5G wireless network slicing for eMBB, URLLC, and mMTC: A communication-theoretic view,"
{\em IEEE Access}, vol. 6, pp. 55765-55779, Sept. 2019.

\bibitem{Ref_Access_Pokhrel'20}
S. R. Pokhrel, J. Ding, J. Park, O.-S Park, and J. Choi,
``Towards enabling critical mMTC: A review of URLLC within mMTC,"
\emph{IEEE Access}, vol. 8, pp. 131796-131813, Jul. 2020.
	
\bibitem{Ref_JSAC_Chen'21}
X. Chen, D. W. K. Ng, W. Yu, E. G. Larsson, N. Al-Dhahir, and R. Schober,
``Massive access for 5G and beyond,"
\emph{IEEE J. Sel. Areas Commun.}, vol. 39, no. 3, pp. 615-637, Mar. 2021.

\bibitem{Ref_WCM_Wu'20}
Y. Wu, X. Gao, S. Zhou, W. Yang, Y. Polyanskiy, and G. Caire,
``Massive access for future wireless communication systems,"
\emph{IEEE Wireless Commun.}, vol. 27, no. 4, pp. 148-156, Aug. 2020.	
	
\bibitem{Ref_CS&T_Laya'18}
A. Laya, L. Alonso, and J. Alonso-Zarate,
``Is the random access channel of LTE and LTE-A suitable for M2M communications? A survey of alternatives,"
\emph{IEEE Commun. Surveys Tuts.}, vol. 16, no. 1, pp. 4-16, 1st Quart., 2018.	

\bibitem{Ref_CL_Zhang'16}
Z. Zhang, X. Wang, Y. Zhang, and Y. Chen,
``Grant-free rateless multiple access: A novel massive access scheme for Internet-of-Things,"
\emph{IEEE Commun. Lett.}, vol. 20, no. 10, pp. 2019-2022, Oct. 2016.

\bibitem{Ref_IoTJ_Shao'19}
X. Shao, X. Chen, C. Zhong, J. Zhao, and Z. Zhang,
``A unified design of massive access for cellular Internet-of-Things,"
\emph{IEEE Internet Things J.}, vol. 6, no. 2, pp. 3934-3947, Apr. 2019.

\bibitem{Ref_TSP_Shao'20}
X. Shao, X. Chen, D. W. K. Ng, C. Zhong, and Z. Zhang,
``Cooperative activity detection: Sourced and unsourced massive random access paradigms,"
\emph{IEEE Trans. Signal Process.}, vol. 68, pp. 6578-6593, Nov. 2020.

\bibitem{Ref_WCM_Ke'21}
M. Ke, Z. Gao, Y. Huang, G. Ding, D. W. K. Ng, Q. Wu, J. Zhang,
``An edge computing paradigm for massive IoT connectivity over high-altitude platform networks,"
\emph{IEEE Wireless Commun.}, vol. 28, no. 5, pp. 102-109, Oct. 2021.

\bibitem{Ref_CL_Shim'12}
B. Shim and B. Song, 
``Multiuser detection via compressive sensing,” 
\emph{IEEE Commun. Lett.}, vol. 16, no. 7, pp. 972–974, July 2012.

%\bibitem{Ref_ISWCS_Schepker'13}
%H. F. Schepker, C. Bockelmann, and A. Dekorsy,
%``Exploiting sparsity in channel and data estimation for sporadic multi-user communication,"
%in \emph{Proc. IEEE Intern. Sympos. Wireless Commun. Systems (ISWCS)}, Ilmenau, Germany, Aug. 2013, pp. 1-5.

\bibitem{Ref_TSP_Liu'18}
L. Liu and W. Yu,
``Massive connectivity with massive MIMO-Part I: Device activity detection and channel estimation,"
\emph{IEEE Trans. Signal Process.}, vol. 66, no. 11, pp. 2933-2946, Jun. 2018.

\bibitem{Ref_TSP_Shao'20-2}
X. Shao, X. Chen, and R. Jia,
``A dimension reduction-based joint activity detection and channel estimation algorithm for massive access,"
\emph{IEEE Trans. Signal Process.}, vol. 68, pp. 420-435, Dec. 2020.

\bibitem{Ref_TSP_Ke'20}
M. Ke, Z. Gao, Y. Wu, X. Gao, and R. Schober,
``Compressive sensing-based adaptive active user detection and channel estimation: Massive access meets massive MIMO,"
\emph{IEEE Trans. Signal Process.}, vol. 68, pp. 764-779, Jan. 2020.

%\bibitem{Ref_JSAC_Ke'21}
%M. Ke, Z. Gao, Y. Wu, X. Gao, and K. K. Wong,
%``Massive access in cell-free massive MIMO-based Internet-of-Things: Cloud computing and edge computing paradigms,"
%\emph{IEEE J. Sel. Areas Commun.}, vol. 39, no. 3, pp. 756-772, Mar. 2021.

\bibitem{Ref_TSP_Gao'15}
Z. Gao, L. Dai, Z. Wang, and S. Chen,
``Spatially common sparsity based adaptive channel estimation and feedback for FDD massive MIMO,"
\emph{IEEE Trans. Signal Process.}, vol. 63, no. 23, pp. 6169-6183, Dec. 2015.

\bibitem{Ref_ISIT_Polyanskiy'17}
Y. Polyanskiy,
``A perspective on massive random-access,"
in \emph{Proc. IEEE Int. Symp. Inf. Theory (ISIT)}, Aachen, Germany, Jun. 2017, pp. 2523-2527.

\bibitem{Ref_ISIT_Ordentlich'17}
O. Ordentlich and Y. Polyanskiy,
``Low complexity schemes for the random access Gaussian channel,"
in \emph{Proc. IEEE Int. Symp. Inf. Theory (ISIT)}, Aachen, Germany, Jun. 2017, pp. 2528-2532.

\bibitem{Ref_Tcom_Vem'19}
A. Vem, K. R. Narayanan, J. F. Chamberland, and J. Cheng,
``A user-independent successive interference cancellation based coding scheme for the unsourced random access Gaussian channel,"
\emph{IEEE Trans. Commun.}, vol. 67, no. 12, pp. 8258-8272, Dec. 2019.

\bibitem{Ref_TIT_Amalladinne'20}
V. K. Amalladinne, A. Vem, D. K. Soma, K. R. Narayanan, and J. F. Chamberland,
``A coded compressed sensing scheme for unsourced multiple access,"
\emph{IEEE Trans. Inf. Theory}, vol. 66, no. 10, pp. 6509-6533, Oct. 2020.

\bibitem{Ref_arXiv_Fenler'19}
A. Fengler, G. Caire, P. Jung, and S. Haghighatshoar,
``Massive MIMO unsourced random access,"
[Online]. Available: https://arxiv. org/abs/1901.00828, Jan. 2019.

\bibitem{Ref_TIT_Fengler'21}
A. Fengler, S. Haghighatshoar, P. Jung, and G. Caire,
``Non-bayesian activity detection, large-scale fading coefficient estimation, and unsourced random access with a massive MIMO receiver,"
\emph{IEEE Trans. Inf. Theory}, vol. 67, no. 5, pp. 2925-2951, May 2021.

\bibitem{Ref_JSAC_Shyianov'21}
V. Shyianov, F. Bellili, A. Mezghani, and E. Hossain,
``Massive unsourced random access based on uncoupled compressive sensing: Another blessing of massive MIMO,"
\emph{IEEE J. Sel. Areas Commun.}, vol. 39, no. 3, pp. 820-834, Mar. 2021.

\bibitem{Ref_TSP_Parker'14}
J. T. Parker, P. Schniter, and V. Cevher,
``Bilinear generalized approximate message passing-Part I: Derivation,"
\emph{IEEE Trans. Signal Process.}, vol. 62, no. 22, pp. 5839-5853, Nov. 2014.	

\bibitem{Ref_IoTJ_Ding'22}
J. Ding, M. Nemati, S. R. Pokhrel, O.-S. Park, J. Choi, and F. Adachi,
``Enabling grant-free URLLC: An overview of principle and enhancements by massive MIMO,"
{\em IEEE Internet Things J.}, vol. 9, no. 1, pp. 384-400, Jan. 2022.
	
\bibitem{Ref_TSP_Parker'14-2}
J. T. Parker, P. Schniter, and V. Cevher,
``Bilinear generalized approximate message passing-Part II: Applications,"
\emph{IEEE Trans. Signal Process.}, vol. 62, no. 22, pp. 5854-5867, Nov. 2014.		

\bibitem{Ref_Access_Yan'19}
W. Yan and X. Yuan,
``Semi-blind channel-and-signal estimation for uplink massive MIMO with channel sparsity,"
\emph{IEEE Access}, vol. 7, pp. 95008-95020, July 2019.

\bibitem{Ref_Book_Wen'12}
Z. Wen, W. Yin, and Y. Zhang,
``Solving a low-rank factorization model for matrix completion by a nonlinear successive over-relaxation algorithm,"
\emph{Math. Programm. Comput.}, vol. 4, pp. 333-361, July 2012. [Online]. Available: http://dx.doi.org/10.1007/s12532-012-0044-1.

\bibitem{Ref_TIT_Frank'01}
F. R. Kschischang, B. J. Frey, and H-A. Loeliger,
``Factor graphs and the sum-product algorithm,"
\emph{IEEE Trans. Inform. Theory}, vol. 47, no. 2, pp. 498-519, Feb. 2001.

\bibitem{Ref_Book_Vaart'98}
A. W. Van der Vaart,
``Asymptotic Statistics,"
\emph{Cambridge University Press}, 1998.

\bibitem{Ref_ITW_Rangan'10}
D. L. Donoho, A. Maleki, and A. Montanari,
``Message passing algorithms for compressed sensing: I. Motivation and construction,"
in \emph{Proc. IEEE Inf. Theory Workshop. (ITW)}, Jan. 2010, pp. 1-5.

\bibitem{Ref_JRSC_Dempster'77}
A. Dempster, N. M. Laird, and D. B. Rubin,
``Maximum-likelihood from incomplete data via the EM algorithm,"
\emph{J. Roy. Statist. Soc.}, vol. 39, pp. 1-17, 1977.

\bibitem{Ref_Book_Neal'98}
R. M. Neal and G. E. Hinton,
``A view of the EM algorithm that justifies incremental, sparse, and other variants,"
in \emph{Learning in Graphical Models.}, Springer, 1998, pp. 355-368.

\bibitem{Ref_ICASSP_Ngo'12}
H. Q. Ngo and E. G. Larsson,
``EVD-based channel estimation in multicell multiuser MIMO systems with very large antenna arrays,"
in \emph{Proc. IEEE Int. Conf. Acoust., Speech, Signal Process.}, Mar. 2012, pp. 3249-3252.

\bibitem{Ref_Tcom_Xie'22}
X. Xie, Y. Wu, J. An {\em et al.} ,
``Massive unsourced random access: Exploiting angular domain sparsity,"
{\em IEEE Trans. Commun.}, vol. 70, no. 4, pp. 2480-2498, Apr. 2022.

\bibitem{Ref_ISWCS_Li'18}
Z. Li, M. A. Uusitalo, H. Shariatmadari, and B. Singh,
``5G URLLC: Design challenges and system concepts,"
in {\em Proc. 15th Int. Symp. Wireless Commun. Syst. (ISWCS)}, Lisbon, Portugal, Aug. 2018, pp. 1-6.

\bibitem{Ref_JMLR_Spielman'12}
D. A. Spielman, H. Wang, and J. Wright,  
``Exact recovery of sparsely-used dictionaries," 
in \emph{Proc. JMLR: Workshop Conf. Proc. 25th Annu. Conf. Learn. Theory}, 2012, vol. 23, pp. 37.1-37.18.
%[Online]. Available: http://jmlr.org/proceedings/papers/v23/spielman12/spielman12.pdf.

\bibitem{Ref_Tcom_Zhang'18}
J. Zhang, X. Yuan, and Y. J. A. Zhang,
``Blind signal detection in massive MIMO: Exploiting the channel sparsity,"
\emph{IEEE Trans. Commun.}, vol. 66, no. 2, pp. 700-712, Feb. 2018.

\bibitem{Ref_Tcom_Baicheva'08}
T. S. Baicheva,  
``Determination of the best CRC codes with up to 10-bit redundancy,"  
\emph{IEEE Trans. Commun.}, vol. 56, no. 8, pp. 1214-1220, Aug. 2008.

%\bibitem{Jabu_TVT'01}
%B. El-Jabu and R. Steele,
%``Cellular communications using aerial platforms,"
%\emph{IEEE Trans. Veh. Technol.}, vol. 50, no. 3, pp. 686-700, May 2001.
%
%\bibitem{Liu_TSP'19}
%H. Liu, X. Yuan, and Y. Zhang,
%``Super-resolution blind channel-and-signal estimation for massive MIMO with one-dimensional antenna array,"
%\emph{IEEE Trans. Signal Process.}, vol. 67, no. 17, pp. 4433-4448, Sept. 2019.
%
%\bibitem{Lin_CL'17}
%X. Lin, S. Wu, L. Kuang, Z. Ni, X. Meng, and C. Jiang,
%``Estimation of sparse massive MIMO-OFDM channels with approximately common support,"
%\emph{IEEE Commun. Lett.}, vol. 21, no. 5, pp. 1179-1182, May 2017.
%
%\bibitem{Lin_TWC'18}
%X. Lin, S. Wu, C. Jiang, L. Kuang, J. Yan, and L. Hanzo,
%``Estimation of broadband multiuser millimeter wave massive MIMO-OFDM channels by exploiting their sparse structure,"
%\emph{IEEE Trans. Wireless Commun.}, vol. 17, no. 6, pp. 3959-3973, Jun. 2018.
%
%\bibitem{Cover_TIT'67}
%T. M. Cover and P. E. Hart,
%``Nearest neighbor pattern classification,"
%\emph{IEEE Trans. Inf. Theory}, vol. 13, no. 1, pp. 21-27, 1967.

\end{thebibliography}
\end{document}